\RequirePackage{fix-cm}
\documentclass[fullpage]{article}       % onecolumn (second format)

\usepackage{graphicx}
\usepackage{authblk}
\let\vec\relax % nullify the horrible definition of \vec
\DeclareMathAccent{\vec}{\mathord}{letters}{"7E} % restore the standard

\usepackage{lineno}
\modulolinenumbers[1]
\usepackage[title]{appendix}
\usepackage{listings}
\usepackage{mathtools}
\usepackage{booktabs}
\usepackage{bm}
\usepackage[font=small,skip=0pt]{caption}
\usepackage{color}
\usepackage{enumitem}
\usepackage{float}
\usepackage{placeins}
\usepackage[top=1in,bottom=1in,right=1in,left=1in]{geometry}
\usepackage{graphicx}
\usepackage[hidelinks]{hyperref}
\usepackage{MnSymbol}
\usepackage{multirow}
\usepackage{siunitx}
\usepackage{subfig}
\usepackage{tikz}
\usepackage{todonotes}
\usepackage{verbatim}
\usepackage{wrapfig}

\graphicspath{{Images/}}
\newcommand{\CHRONO}{{\sffamily{{Chrono}}}}

\newcommand{\body}[1]{{b_{#1}}}
\newcommand{\tngPlane}[1]{{\tau_{#1}}}
\newcommand{\cPoint}[2]{{C_{#1}^{#2}}}
\newcommand{\prjCPoint}[2]{{{\bar C}_{#1}^{#2}}}
\newcommand{\nAxis}[1]{{\bf n}_{#1}}
\newcommand{\uAxis}[1]{{\bf u}_{#1}}
\newcommand{\wAxis}[1]{{\bf w}_{#1}}
\newcommand{\nAxTwo}[2]{{\bf n}_{#1}^{#2}}
\newcommand{\uAxTwo}[2]{{\bf u}_{#1}^{#2}}
\newcommand{\wAxTwo}[2]{{\bf w}_{#1}^{#2}}
\newcommand{\uBarAxis}[1]{{\bar{\bf u}}_{#1}}
\newcommand{\wBarAxis}[1]{{\bar{\bf w}}_{#1}}
\newcommand{\dirArc}[1]{{\vec{s}}_{#1}}

\DeclareMathOperator*{\argmax}{arg\,max}

\newcommand{\EVal}{{\bf S}_{ij}}
\newcommand{\dltaEVal}{{\Delta{\bf S}_{ij}}}
\newcommand{\EValSize}{\|{\bf S}_{ij}\|}
\newcommand{\SlideFForceE}{{\bf E}_{f}}
\newcommand{\SlideFForceD}{{\bf D}_{f}}
\newcommand{\SlideFForce}{{\bf F}_{f}}
\newcommand{\CurvRoll}{{\kappa}}
\newcommand{\RollFricTorque}{{\bm{\mathcal{T}}}}
\newcommand{\SlackSsTh}{{\mathcal{S}_s^S}}
\newcommand{\SlackSkTh}{{\mathcal{S}_k^S}}
\newcommand{\SlackS}{{\mathcal{S}^S}}

\newcommand{\SlackPsisTh}{{\mathcal{S}_s^\psi}}
\newcommand{\SlackPsikTh}{{\mathcal{S}_k^\psi}}

\newcommand{\SpinFunction}[2]{{\mathcal{F}^\psi_{#1}({#2})}}
\newcommand{\SpinFunctionDamping}[1]{{\mathcal{F}^\psi_D({#1})}}
\newcommand{\SpinFricTorque}{{\mathcal{T}}^\psi}
\newcommand{\SpinFricTorqueE}{{\mathcal{T}}^\psi_E}
\newcommand{\SpinFricTorqueD}{{\mathcal{T}}^\psi_D}
\renewcommand{\vec}[1]{\mathbf{#1}} % like Springer likes

\begin{document}
		
		\title{\LARGE Producing 3D Friction Loads by Tracking the Motion of the Contact Point on Bodies in Mutual Contact}

%		\author{Luning Fang         \and
%			Dan Negrut %etc.
%		}
	
%	\date{$^1$ University of Wisconsin-Madison, 1513 University Avenue, Madison, WI-53706}
		\author{\Large Luning Fang}
		\author{\Large Dan Negrut \thanks{corresponding author: negrut@wisc.edu}}
		\affil{\small Department of Mechanical Engineering \\
			University of Wisconsin-Madison\\
			1513 University Avenue, Madison, WI-53706}
		
%		\institute{Luning Fang \at
%			 \\
%			\email{lfang9@wisc.edu}           %  \\
%			\and
%			Dan Negrut \at
%			University of Wisconsin-Madison, 1513 University Avenue, Madison, WI-53706 \\
%			\email{negrut@wisc.edu}
%		}

		\maketitle

\begin{abstract}
	We outline a phenomenological model to assess friction at the interface between two bodies in mutual contact. Although the approach is general, the application inspiring the approach is the Discrete Element Method. The kinematics of the friction process is expressed in terms of the relative 3D motion of the contact point on the two surfaces in mutual contact. The model produces expressions for three friction loads: slide force, roll torque, and spin torque. The cornerstone of the methodology is the process of tracking the evolution/path of the contact point on the surface of the two bodies in mutual contact. The salient attribute of the model lies with its ability to simultaneously compute, in a 3D setup, the slide, roll, and spin friction loads for smooth bodies of arbitrary geometry while accounting for both static and kinematic friction coefficients.
\end{abstract}

	%\begin{keyword}
	%    slide friction \sep roll friction \sep spin friction \sep contact history
	%    \MSC[2010] 74A55 \sep  70F40 \sep 74M10
	%\end{keyword}

	%\linenumbers
	
	% ====================================================
	\section{Introduction}
	\label{sec:problemsCurrent}
	This contribution proposes a phenomenological model to compute friction loads acting between two bodies that experience both stick and slip as part of their anticipated mode of operation. In motivating this effort, we look beyond systems engineered to operate in stick mode only. For a recent overview that focuses on stick-only regimes, see \cite{elastoPlasticCntctMdls2017}. Therein, the interest is in preventing slip, which is regarded as a sign of design failure. 
	
	Herein, the interest is in applications for which the relative motion between two bodies is not a sign of failure, but a normal part of the system's dynamics. As such, one needs to be able to assess the friction loads at zero relative velocity (stick mode) as well as when there is relative motion between the bodies (slip mode). Generally, at a mutual contact point between two bodies neither the normal force $N$ nor the tangential, friction force $F_f$ are known and they have to be computed as part of the numerical solution. 
	The focus of this contribution is on computing $F_f$ as well as its direction in the tangent plane under a dry friction assumption. Herein, $N$ is considered known.

	Several aspects make the task of producing a friction model challenging: ($i$) handling 3D geometries is difficult, particularly so if the shapes are non-trivial. In many cases, e.g. \cite{cundall79,kruggelEmdenDEM08,iwashitaRolling1998}, the discussion is carried out in the context of 2D geometries, which eschews kinematic challenges associated with 3D friction, particularly handling changes in orientation for the bodies in contact; ($ii$) it is difficult to handle the set-valued function nature of the friction loads. In the presence of macro-scale sliding between the bodies, the friction force is expected to assume the value $\mu_k N$. However, in stick mode, the friction force can assume any value between $0$ and $\mu_s N$. The fact that there are two friction coefficients $\mu_s$ and $\mu_k$; i.e., static and kinetic, is often times ignored in practice; and, ($iii$) in 3D dynamics, spin-friction and roll-friction are often disregarded. They typically account for less energy dissipation than normally associated with slide-friction. Yet there are scenarios; i.e., spin tops, tippie tops, bowling balls, etc., when roll-friction and/or spin-friction play an important roll in the overall dynamics. 
	
	This contribution seeks to establish a {\textit{phenomenological}} friction model that addresses ($i$) through ($iii$). To that end, section~\ref{sec:preamble} introduces notions of contact kinematics, subsequently used in section~\ref{sec:frictionModel}, where models for slide, roll, and spin friction are outlined. Numerical experiments are discussed in section~\ref{sec:modelTesting}. The contribution concludes by highlighting key model features and outlining directions of future work. To maintain a flow in the presentation, several pieces of information were moved to the Appendix section. Thus, the review of literature in done in Appendix~\ref{sec:discussion}, once this model and the required notation have been introduced. Appendices \ref{sec:appendix_global_frame} through \ref{sec:appendix_sphere_incline} contain technical details related to the model discussed and numerical experiments.

	% ====================================================
	\section{Preamble: Kinematics Aspects}
	\label{sec:preamble}
	\subsection{The Kinematics of the Contact Point}
	\label{subsec:contactKinematics}
	Let ${\body{i}}$ and ${\body{j}}$ be two bodies in mutual contact. Although not called for by the model presented here, to remove any notational ambiguity, we assume that ${\body{i}}$ and ${\body{j}}$ have only one contact point in common and that the bodies are different ($i<j$). At time $t_0$ let ${\cPoint{i}{0}}$ be the contact point on the surface of ${\body{i}}$ (imagine a cross painted on body $\body{i}$'s surface) and ${\cPoint{j}{0}}$ be the contact point on the surface of ${\body{j}}$ (imagine a dot painted on body $\body{j}$'s surface). At time $t_0$ the cross and the dot coincide; the focus in this section is on how the cross and dot change their locations on the surface of the two bodies in mutual contact as time advances from $t_0$ to $t_1$. At time $t_1$, the new contact point between the two bodies is registered as ${\cPoint{i}{1}}$ (the new cross) and ${\cPoint{j}{1}}$ (the new dot), on bodies ${\body{i}}$ and ${\body{j}}$, respectively.
	
	%At each mutual contact point, a unit normal ${\nAxis{0}}$ is defined as acting from ${\body{j}}$ to ${\body{i}}$ (the dependency of ${\nAxis{0}}$ on $i$ and $j$ was dropped for convenience). For instance, if spheres 2 and 6 are in contact, the unit normal would be located at the point of contact between the spheres and pointing to the center of sphere 2 (as the body of smaller index). At the contact point one can define a tangent plane that passes through ${\cPoint{i}{0}}$ and is perpendicular to ${\nAxis{0}}$. We assume that at $t_0$ two mutually orthogonal unit vectors ${\uAxis{0}}$ and ${\wAxis{0}}$ span the tangent contact plane.
	
\begin{figure}[!ht]
		\vspace{-5pt}
		\centering
		\subfloat[]{{\includegraphics[width=1.3in]{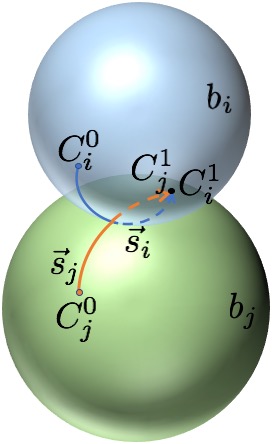} }} 
		\quad \quad`
		\subfloat[]{{\includegraphics[width=1.3in]{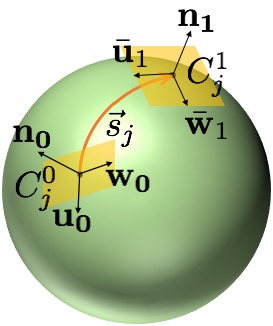} }}%
		\caption{Several kinematic quantities that come into play when the contact point moves on the surfaces of $\body{i}$ and $\body{j}$ as the time advances from $t_0$ to $t_1$. The length of the arcs ${\dirArc{i}}$ and ${\dirArc{j}}$ exaggerated; in simulation ${\cPoint{j}{0}}$ and ${\cPoint{j}{1}}$ are close to each other.}
		\label{fig:contactKinematic}
		\vspace{-10pt}
	\end{figure}

	In what follows, the times $t_0$ and $t_1$ are assumed close to each other and separated by a small simulation step size $\Delta t$ as in $t_1 = t_0 + \Delta t$. Typically, $\Delta t=$\numrange[range-phrase = --]{e-6}{e-3}\SI{}{s}. As far as body $i$ is concerned, during time integration from $t_0$ to $t_1$, we only register the locations ${\cPoint{i}{0}}$ and ${\cPoint{i}{1}}$. Thus, in scenarios in which ${\cPoint{i}{0}}\neq{\cPoint{i}{1}}$, there is no information regarding the trajectory followed by the contact point on the surface of ${\body{i}}$ while moving between ${\cPoint{i}{0}}$ and ${\cPoint{i}{1}}$. When ${\cPoint{i}{0}} \neq {\cPoint{i}{1}}$, we assume that the contact point moves on the surface of ${\body{i}}$ along the geodesic, although this will be subsequently relaxed. Tracing this contact point on the surface describes a ${\textit{directed arc}}$, from ${\cPoint{i}{0}}$ to ${\cPoint{i}{1}}$, which is called ${\dirArc{i}}$, see Fig.~\ref{fig:contactKinematic}. Analogously, there is an arc ${\dirArc{j}}$ defined by the trajectory of the contact point when moving on the surface of ${\body{j}}$ while the time passes from $t_0$ to $t_1$. 
	
	During one time step $\Delta t$, given its short duration, the motion of the contact point on $\body{i}$ and $ \body{j} $ allows for the introduction of three kinematic quantities. First, there is {\textit{relative spin}}, in which body ${\body{j}}$ has a spinning motion relative to ${\body{i}}$ around the instantaneous contact normal ${\nAxis{}}$. Second, there is {\textit{relative roll}} experienced by the two bodies. In pure rolling, one point on ${\body{i}}$ touches exactly one point on ${\body{j}}$, and vice-versa. In other words, the length of ${\dirArc{i}}$ is identical to that of ${\dirArc{j}}$. During one time step, this rolling motion is associated with a curvature $\kappa_i$ and $\kappa_j$, for ${\body{i}}$ and ${\body{j}}$, respectively. 
	%Note that for spheres, these curvatures are tied to the radii as in $\kappa_i = R^{-1}_i$ and $\kappa_j = R^{-1}_j$. 
	Finally, if there is a difference between the lengths of ${\dirArc{i}}$ and ${\dirArc{j}}$, this value is taken to be the {\textit{relative slip}}:
	\begin{equation*}
	s = | \|\dirArc{i}\| - \|\dirArc{j}\| |\; .
	\end{equation*}
	
	\subsection{Contact Frame Related Issues}
	\label{subsec:contactFrameIssues}
	All reference frames considered in this discussion obey the right hand-rule. The normal ${\bf n}$ of the contact reference frame is determined under the assumption that one can construct a tangent plane at the contact point, which is the case for any smooth surface \cite{geometricModeling2006}. Then, the unit vector ${\bf n}$ is chosen to be perpendicular to the tangent plane at the contact point. Its direction is chosen to point towards the interior of $\body{i}$, see Fig.~\ref{fig:contactKinematic}. Producing the ${\bf u}$ and ${\bf w}$ unit vectors relies on a continuation approach. To this end, at the onset of the $\body{i}$-$\body{j}$ contact, ${\uAxis{0}}$ and ${\wAxis{0}}$ are chosen arbitrarily so that $({\nAxis{0}},{\uAxis{0}},{\wAxis{0}})$ makes up a proper reference frame with three mutually orthonormal vectors. While the representation of the unit vectors ${\nAxis{0}}$, ${\uAxis{0}}$, and ${\wAxis{0}}$ in the {\textit{local}} reference frame associated with body $i$ does not change from $t_0$ to $t_1$, their representation in the {\textit{global}} reference frame can change. The latter is denoted by $({\nAxis{0}},{\uAxis{0}},{\wAxis{0}})^{i,1}$, with ``1'' added to emphasize that this representation in the global reference frame is done at $t_1$. Similarly, one has $({\nAxis{0}},{\uAxis{0}},{\wAxis{0}})^{j,1}$, which leads to the following notation for the associated axes: ${\nAxTwo{0}{j,1}}$, ${\uAxTwo{0}{j,1}}$ and ${\wAxTwo{0}{j,1}}$.

	The contact reference frame $({\nAxis{1}},{\uAxis{1}},{\wAxis{1}})$, expressed in the global reference frame at $t_1$ is located at point ${\cPoint{i}{1}}$ on $i$ and ${\cPoint{j}{1}}$ on $j$, and is defined as follows: ${\nAxis{1}}$ is the unit normal at the point of contact, pointing towards the interior of body $i$ (lower index body). The unit vectors ${\uAxis{1}}$ and ${\wAxis{1}}$ are chosen to belong to the tangent plane at ${\cPoint{i}{1}}$, be mutually orthogonal, and maximize the linear cost function $\uAxTwo{0}{i,1}\cdot{\uAxis{}}+\wAxTwo{0}{i,1}\cdot{\wAxis{}}$; i.e., $({\uAxis{1}},{\wAxis{1}})=\argmax_{({\bf u},{\bf w})}(\uAxTwo{0}{i,1} \cdot {\bf u} + \wAxTwo{0}{i,1}\cdot {\bf w})$. This is a convex optimization problem that has a global and unique solution (see Appendix~\ref{sec:appendix_global_frame}). A similar approach is followed to determine $({\nAxis{1}},{\uBarAxis{1}},{\wBarAxis{1}})$ at ${\cPoint{j}{1}}$; i.e.,  $({\uBarAxis{1}},{\wBarAxis{1}})=\argmax_{({\bf u},{\bf w})}(\uAxTwo{0}{j,1} \cdot {\bf u} + \wAxTwo{0}{j,1}\cdot {\bf w})$. The smallest rotation angle $\psi$ required to align $({\nAxis{1}},{\uBarAxis{1}},{\wBarAxis{1}})$, which is tied to ${\body{j}}$, and $({\nAxis{1}},{\uAxis{1}},{\wAxis{1}})$, which is tied to ${\body{i}}$, is called the spin angle. By convention, $\psi$ is considered positive if getting the former reference frame over the latter reference frame following the shortest way calls for a counterclockwise rotation.  
	
	Note that for a spinning top, in an idealized ``spinning-in-place'' situation, the contact points ${\cPoint{i}{0}}$ and ${\cPoint{i}{1}}$ (on the ground), and ${\cPoint{j}{0}}$ and ${\cPoint{j}{1}}$ (on the spinning top) would coincide. The spin angle $\psi$ in this scenario is the rotation angle of the spinning top as measured after one time step $\Delta t$; i.e., the shortest rotation that takes ${\uBarAxis{1}}$ over ${\uAxis{1}}$ (see also Fig.~\ref{fig:pipj}). 
	
	%A second issue related to the contact frame pertains a frame-duplication procedure. The latter refers to  producing at {\cPoint{i}{1}} a helper frame that is a close replica of the old frame from $t_0$. In other words, given $(${\nAxis{i,j}{0}},{\uAxis{i,j}{0}},{\wAxis{i,j}{0}}$)$, we are interested in duplicating this frame from {\cPoint{i}{0}} to the new contact point {\cPoint{i}{1}} to produce a new orthonormal reference frame $(${\nAxis{i,j}{1}},{\uBarAxis{i,j}{0}},{\wBarAxis{i,j}{0}}$)$ that is in some sense a close replica, if possible a clone, of $(${\nAxis{i,j}{0}},{\uAxis{i,j}{0}},{\wAxis{i,j}{0}}$)$. Given that the helper reference frame share the normal with the nominal reference frame $(${\nAxis{i,j}{1}},{\uAxis{i,j}{1}},{\wAxis{i,j}{1}}$)$ at {\cPoint{i}{1}}, it is not always possible to have in the helper reference frame $(${\nAxis{i,j}{1}},{\uBarAxis{i,j}{0}},{\wBarAxis{i,j}{0}}$)$ a clone of $(${\nAxis{i,j}{0}},{\uAxis{i,j}{0}},{\wAxis{i,j}{0}}$)$. Yet there are cases in which this does happen, for instance scenarios in which {\cPoint{i}{0}}$=${\cPoint{i}{1}}.
	
	%=================================================
	%=================================================
	\section{The Friction Model}
	\label{sec:frictionModel}
	The model proposed is not derived from first principles; instead, it is phenomenological. It is informed by insights into micro-scale phenomena and anchored by three assumptions: small displacements, decoupled dissipation mechanisms, and governing of the stick mode by micro-scale elasticity only.

	First, given that $\Delta t$ is small, e.g., $10^{-6}$ to $10^{-3}$ s, the model builds off a small relative displacements assumption; i.e., small translations and small rotations. This opens the door for relative rotations to be treated like vectors. Note that the overall motion of any body ${\body{i}}$ may experience large translations and large rotations with respect to the global reference frame. Yet, when two bodies are in contact, the integration time step is small enough to render the {\textit{relative}} displacement manifest between bodies ${\body{j}}$ and ${\body{i}}$ small.
	
	The second assumption embraced is that the energy dissipation through friction takes place via three {\textit{decoupled}} mechanisms: relative slip of the two surfaces, relative spin, and rolling. Although in many cases there is an interplay between these three dissipation mechanisms, here they are considered decoupled. 
	%Typically, although not always, the amount of energy dissipated through these mechanisms is such that if slip is present, the energy dissipated by both the spin- and roll-friction is smaller than the energy dissipated through slip-friction, particularly when bodies ${\body{i}}$ and ${\body{j}}$ are very stiff. Indeed, consider the case of a bowling ball in three scenarios: pure slide, pure spin, and pure roll. In all scenarios there is no force except the weight of the ball; the ground is assumed quasi-rigid. Initiating motion in these three scenarios under the presumption of equal kinetic energy will see the ball come to a rest after times $T_{slide}$, $T_{spin}$ and $T_{roll}$. By and large, $T_{slide} \ll T_{roll}, T_{spin}$. However, particularly in granular dynamics, under certain loading conditions, rolling rather than sliding is a dominant energy dissipation micro-mechanism \cite{rollingFrictionOda2000} and as such rolling becomes central to the overall dynamics of the system. Such a scenario is encountered, for instance, in shear bands \cite{iwashitaRolling1998}.
	
	The third assumption pertains the governing of the stick mode by micro-scale elasticity only. The Coulomb friction model caps the friction force as in $F_t \leq \mu_s N$. Herein, what is capped by $\mu_s N$ is not the entire friction force, but rather the component that is produced by micro-scale elastic deformation, called herein micro-deformation. Any contribution associated with a damping component, while included in the value of $F_t$, is ignored in deciding whether yielding (onset of the slip mode) takes place.
	
	Beyond these three assumptions, the model is informed by Amontons' laws (slide friction force is directly proportional to applied load when in slip mode; friction force is independent of apparent area of contact) \cite{amontons1699} and Coulomb's law (kinetic friction is independent of the slide velocity) \cite{coulomb1821}. 
	
	The remainder of this section concentrates on computing slide, roll, and spin friction loads at $t_1$ under the presumption that the normal load $N$ is known. Also known is the kinematic information at $t_0$. This is always the case, except at the onset of contact when there is no ``history'' to draw on and the friction loads are automatically set to zero. 
	
	\subsection{The Slide-friction Force Model}
	\label{subsec:slidingFriction}
\begin{figure}[!ht]
		\begin{center}
			\includegraphics[width=1.5in]{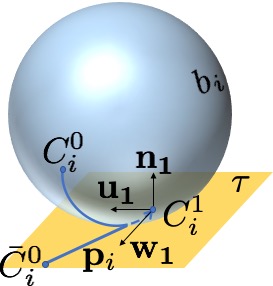}
		\end{center}
		\caption{The projection process that leads to ${\prjCPoint{i}{0}}$.}
		\label{fig:projProcess}
	\end{figure}

	Assume that friction is present at the interface, and $\mu_s \ge \mu_k > 0$. To start off, at $t_1$, ${\cPoint{i}{0}}$ is projected onto the contact tangent plane ${\tngPlane{}}$ spanned by ${\uAxis{1}}$ and ${\wAxis{1}}$. The unique point is called ${\prjCPoint{i}{0}}$ and is defined as the closest point in ${\tngPlane{}}$ to ${\cPoint{i}{0}}$ that has the property that the length of the vector from ${\prjCPoint{i}{0}}$ to ${\cPoint{i}{0}}$ is equal to the length of the arc ${\dirArc{i}}$, see Fig.~\ref{fig:projProcess}. A similar projection is used to determine ${\prjCPoint{j}{0}}$. Two vectors are defined in the tangent plane at $t_1$: one from ${\prjCPoint{i}{0}}$ to ${\cPoint{i}{1}}$, called ${\bf p}_i$; and ${\bf p}_j$, defined from ${\prjCPoint{j}{0}}$ to ${\cPoint{j}{1}}$. 
	
\begin{figure}[!ht]
	\begin{center}
		\includegraphics[width=3.34in]{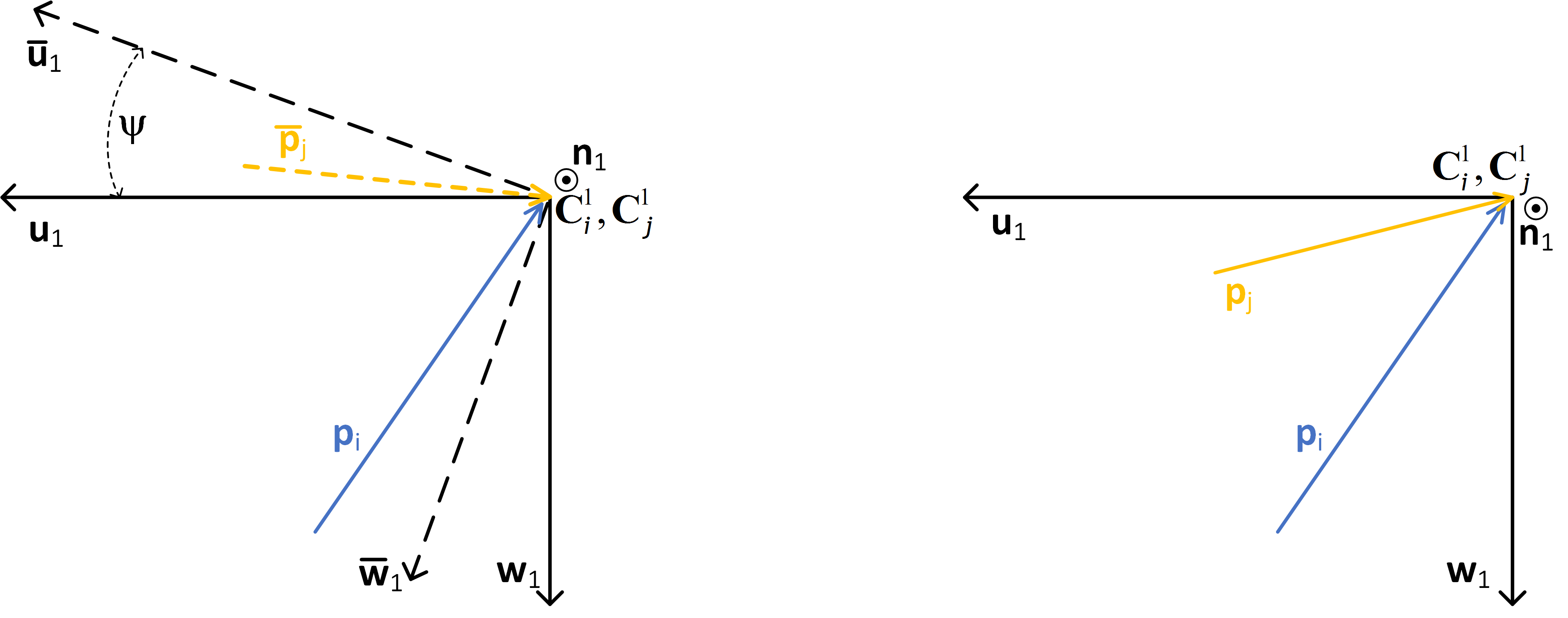}
	\end{center}
	\caption{The process of realigning the $({\nAxis{1}},{\uBarAxis{1}},{\wBarAxis{1}})$ to the $({\nAxis{1}},{\uAxis{1}},{\wAxis{1}})$ reference frames. Note that the realigning process takes ${\bar {\bf p}}_j$ into ${\bf p}_j$ and that the angle between ${\bar {\bf p}}_j$ and ${\uBarAxis{1}}$ is equal to the angle between ${\bf p}_j$ and ${\uAxis{1}}$. The image also illustrates the spin angle $\psi$. In the example shown, $({\nAxis{1}},{\uBarAxis{1}},{\wBarAxis{1}})$ undergoes a counterclockwise rotation of angle $\psi$ to overlap $({\nAxis{1}},{\uBarAxis{1}},{\wBarAxis{1}})$.}
	\label{fig:pipj}
\end{figure}

	The slide-friction force computation proceeds by aligning the $j$ contact reference frame $({\nAxis{1}},{\uBarAxis{1}},{\wBarAxis{1}})$ to the $i$ contact reference frame $({\nAxis{1}},{\uAxis{1}},{\wAxis{1}})$ by a rotation of angle $\psi$ of the $({\nAxis{1}},{\uBarAxis{1}},{\wBarAxis{1}})$ reference frame; i.e., by applying first a ``counter-spin'' step, see Fig.~\ref{fig:pipj}. At this point the two contact reference frames coincide thus providing the contact reference frame for the next time step and also the setup for computing the slide-friction force in the configuration at $t_1$.
	
	As an example for how ${\bf p}_i$ and ${\bf p}_j$ are defined, consider a sphere moving on a rigid horizontal surface; assume ${\body{i}}$ is the ground and ${\body{j}}$ is the sphere. One scenario in which ${\bf p}_i = {\bf p}_j$ is when the sphere's center moves in a vertical plane ${\mathcal{V}}$ rolling without slip on the horizontal surface with an angular velocity perpendicular to ${\mathcal{V}}$. In another case, consider the sphere being dragged, with zero angular velocity, over the rigid surface with its center moving in a straight vertical plane ${\mathcal{V}}$. This latter scenario is similar to a vehicle with its wheels locked as a result of vigorous braking: the vehicle still moves forward (slides) although the wheels don't rotate at all. Finally, one has the case of a sphere that rotates in place; zero translational velocity, nonzero angular velocity. In the vehicle analogy, this represents the case when the wheels rotate in place and the vehicle doesn't move forward (vehicle stuck in snow, for instance). For the ``locked sphere/wheels'' example $\| {\bf p}_j \|=0$, while $\| {\bf p}_i\| >0$. In the second case (sphere spinning in place), $\| {\bf p}_i \|=0$, while $\| {\bf p}_j\| >0$. 
	The slide-friction force has two components. One is incremental in nature and has a saturation value; in that sense, it has memory. The other has no memory and its value depends on the rate of relative, micro-scale sliding. In going from $t_0$ to $t_1$, referring to Fig.~\ref{fig:pipj}, a variable $\EVal$ is updated as 
	\begin{subequations}
		\begin{equation}
		\EVal = \EVal +  \dltaEVal \; , \qquad \dltaEVal \equiv {\bf p}_i - {\bf p}_j \; .
		\end{equation}
		The elastic component of the slide-friction force is then computed as
		\begin{equation*}
		\SlideFForceE = K_E \cdot \EVal \; .
		\end{equation*}
		One must ensure that the elastic component of the slide-friction force is $\mu N$-capped, a step that relies on two quantities: the current ``slide-mode'' and the ``slide-microdeflection''. Thus, a slide-mode variable $s_m$ is established to assume one of two values: static $s$ (stick mode); or kinetic $k$ (slip mode). The slide-microdeflection is defined as $\SlackS = \EValSize$. Its value is capped in stick slide-mode by $\SlackSsTh \equiv \mu_s N/K_E$; in slip slide-mode it is capped by $\SlackSkTh \equiv \mu_k N/K_E$. In terms of nomenclature, $\SlackSsTh$ and $\SlackSkTh$ are called the static and, respectively, kinetic, threshold slide-microdeflections. The capping is enforced as $0 \leq \SlackS \leq \SlackSsTh$ in stick mode, and $0 \leq \SlackS \leq \SlackSkTh$ in slip mode. Importantly, these two sets of inequalities are brought into play at $t_1$; i.e., at the end of the time step once the ``new state''(positions, velocities, and $N$) has been established. At that point, we adjust the slide-microdeflection $\SlackS$ and update the slide mode $s_m$ as follows:
		
		\begin{description}[itemsep=-0.5ex]
			\item [If in the ``stick case'' ($s_m==s$):] Compute the scaling factor $\alpha_s = \SlackS/\SlackSsTh$. If $\alpha_s>1$, then set $\EVal = \EVal/\alpha_s$ and $s_m = k$.    
			\item [If in the ``slip case'' ($s_m==k$):] Compute the scaling factor $\alpha_k = \SlackS/\SlackSkTh$. If $\alpha_k>1$, then $\EVal = \EVal/\alpha_k$. Else, set $s_m = s$.
		\end{description}
		The model proposed accommodates two friction coefficients -- one static and one kinetic. Additionally, it does not seek to enforce strict capping at $\mu N$; rather, it a-posteriorly adjusts $\EVal$.

		The damping component for the slide-friction force is added as
		\begin{equation*}
		\SlideFForceD = K_D \cdot \frac{\dltaEVal}{\Delta t}  \; .
		\end{equation*}
		The slide-friction force is then obtained as 
		\begin{equation}
		\label{eq:slideForceExpression}    
		\SlideFForce = \SlideFForceE + \SlideFForceD = K_E \EVal + \frac{K_D}{\Delta t}\dltaEVal \; .
		\end{equation}
		
		Selecting a value for $K_D$ is rarely discussed in the literature. The physical insight guiding its selection is that it represents a ``knob'' for controlling energy dissipation. The value of $K_D$ is typically regarded as a constant dictated by the materials in contact. As such, it is problem dependent and determined through a parameter identification process using experimental data. In the absence of experimental data, ad-hoc nonzero values are still used given that the ensuing energy dissipation leads to a desirable stabilization of the numerical solution by damping out the high frequency oscillations caused by the stiffness injected into the problem via $K_E$. When calibration data is missing, we suggest a $K_D$ ``default'' value that indirectly imparts a particular critical damping regime. Specifically, a fictitious mass $m_{ij}$ is chosen as the average of the $\body{i}$ and $\body{j}$ masses or some other convenient values (for instance, if $\body{i}$ is the ground, one can take $m_{ij} = m_j$). With this,
		\begin{equation}
		\label{eq:criticalDampingCoeff}
		K_D \equiv 2 \sqrt{m_{ij} K_E} \; .
		\end{equation}
	\end{subequations}
	%The resetting of the elastic component takes place as follows: if the slide friction force has not reach saturation and the elastic strain changes sign, the history is reset. In other words, if there is no relative sliding between the two bodies and the internal strain changes direction, the elastic force is reset and the slide friction will change sign thus not violating the second principle of thermodynamics. However, if the two bodies slide to each other, the elastic deformation is not reset upon the internal strain changing sign. Instead, first the sliding stops, which provides the opportunity for the static friction coefficient to come into play in the definition of the slide friction force.
	\subsection{The Roll-friction Model}
	\label{subsec:rollingFriction}
\begin{figure}[!ht]
	\begin{center}
		\includegraphics[width=2.5in]{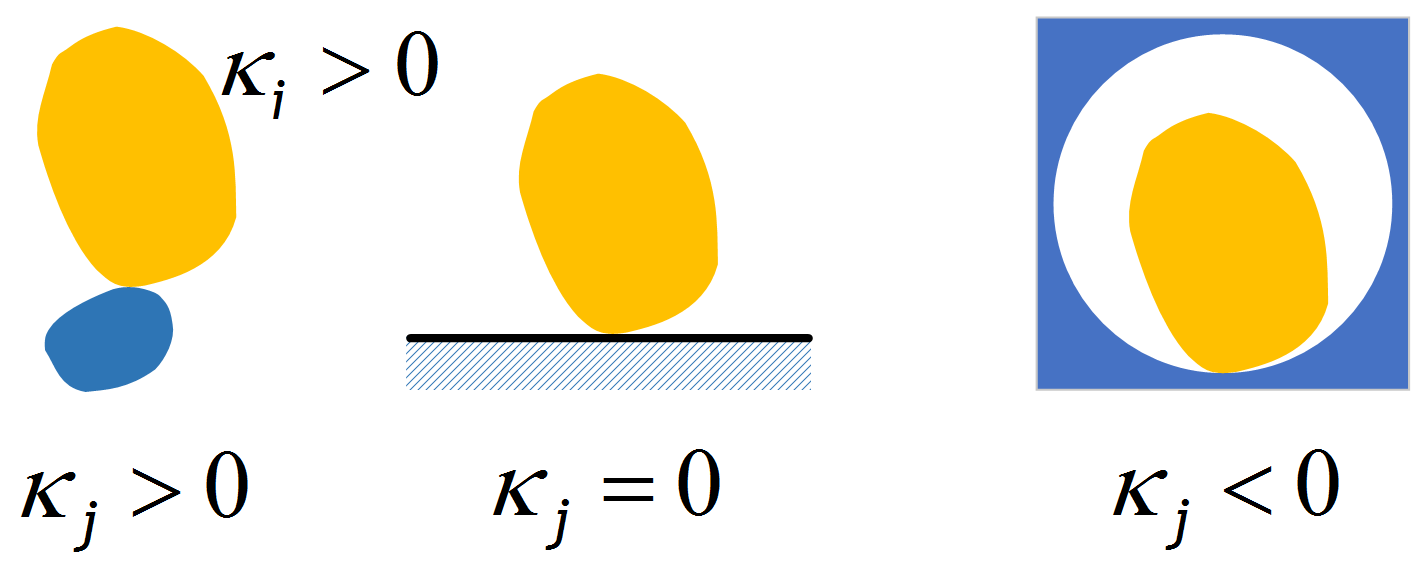}
	\end{center}
	\caption{Body $\body{i}$ on top, with $\CurvRoll_i>0$. Three scenarios shown for body $\body{j}$: positive, zero, and negative curvature.}
	\label{fig:rollingCases}
\end{figure}

	We assume that the rolling of ${\body{i}}$ takes place in the plane defined by the normal ${\nAxis{1}}$ and vector ${\bf p}_i$, and the rolling resistance torque is acting in this plane. The roll-friction model builds on the assumption that over one time step $\Delta t$ an average curvature $\kappa_i$ can be defined for ${\body{i}}$ along the ${\bf p}_i$ direction. Considering that a curvature can be positive or negative (see Fig.~\ref{fig:rollingCases}), a radius is associated with it and evaluated as $R_i = 1/{{|\kappa_i}|}$. A similar argument is made for body $\body{j}$ -- which leads to $\kappa_j$ and $R_j$. As far as body $\body{i}$ is concerned, its rolling motion over one time step $\Delta t$ is assumed 2D in nature; i.e., a 3D motion is represented as a sequence of 2D rolling stretches, each of duration $\Delta t$. Body $\body{i}$ of radius $R_i$ rolls over body $\body{j}$ of radius $R_j$ and in this process the contact point moves on $\body{i}$ a distance $\|{\bf p}_i\|$. Note that the rolling of ${\body{j}}$ takes place in the plane defined by the normal ${\nAxis{1}}$ and vector ${\bf p}_j$. Thus, the rolling resistance torques might not act in the same plane and might have different magnitudes.
	
	From here, the model follows in the steps of the slide-friction case: an elastic component is evaluated and capped to a threshold/saturation value; and, a damping mechanism is factored into the model and tied to the microdeflection rate of change. In this phenomenological model, a small forward-backward rocking of angle $\theta_i$ about the contact point leads to small excursions about this point. The angle and excursion are related as in
	\begin{subequations}
		\begin{equation*}
		\theta_i = \CurvRoll_i {\bf p}_i \; .
		\end{equation*}
		Subsequently, the rolling friction history, which is a vector quantity, is updated as in
		\begin{equation*}
		{\bm{\Theta}}_i = {\bm{\Theta}}_i + \CurvRoll_i {\bf p}_i.
		\end{equation*}
		Finally, the rolling friction torque is evaluated as 
		\begin{equation*}
		\RollFricTorque_{E,i} = K_{R,i} \: {\bm{\Theta}}_i \; .
		\end{equation*}
		The roll-friction is capped using the approach employed for the slide-friction. This involves a ``roll-mode'' variable $r_m$, which assumes one of two values: static $s$ (stick mode), or kinetic $k$ (slip mode). The roll-mode is kept constant during one time step; the $r_m$ value doesn't change while ``in flight,'' from $t_0$ to $t_1$. After accepting the new state at $t_1$, to prep the next time step, one would go through the following process to update the roll state $r_m$ and adjust the microdeflection $\Theta_i$:

		\begin{description}[itemsep=-0.5ex]
			\item [If in the ``stick case'' ($r_m==s$):] Compute the scaling factor $\alpha_s = {\Theta_i}/{\Theta_i^s}$. If $\alpha_s>1$, then set $\RollFricTorque_{E,i} = \RollFricTorque_{E,i}/{\alpha_s}$, $\Theta_i = \Theta^s_i$ and $r_m = k$.
			\item [If in the ``slip case'' ($r_m==k$):] Compute the scaling factor $\alpha_k = {\Theta_i}/{\Theta_i^d}$. If $\alpha_k>1$, then $\RollFricTorque_{E,i} = \RollFricTorque_{E,i}/{\alpha_k}$ and $\Theta_i = \Theta^d_i$. Else, set $r_m = s$.
		\end{description}
		
		There is little to draw on when choosing stiffnesses for roll-mode friction, see, for instance \cite{iwashitaRolling1998}. We suggest two strategies when it comes to selecting the stiffness $K_{R}$ and threshold roll-microdeflections $\Theta^s$ and $\Theta^d$. One could compare results to experimental data to produce best fits for these values. This can be challenging -- consider the case of sand grains when measuring directly the roll-friction torque is not straightforward. There are cases in which one can focus on macroscale behavior, for instance, the rolling to a stop of a ball/sphere. The distance required by the sphere to stop is easily measurable and can be used to tune the model parameters. Generalizing this methodology to non-trivial body shapes that have different curvatures at different locations of the body surface is not straightforward. A second approach to produce these values involves heuristics. Herein, they draw on an energy argument when determining $\Theta^s$ (for convenience, the subscript $i$ is dropped; $\Theta^d$ is similarly evaluated). The main point is that the threshold-microdeflection potential energy for roll-friction is a fraction $\eta_e$ of the threshold-microdeflection potential energy for slide-friction:
		
		\begin{equation}
		\frac{1}{2}K_R (\Theta^s)^2 = \eta_e \frac{1}{2}K_E \left(\SlackSsTh\right)^2 \; ,
		\end{equation}
		where the subscript $e$ is chosen to highlight the energy tie. Relating the two microdeflections as in $\Theta^s = \eta_\theta \CurvRoll \SlackSsTh$ via an average curvature $\CurvRoll \equiv (\CurvRoll_i+\CurvRoll_j)/2$ and a scaling coefficient $\eta_\theta$ leads to 
		\begin{equation}
		\label{eq:K_R}
		K_R = \frac{4 \eta_r}{(\CurvRoll_i + \CurvRoll_j)^2} \; K_E \; ,
		\end{equation}
		where $\eta_r \equiv \eta_e/\eta_\theta^2$. Given that the slide threshold-deflection satisfies $K_E \SlackSsTh = \mu_s N$, one obtains
		\begin{equation}
		\label{eq:thetaMax}
		\Theta^s = \frac{\mu_s N}{K_E} \CurvRoll = \frac{\mu_s N}{2 K_E} (\CurvRoll_i + \CurvRoll_j) \; .
		\end{equation}
		Table \ref{table:rollingRelated} reports sample values for $K_{R}$ and $\Theta^\blacksquare_i$ when body $\body{j}$ assumes several representative cases. Note that $\CurvRoll_j$ cannot drop below $-\CurvRoll_i$; i.e., $\CurvRoll_j \in [-\CurvRoll_i, \: \infty)$. 
\begin{table}[H]
	\centering
	{\renewcommand{\arraystretch}{1.2}
		\caption{Various values of the body $\body{j}$ curvature $\CurvRoll_j$ lead to different rolling friction parameters $K_R$ and $\Theta^\blacksquare_i$ for body $\body{i}$, see Eqn.~(\ref{eq:K_R}) and Eqn.~(\ref{eq:thetaMax}). For $r_m==s$, replace the black square $\blacksquare$ with $s$; for $r_m==d$, replace it with $d$.}
		\label{table:rollingRelated}    	
		\begin{tabular}{cccccc}
			\toprule
			{} &
			\multicolumn{1}{c}{$\CurvRoll_j$} &
			\multicolumn{1}{c}{$K_{R}$} &
			\multicolumn{3}{c}{$\Theta^\blacksquare_i$} \\
			\midrule
			Tip of pin & $\infty$ & 0 & $\infty$ & $\infty$ & $\infty$ \\[3pt]
			Identical  & $\CurvRoll_i$ & $\eta_r R_i^2 K_E$ & $\frac{\mu_\blacksquare N}{R_i K_E}$ & $\frac{{\mathcal{S}^S_\blacksquare}}{R_i}$ & $\eta_r \frac{\mu_\blacksquare R_i}{K_R} N$ \\
			Plane & 0 & $4 \eta_r R_i^2 K_E$ & $\frac{\mu_\blacksquare N}{2 R_i K_E}$ & $\frac{{\mathcal{S}^S_\blacksquare}}{2 R_i}$ & $2\eta_r \frac{\mu_\blacksquare R_i}{K_R} N$ \\
			Mirror & $-\CurvRoll_i$ & $\infty$ & 0 & 0 & 0 \\
			\bottomrule
	\end{tabular}}
	\vspace{10pt}
\end{table}
		
		A roll-friction damping torque can be added to the model,
		\begin{equation*}
		\RollFricTorque_{D,i} = D_{R} \frac{\theta_i}{\Delta t},
		\end{equation*}
		which requires a damping coefficient $D_R$. Given the lack of an established procedure to select $D_R$, herein one is determined by imitating a critical damping response, which yields, similarly to Eqn.~(\ref{eq:criticalDampingCoeff}),
		\begin{equation*}
		D_R = 2 \sqrt{I_{ij}K_R}.
		\end{equation*}
		With this, the roll-slide torque on body $\body{i}$ is computed as
		\begin{equation*}
		\RollFricTorque_{i} = \RollFricTorque_{E,i} + \RollFricTorque_{D,i} \;.
		\end{equation*}
	\end{subequations}
	
	%=================================================
	\subsection{The Spin-friction Model}
	\label{subsec:spinningFriction}
	The spin-friction torque has two components: one elastic and one dissipative. The model ensures that the elastic component of the spin-friction torque is capped, a step that relies on two quantities: the current ``spin mode'' and the ``spin-microdeflection''. The spin-mode variable $\psi_m$ assumes one of two values: static $s$ (stick mode), or kinetic $k$ (slip mode). The spin-microdeflection $\Psi_{ij}$ is updated at each time step as
	
	\begin{subequations}
		\label{subeq:spinFriction}
		\begin{equation*}
		%\label{eq:slackSpin}
		\Psi_{ij} = \Psi_{ij} + \psi_{ij} \; .
		\end{equation*}
		The microdeflection value is capped in stick spin-mode by $\SlackPsisTh$; in slip spin-mode it is capped by $\SlackPsikTh$. In terms of nomenclature, $\SlackPsisTh$ and $\SlackPsikTh$ are called the static and, respectively, kinetic, threshold spin-microdeflections. It follows that $0 \leq \Psi_{ij} \leq \SlackPsisTh$ in stick mode, and $0 \leq \Psi_{ij} \leq \SlackPsikTh$ in slip mode. A negative spin angle $\Psi_{ij}$ leads to a positive spin-friction torque acting on body $\body{i}$. Just like before, it is assumed that the spin-friction torques in stick and slip modes are evaluated using two suitably defined functions $\SpinFunction{s}{\Psi_{ij}}$ and $\SpinFunction{d}{\Psi_{ij}}$, respectively. Herein, they are taken to be the same and assume the expression
		\begin{equation*}
		%\label{eq:funcSpin}
		\SpinFunction{s}{\Psi_{ij}} = \SpinFunction{d}{\Psi_{ij}} \equiv \SpinFunction{}{\Psi_{ij}} = K_\psi \Psi_{ij} \; .
		\end{equation*}
		
		Note that the spin-mode $\psi_m$ between $\body{i}$ and $\body{j}$ is kept constant in advancing the solution from $t_0$ to $t_1$; i.e., $\psi_m$ doesn't change during one time step. After accepting the new state at $t_1$, to prep the next time step, one would go through the following process to update the $\psi_m$ state, adjust $\Psi_{ij}$, and set the spin-friction $\SpinFricTorque$ torque for the new time step:
		
		\begin{description}[itemsep=-0.5ex]
			\item [If in the ``stick case'' ($\psi_m==s$):] Compute the scaling factor $\alpha_s = \Psi_{ij}/\SlackPsisTh$. If  $\alpha_s>1$, then set $\Psi_{ij} = \SlackPsisTh$, $\SpinFricTorqueE = \SpinFunction{s}{\Psi_{ij}}$, and $\psi_m=k$.
			\item [If in the ``slip case'' ($\psi_m==k$):] Compute the scaling factor $\alpha_k = \Psi_{ij}/\SlackPsikTh$. If  $\alpha_k>1$, then set $\Psi_{ij} = \SlackPsikTh$ and $\SpinFricTorqueE = \SpinFunction{d}{\Psi_{ij}}$. Else, set $\psi_m=s$.
		\end{description}
		
		A damping term $\SpinFricTorqueD$ can be considered via a suitably defined function $\SpinFunctionDamping{\cdot}$, which here is taken as
		\begin{equation}
		%\label{eq:DampSpinFunc}
		\SpinFunctionDamping{\omega} = D_\psi \: \frac{\psi_{ij}}{\Delta t} \; .
		\end{equation}
		With this 
		\begin{equation}
		\label{eq:spinningTorque1}
		\SpinFricTorque = \SpinFricTorqueE + \SpinFricTorqueD = \SpinFricTorqueE + D_\psi \: \frac{\psi_{ij}}{\Delta t}\: .
		\end{equation}
		
	For closure, one needs to provide the threshold values $\SlackPsisTh$ and $\SlackPsikTh$; and coefficients $K_\psi$ and $D_\psi$ (or the corresponding functions $\SpinFunction{s}{\Psi_{ij}}$, $\SpinFunction{d}{\Psi_{ij}}$, and $\SpinFunctionDamping{\cdot}$). In our numerical experiments, the threshold microdeflection values have been set to small angle values, where $\mathcal{S}_s^{\psi} = \mathcal{K}\SlackSsTh$ and $\mathcal{S}_k^{\psi} = \mathcal{K}\SlackSkTh$. Rather than selecting $K_\psi$ directly, one can equivalently use a coefficient $\eta_\psi$ to tie the threshold spin potential energy to the threshold slide potential energy and thus get some guidance in selecting this quantity:
		\begin{equation}
		\label{eq:gettingKpsi}
		\frac{1}{2}K_\psi \left(\SlackPsisTh\right)^2 = \frac{1}{2}\eta_\psi K_E \left(\SlackSsTh\right)^2 
		\Rightarrow 
		K_\psi = \eta_\psi  \: K_E \; / \mathcal{K}^2 \;.
		\end{equation}
		%\SBELfeedbackONE{Luning: above equation is equivalent to, $K_\psi = \eta_{\psi}   K_E / \mathcal{K}^2$, from which one can get the maximum static/kinetic spinning torque, $K_\psi \mathcal{S}_s^{\psi} = \eta_{\psi} K_E / \mathcal{K}^2  \; (\mu N /K_E) \mathcal{K}  = \mu N \eta_{\psi} / \mathcal{K}$. This, to some extend, has physical meaning, which is slide friction force $\mu N$ summing its torque with respect to the center of contact patch, and the moment of arm can be treated as $\eta_{\psi}/\mathcal{K}$, a fraction of contact radius. This can be put in discussion section when comparing our model to others. }
		Note that curvature $\mathcal{K}$ and coefficient $\eta_{\psi}$ can either be empirical or derived through Hertzian elastic contact theory. For the latter case, the radius of the contact patch between two surfaces is used to derive the curvature, $\mathcal{K} = 1/a$, and Eqn.(\ref{eq:gettingKpsi}) is reduced to $K_\psi = 0.5 a ^ 2 K_E$, which can be further determined based on material properties such as Young's modulus and Poisson ratio, see Appendix \ref{sec:appendix_spinning_kappa} for details.
		Finally, insofar as the selection of $D_\psi$ is concerned, just like for the slide-friction, one can use heuristics tied to the concept of critical damping. We select $D_\psi$ to impart, in some sense, critical damping. To that end, a fictitious mass moment of inertia $I_{ij}$ is chosen as the average of the $\body{i}$ and $\body{j}$ mass moments of inertia or some other convenient value (for instance, if $\body{i}$ is the ground, one can take $m_{ij} = m_j$). With this,
		\begin{equation}
		\label{eq:criticalDampingSpinCoeff}
		D_\psi \equiv 2 \sqrt{I_{ij} K_\psi} \; .
		\end{equation}
	\end{subequations}

	\section{Numerical Experiments}
	\label{sec:modelTesting}
	The results reported detail the slide, roll, and spin friction loads for several test cases: brick on incline, 2D disk rolling, 3D sphere up on incline, 3D sphere spinning, 3D ellipsoid, and a collection of three interacting bodies (stacking problem). The tests herein are a subset of a larger suite discussed in more detail in \cite{TR-2020-03}. 
	
	All simulations use a time step size of $\Delta t = 10^{-4} \si{sec}$. A first order, half implicit symplectic integrator is used due to its attractive numerical properties; i.e., it preserves a numerical Hamiltonian \cite{hairer2006geometric}, thus reducing the impact that the time integration scheme has on the overall simulation results. Given at time $t_0$ the acceleration $a_0$, velocity $v_0$, and position $x_0$, the integration scheme used updates the system state at $t_1=t_0 + \Delta t$ as $v_1 = v_0 + \Delta t \: a_0$, and $x_1 = x_0 + \Delta t \: v_1$. The acceleration $a_1$ is computed from the momentum balance equation.
	
	\subsection{Brick on an Incline}
	\label{subsec:brickIncline}
\begin{figure}[!ht]
	\begin{minipage}{0.3\textwidth}
		\centering
		\includegraphics[width=1.in]{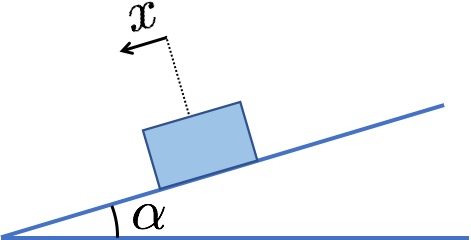}
		\captionsetup{type=figure}
		\caption{Brick on incline.}
		\label{fig:brick_on_incline}
	\end{minipage}
	\begin{minipage}{0.65\textwidth}
		\centering
		\captionsetup{type=table}
		\caption{Brick on incline parameters (all SI units).}
		\label{table:brick_on_incline}        
		\begin{tabular}{c | c c c c c} 
			\hline
			parameter & $m$ & $\mu_k$ & $\mu_s$ & $K_E$ & $K_D$ \\ \hline
			value  & 1 & 0.2  & 0.25 & $10^5$ & 632\\ 
			\hline		
		\end{tabular}
	\end{minipage}
\end{figure}
	
	The goal in this test is to examine the behavior of the slide friction model. To that end, a brick of mass $m$ is initially at rest on an incline as shown in Fig.~\ref{fig:brick_on_incline}. The angle of the incline is $\alpha$. The static and kinetic friction coefficients between the brick and the slope are $\mu_s$ and $\mu_k$, respectively. The stiffness and damping coefficients associated with slide friction force, $K_E$ and $K_D$, are given in Table~\ref{table:brick_on_incline}. The normal force is set to be constant, $N = m g \cos \alpha$.
	Initially, the slope angle $\alpha = 0.18 < tan^{-1} \mu_s$. The static and kinetic slide-microdeflection are capped as $\mathcal{S}_s^{S}=\mu_s N /K_E=2.41\times 10^{-5} \si{m}$ and $\mathcal{S}_k^{S} = \mu_k N/K_E=1.92\times 10^{-5} \si{m}$, respectively. 
	Two tests are run, with and without frictional damping component, $D_f$, see Eqn.~(\ref{eq:slideForceExpression}). The position, velocity, friction force and relative sliding increment and history are reported in Fig.~\ref{fig:brick_incline_small}. If $K_D=0$, the static friction force saturates as the brick slides down, switching from stick mode to slip mode. Additionally, without damping, the brick oscillates within a small magnitude instead of settling. The elastic friction force oscillates around $mg\sin \alpha = 1.75N$. When damping is present; i.e., $K_D=2\sqrt{mK_E}\neq 0$, the brick slides at micrometer level until the elastic friction force $\mathbf{E}_f$ balances out the $m g \sin \alpha$ thus producing results in line with the analytical solution. The static friction force does not saturate and the brick remains in stick mode. 
	\FloatBarrier
\begin{figure}[!ht]
	\centering
	\subfloat[Position (left) and velocity (right).\label{fig:brick_incline_small_kinematics}]{\includegraphics[width=4.34in]{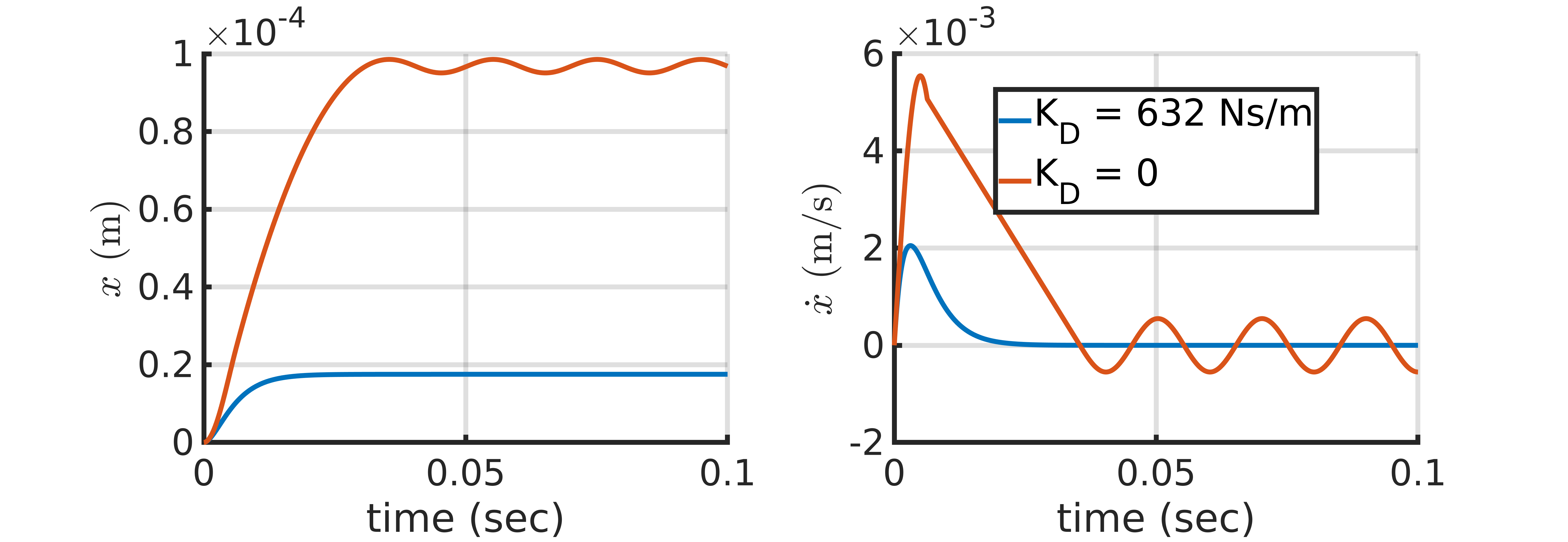}}\\
	\subfloat[Elastic (left) and damping (right) component of the slide friction force. \label{fig:brick_incline_small_friction}]{\includegraphics[width=4.34in]{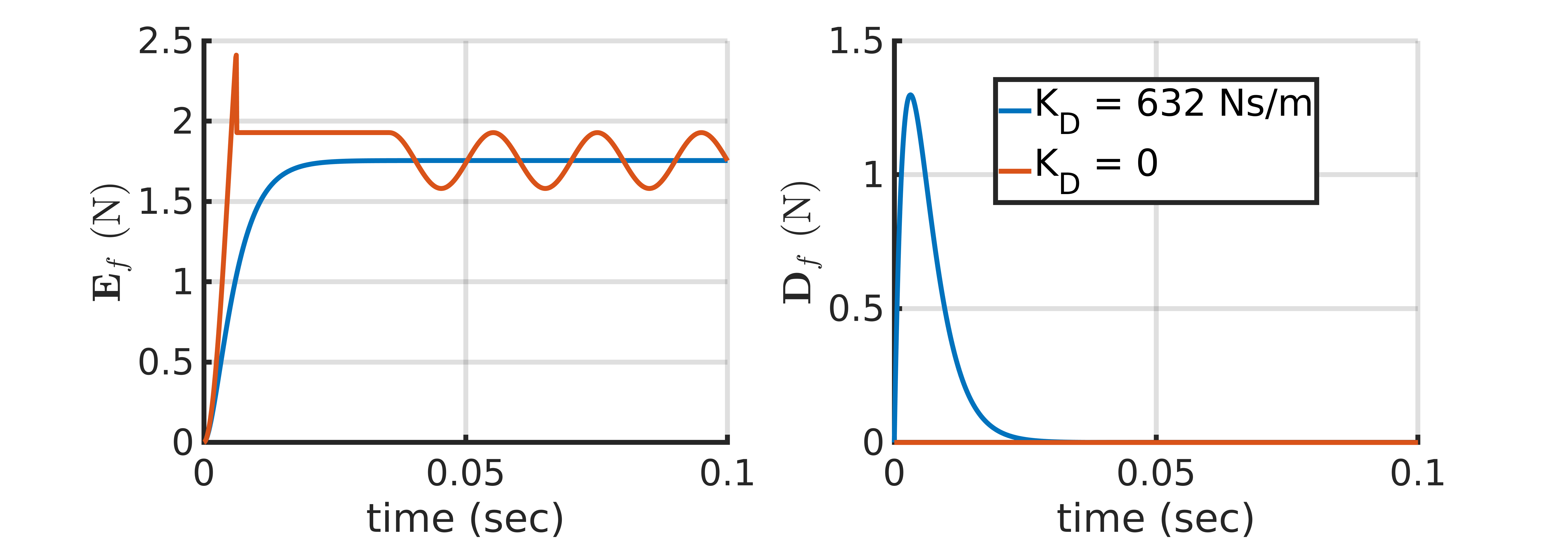}}\\
\end{figure}
\begin{figure}
	\ContinuedFloat
	\centering
	\subfloat[Relative sliding history (left) and increment (right). \label{fig:brick_incline_small_relativeMotion}]{\includegraphics[width=4.34in]{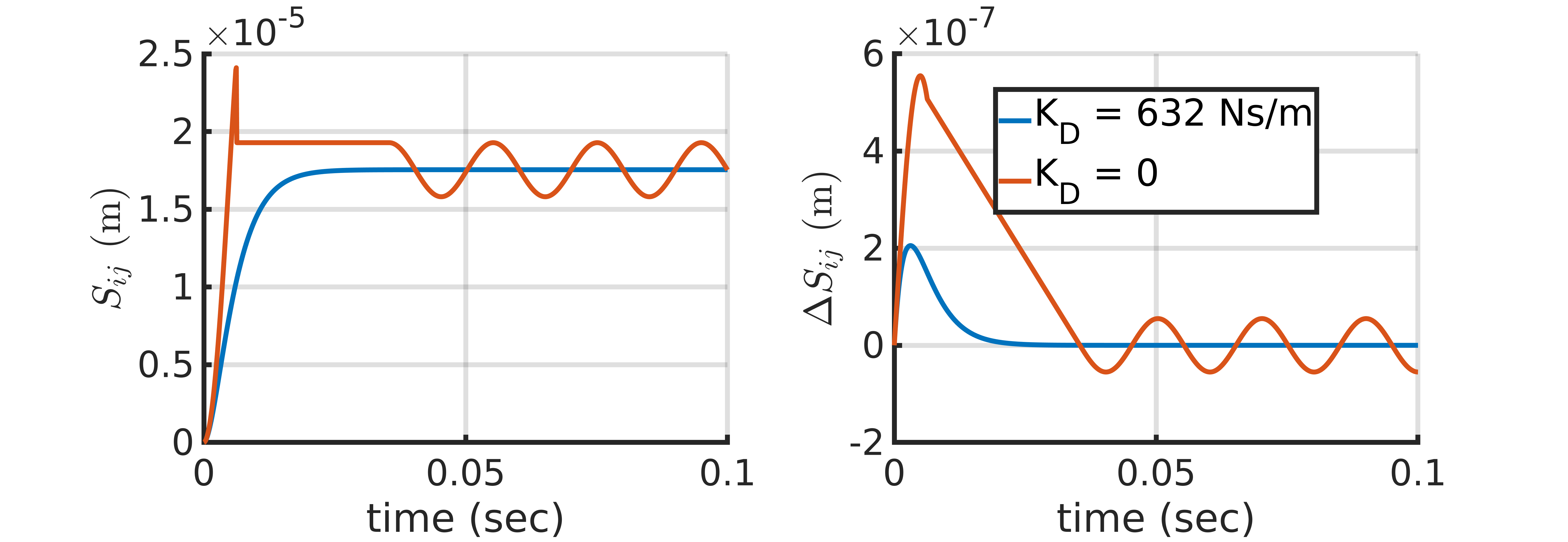}}
	\caption{Brick on incline simulation with and without damping.}
	\label{fig:brick_incline_small}
\end{figure}
	\FloatBarrier
	Next, a set of three incline angles were considered: $\alpha = \tan^{-1} \mu_k$, $\tan^{-1} \mu_s$ and $0.25$. When the incline angle $\alpha \leq \tan^{-1} \mu_s$, the brick should stick. 
	Results in Fig.~\ref{fig:brick_incline_various_angles} indicate that this is the case, and the relative sliding $S_{ij}$ is smaller than, or equal to the stick slide-microdeflection $\mathcal{S}_s^{S}$. 
	As expected, when $\alpha=\tan^{-1} \mu_k$ the brick will stick too, in fact it is further in ``stick territory''. However, when $\alpha > \tan^{-1} \mu_s$, the static friction force saturates and the brick travels down the slope in slip mode with $\mathbf{E}_f= \mu_k m g \cos \alpha =1.899\si{N}$. The brick  acceleration can be evaluated as $g\sin \alpha - \mathbf{E}_f/m =  0.525 \si{m/s^2}$, which corresponds to the kinematics shown in Fig.~\ref{fig:brick_incline_largel_kinematics}.
\begin{figure}[!ht]
	\centering
	\subfloat[Position (left) and velocity (right). \label{fig:brick_incline_largel_kinematics}]{\includegraphics[width=4.34in]{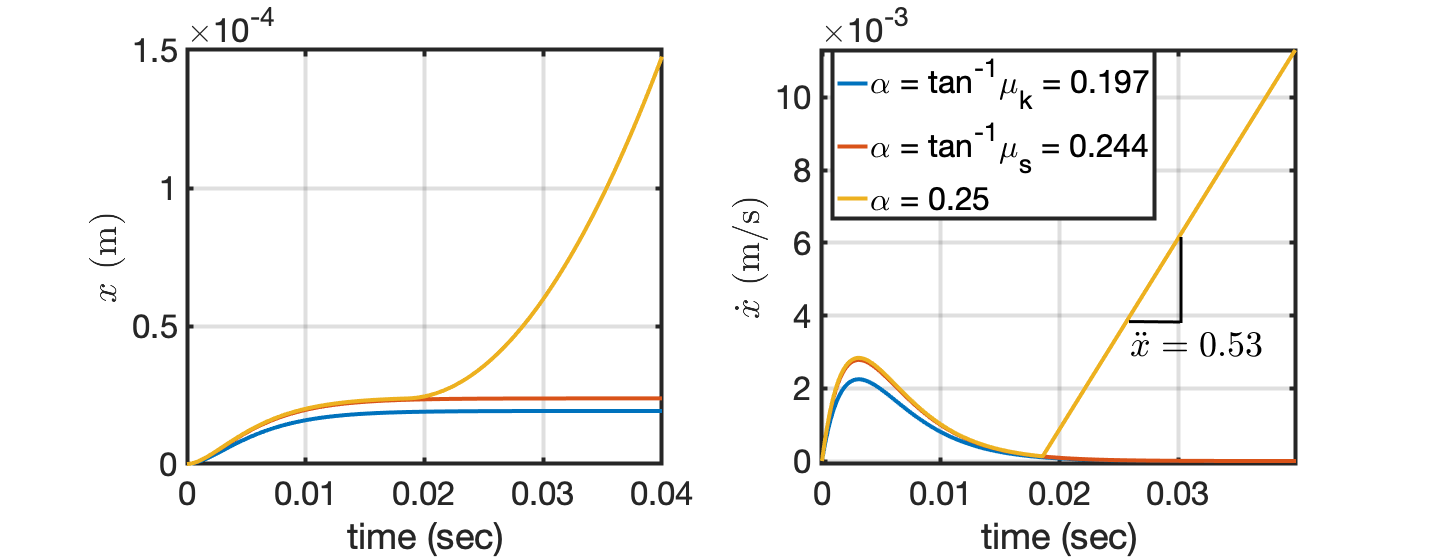}}\\
	\subfloat[Elastic (left) and damping (right) component of the slide friction force. \label{fig:brick_incline_large_friction}]{\includegraphics[width=4.34in]{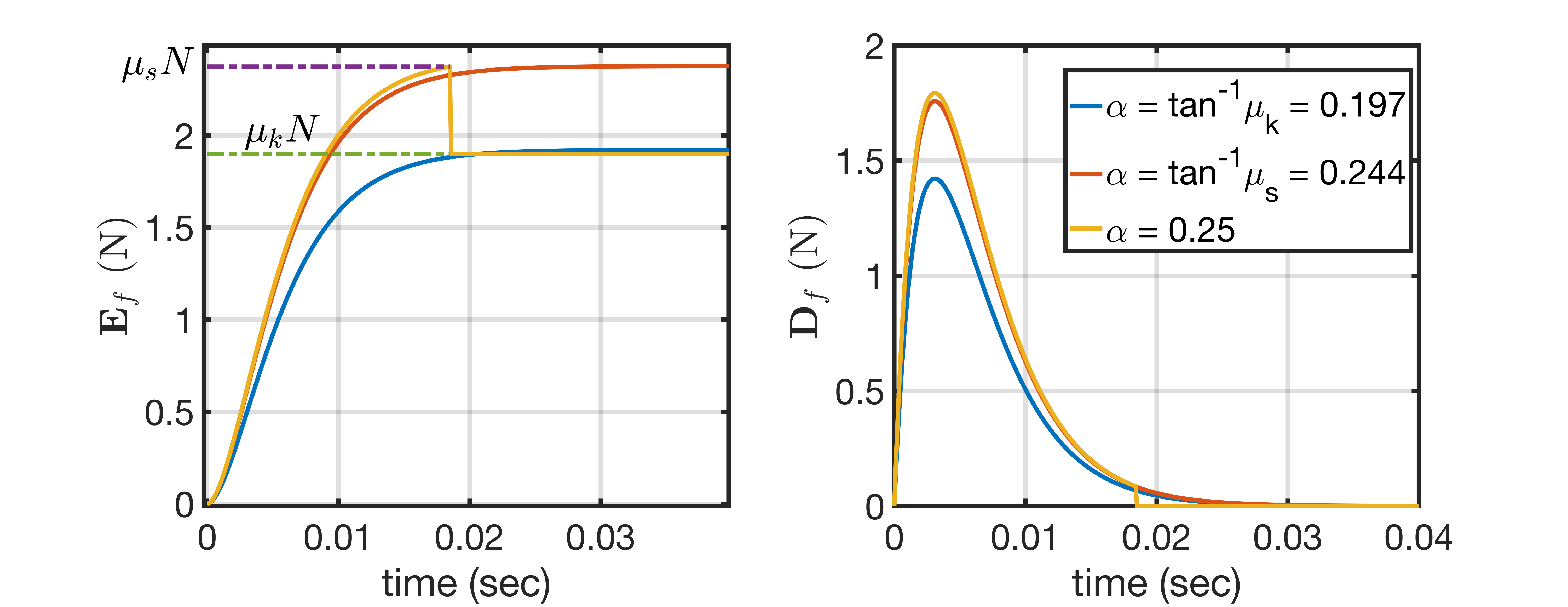}}
\end{figure}
\begin{figure}
	\ContinuedFloat
	\centering
	\subfloat[Relative sliding history (left) and increment (right). \label{fig:brick_incline_large_relativeMotion}]{\includegraphics[width=4.34in]{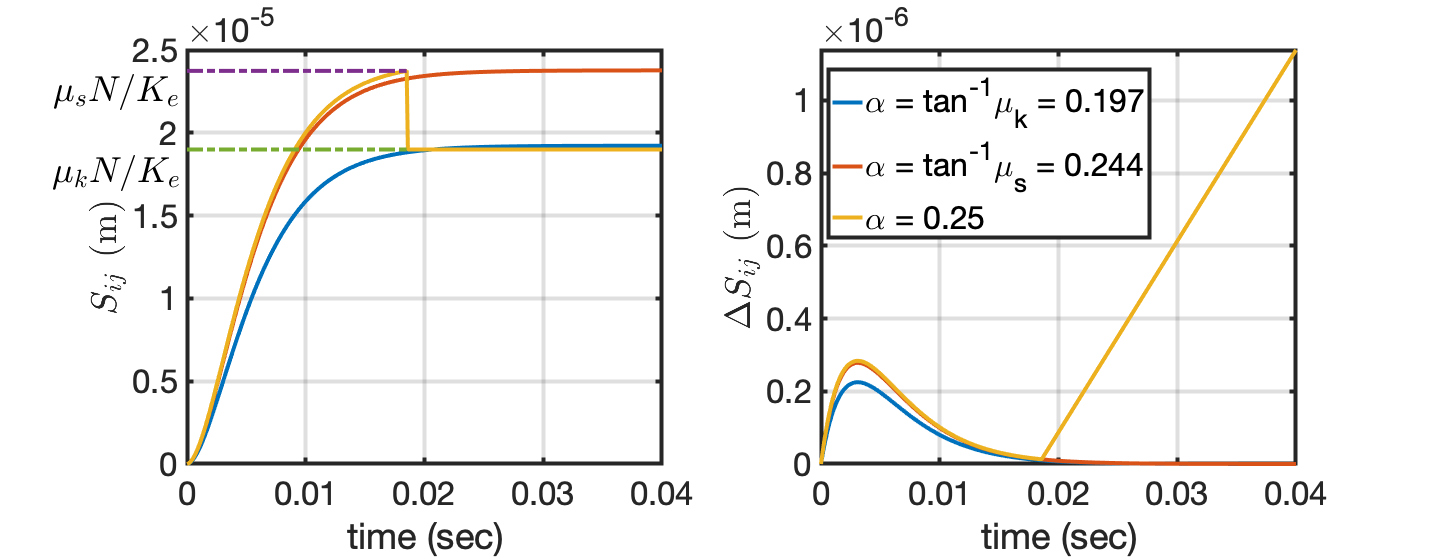}}
	\caption{Brick on incline of different angle $\alpha$.}
	\label{fig:brick_incline_various_angles}
\end{figure}

\FloatBarrier
	
	%%%%%%%
	\subsection{Disk Rolling on a Flat Surface}
	\label{subsec:rollingSphere}
\begin{figure}[!ht]
	\begin{minipage}{0.3\textwidth}
		\centering
		\includegraphics[width=1.in]{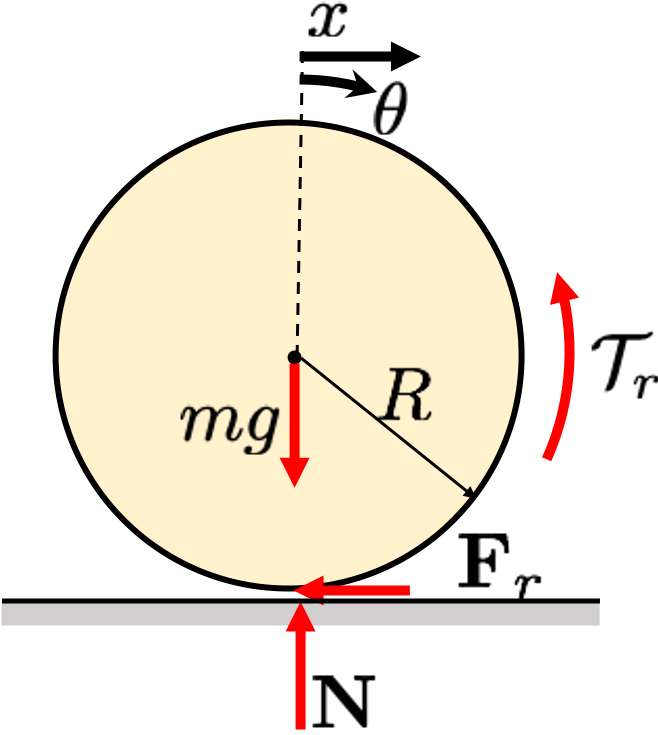}
		\captionsetup{type=figure}
		\caption{Rolling disk.}
		\label{fig:disk_rolling}
	\end{minipage}
	\begin{minipage}{0.7\textwidth}
		\centering
		\captionsetup{type=table}
		\caption{Rolling disk parameters (all SI units).}
		\label{table:rolling_disk}	
		\begin{tabular}{c  c c  c c  c c  c} 
			\hline
			$m$ & $R$ & $I$ & $K_E$ & $K_{cr}$ & $K_r$ & $D_{cr}$ &  $\eta_r$ \\ 
			\hline
			5 &  0.2 &  0.1 &  $10^5$ &  1414.21 &  1600 &  25.30 &  0.4  \\ 
			\hline
		\end{tabular}
	\end{minipage}
	\vspace{-15pt}
\end{figure}

	To investigate the roll friction model, a 2D disk of radius $R$, mass $m$, and inertia $I$ rolls on a flat surface as in Fig.~\ref{fig:disk_rolling}. The translational and rotational coordinates are $x$ and $\theta$, respectively. We set $\dot{x}_0 = 5 \si{m/s}$ and $\dot{\theta}_0 = 0$. The parameters for slide and roll friction models are given in Table~\ref{table:rolling_disk}.
\\	
	This test is challenging owing to the coupling of the slide and roll kinematics, and how the slide and roll micro-deformation condition each other. The kinematic information, elastic and damping components of the slide and roll friction loads, and slide and roll micro-deformations are plotted in Figs.~\ref{fig:disk_rolling_kinematics}-\ref{fig:disk_rolling_relativeMotion}. Quantities associated with coordinate $x$ and slide friction are plotted in blue, whereas ones associated with $\theta$ and roll friction are in red. Considering the initial conditions used, the slide micro-deformation $S_{ij}$ quickly saturates and the disk is in sliding mode with $\mathbf{E}_f = mg\mu_k = 9.8 \si{N}$. The slide friction force opposes the translational motion thus decreasing $\dot{x}$. In the process, it produces a torque greater than the maximum rolling friction, which increases $\dot{\theta}$. At time $t=0.11 \si{sec}$, $\dot{x}$ reaches a value for which $S_{ij}$ no longer saturates, and the slide friction switches to stick mode. Then, $\dot{x} = R \dot{\theta}$, and the slide and roll mechanisms balance out, which leads to rolling without slip. Roll friction remains in kinetic mode, keeping $\mathcal{T}_e$ constant and opposing the disk rotational motion. Both $\dot{x}$ and $\dot{\theta}$ decrease linearly until the disk eventually comes to a full stop. 
\begin{figure}[!ht]
	\centering
	\subfloat[Position-level information (left) and elastic components of the friction loads (right). \label{fig:disk_rolling_kinematics}]{\includegraphics[width=0.75\textwidth]{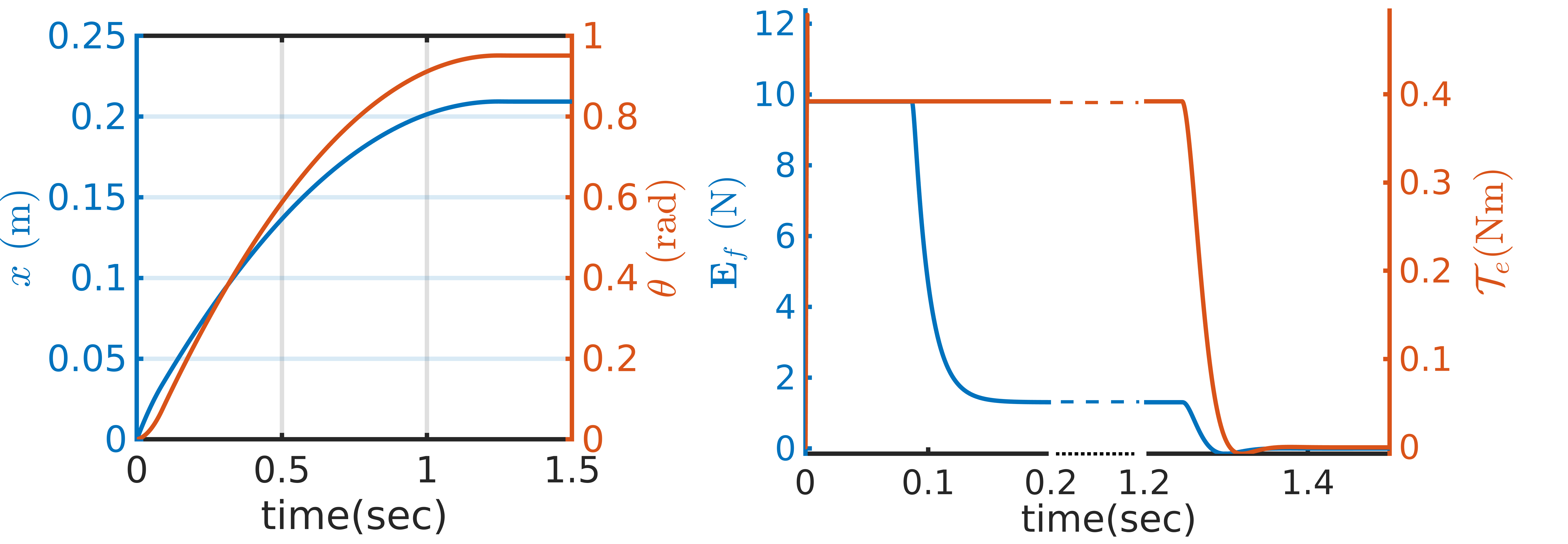}}
\end{figure}
\begin{figure}
	\ContinuedFloat
	\centering
	\subfloat[Velocity-level information (left) and damping component the friction loads (right). \label{fig:disk_rolling_friction}]{\includegraphics[width=0.75\textwidth]{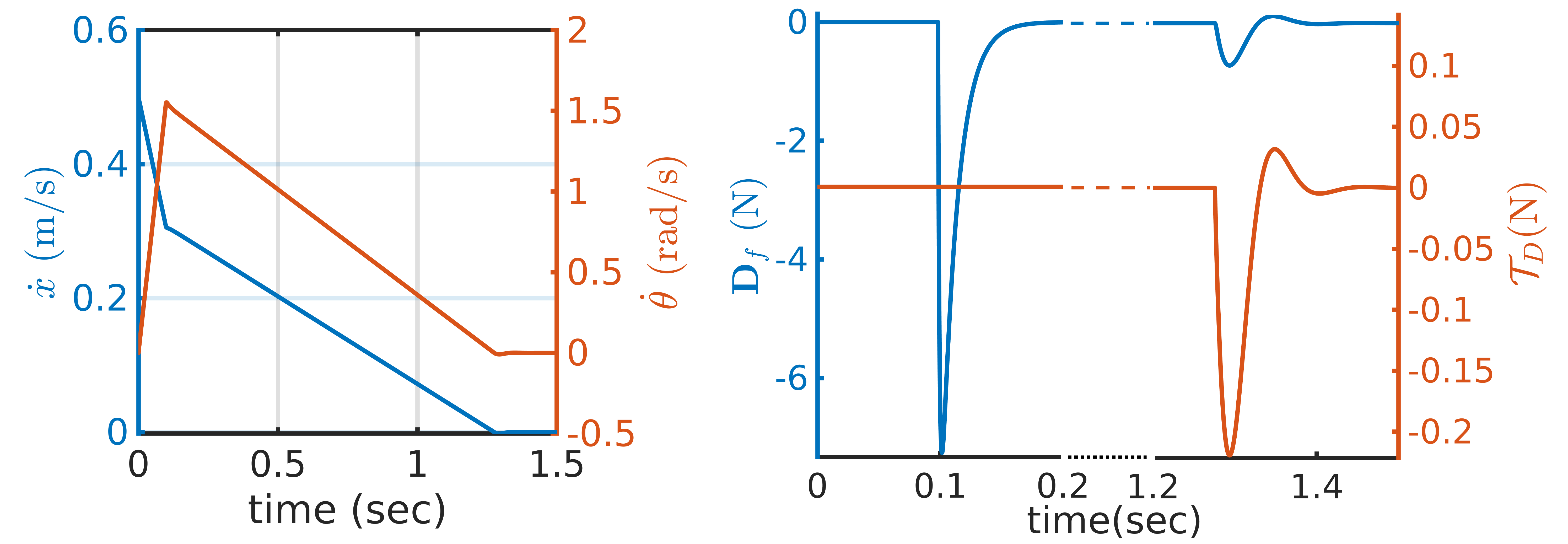}}\\
	\subfloat[Acceleration-level information (left) and micro-deformation for slide and roll (right). \label{fig:disk_rolling_relativeMotion}]{\includegraphics[width=0.75\textwidth]{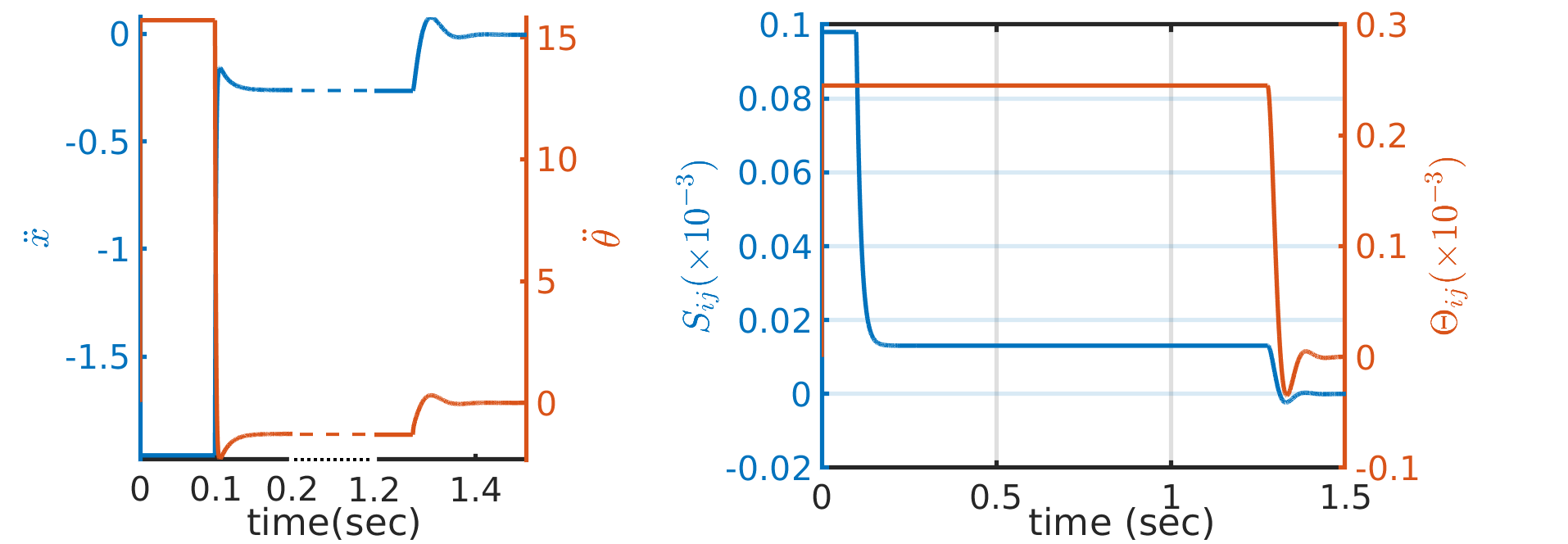}}\\
	\caption{Kinematics and  friction forces of a disk rolling on flat surface.}
\end{figure}
\FloatBarrier
	
\begin{figure}[!ht]
		\centering
		\includegraphics[width=0.75\textwidth]{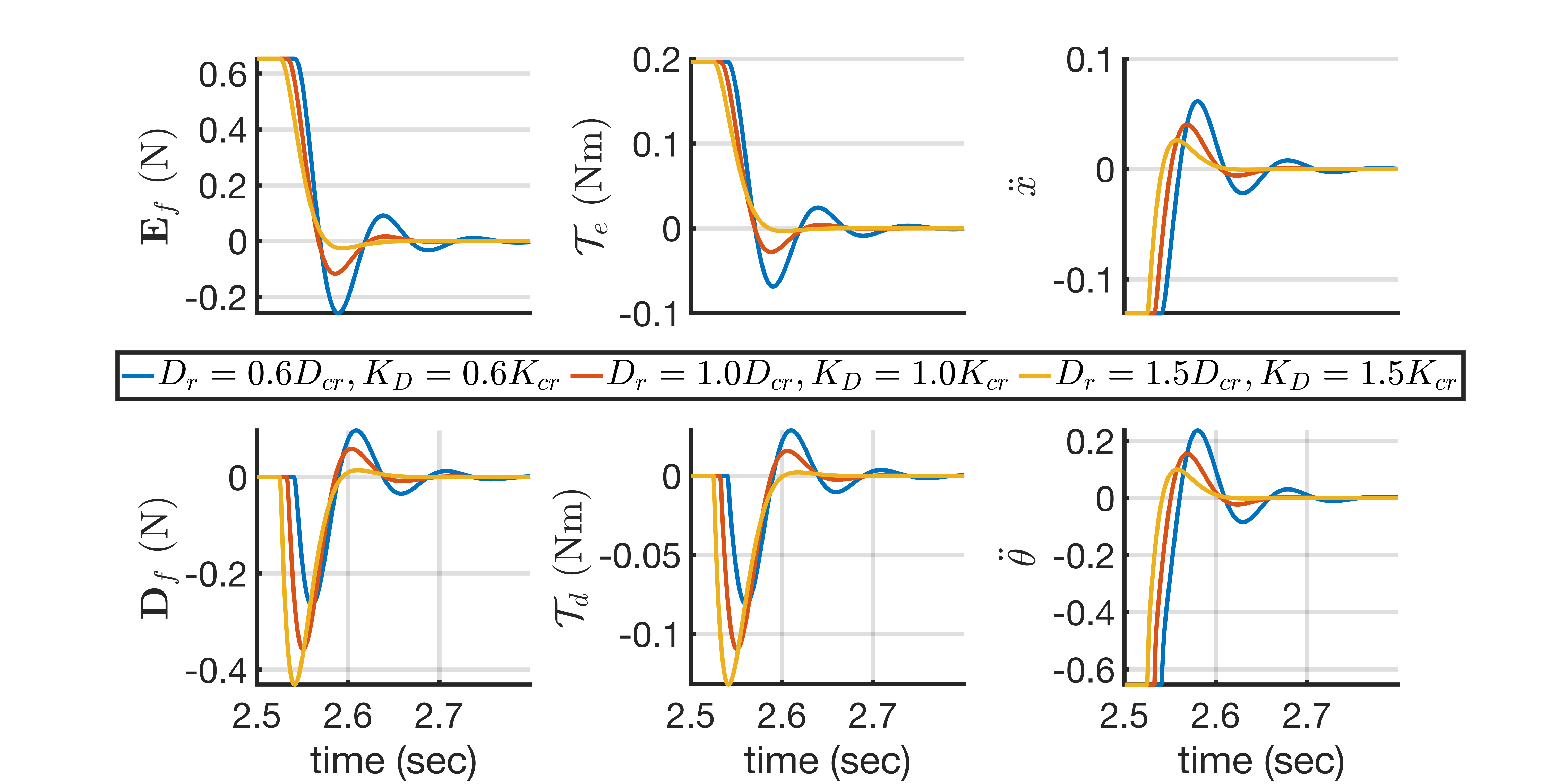}	
		\caption{Comparison using different damping coefficient, ${D}_r$ and $K_D$.}
		\label{fig:disk_rolling_various_Dr}
	\end{figure}

\begin{figure}[!ht]
		\centering
		\subfloat[Legacy, constant torque rolling friction model (Appendix \ref{sec:appendix_rolling_friction_review}, Eqn.~(\ref{eq:rolling_fric_constant_torque_Mr})). \label{fig:disk_rolling_cmp_Schwartz}]{\includegraphics[width=0.75\textwidth]{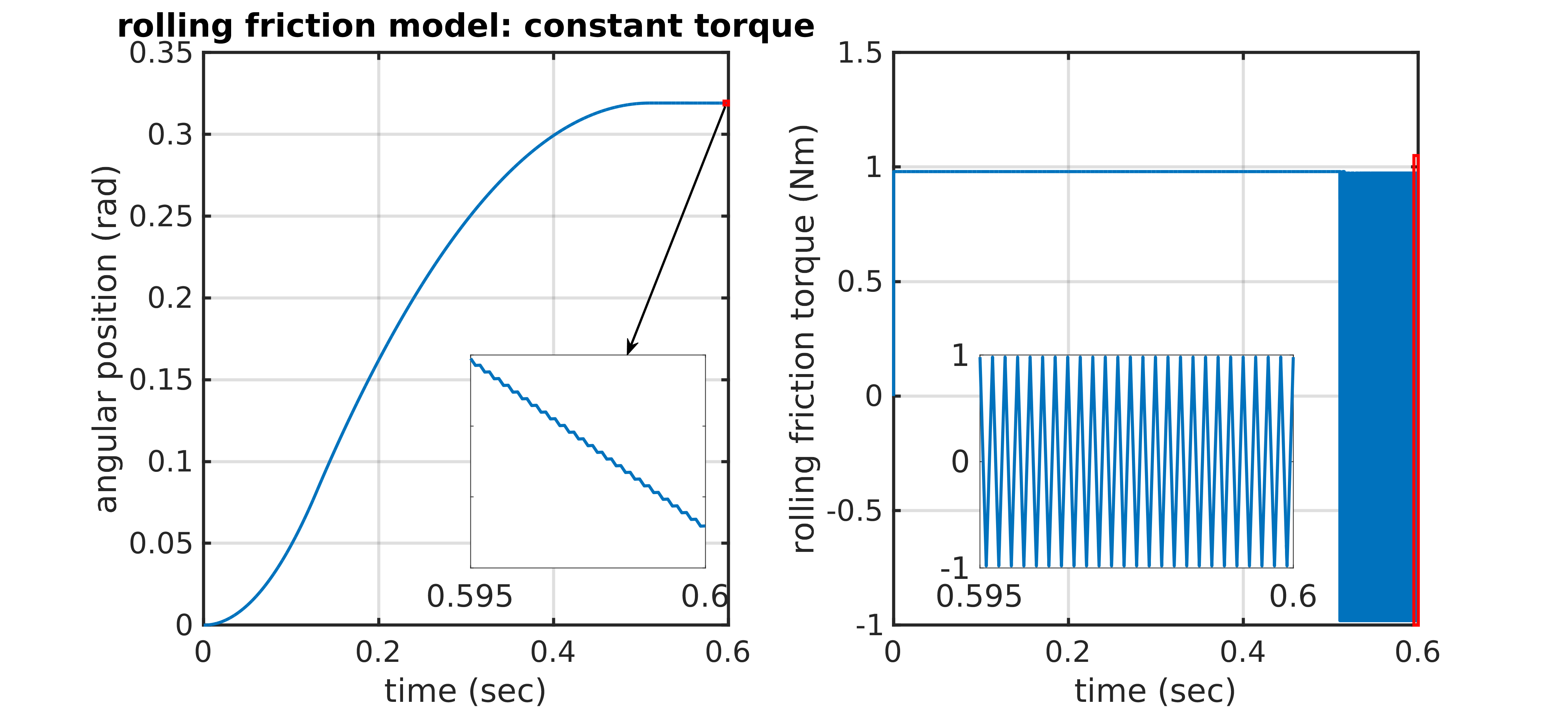}}\\
		\subfloat[Proposed, history-based rolling friction model. \label{fig:disk_rolling_cmp_history}]{\includegraphics[width=0.75\textwidth]{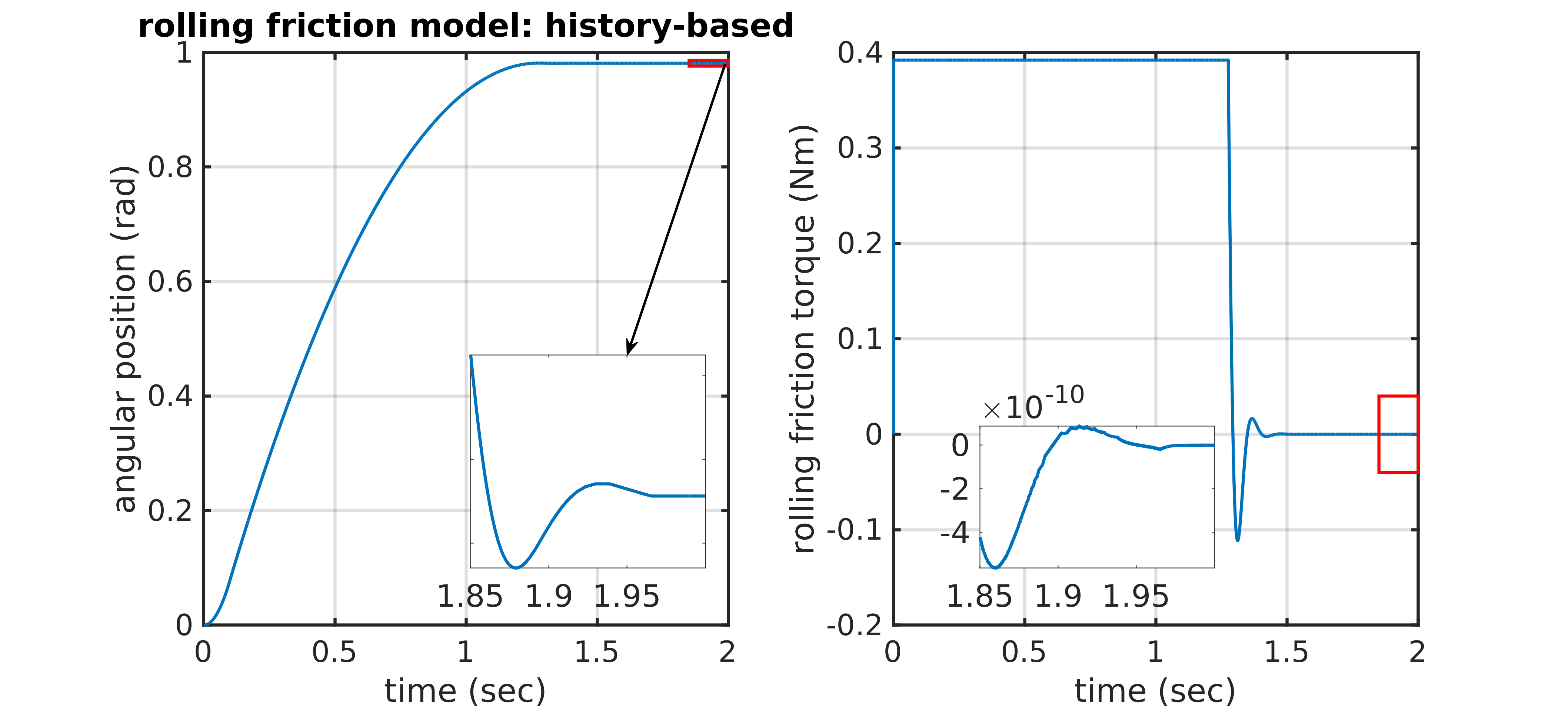}}\\		
		\caption{Angular position (left) and rolling friction torque (right) using different rolling friction models.}
	\end{figure}

	The same test was performed with different damping coefficients, $D_r$ and $K_D$, to further investigate how the disk comes to a stop, see Fig.~\ref{fig:disk_rolling_various_Dr}. The roll and slide critical damping coefficients are $D_{cr}=2\sqrt{IK_r}$ and $K_{cr}=2\sqrt{mK_E}$, respectively. Notice that $\ddot{x}$ and $\ddot{\theta}$ fluctuate/change sign before becoming zero. The overshoot decreases and the disk stops sooner for larger damping coefficients.
	\FloatBarrier
	For comparison, the same scenario is simulated with a different and widely used rolling friction model, \cite{zhou1999rolling,schwartzdem2012}, see Eqn.~(\ref{eq:rolling_fric_constant_torque_Mr})  in Appendix \ref{sec:appendix_rolling_friction_review} and discussion therein. For this test, the rolling friction coefficient $\mu_r$ is set to be $0.1$.  As pointed out in the appendix, in the legacy models the direction of the torque opposes the relative angular velocity, while the magnitude is proportional to the normal contact force and radius of the particle. The simulation is carried out until the disk settles, and the results are plotted in Fig.~\ref{fig:disk_rolling_cmp_Schwartz}. When the disk \textit{seems} to come to a stop, the zoom-in figures display an oscillation of the rolling friction torque between two values. Due to machine precision, angular velocity can not reach exactly zero, therefore, the rolling friction torque will not disappear even when the disk is supposed to settle. This is because the legacy model leads to a ``zero divided by zero'' scenario associated with the angular velocity (see Eqn.~(\ref{eq:rolling_fric_constant_torque_Mr})), which produces the numerical artifacts shown in the inset of Fig.~\ref{fig:disk_rolling_cmp_Schwartz}. In contrast, the rolling friction model presented in this paper, by tracking the history of the changes in relative angular orientation as opposed to angular velocity, comes to a full stop as illustrated in Fig.~\ref{fig:disk_rolling_cmp_Schwartz}. Indeed, the rolling friction torque goes to ``zero'' (machine precision) and the disk stops moving (no rocking behavior).
	\FloatBarrier

	%%%%%%%
	\subsection{Sphere Rolling Up an Incline}
	\label{subsec:sphereIncline}
\begin{figure}[!ht]
	\centering
	\subfloat[Velocities (left), elastic component of slide $F_E$ and roll $T_E$ friction loads (right).\label{fig:sphere_up_incline_35_velo}]{\includegraphics[width=0.8\textwidth]{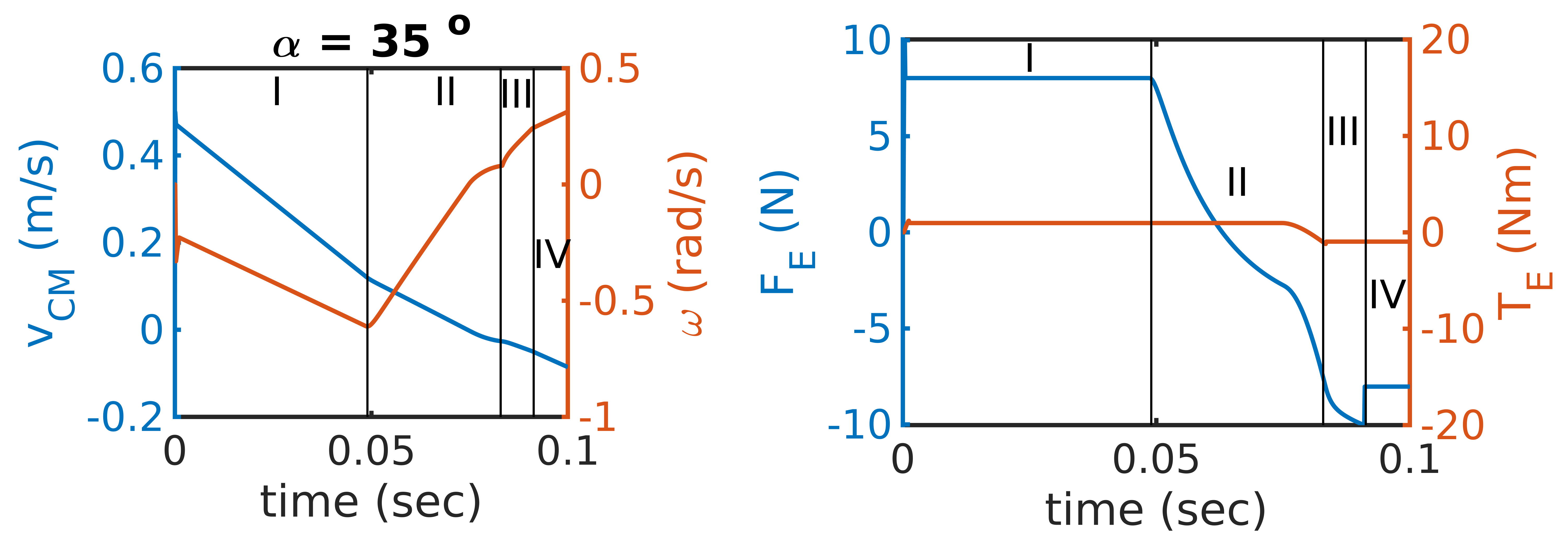}} \\
	\subfloat[Time = $0.004 \si{sec}$ (Phase I)]{{\includegraphics[width=0.4\textwidth]{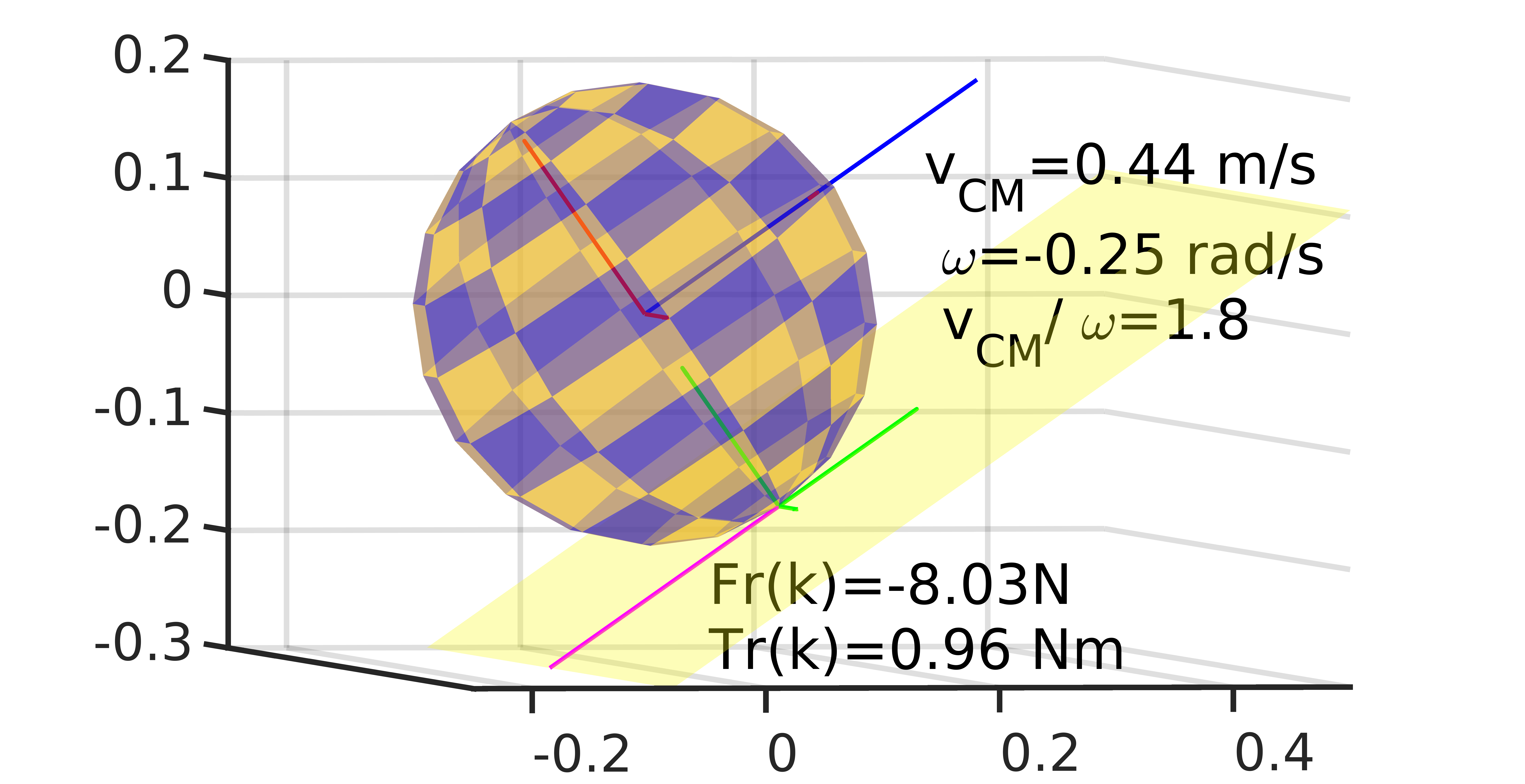} }}%
	\subfloat[time = $0.05  \si{sec}$ (Phase II)]{{\includegraphics[width=0.4\textwidth]{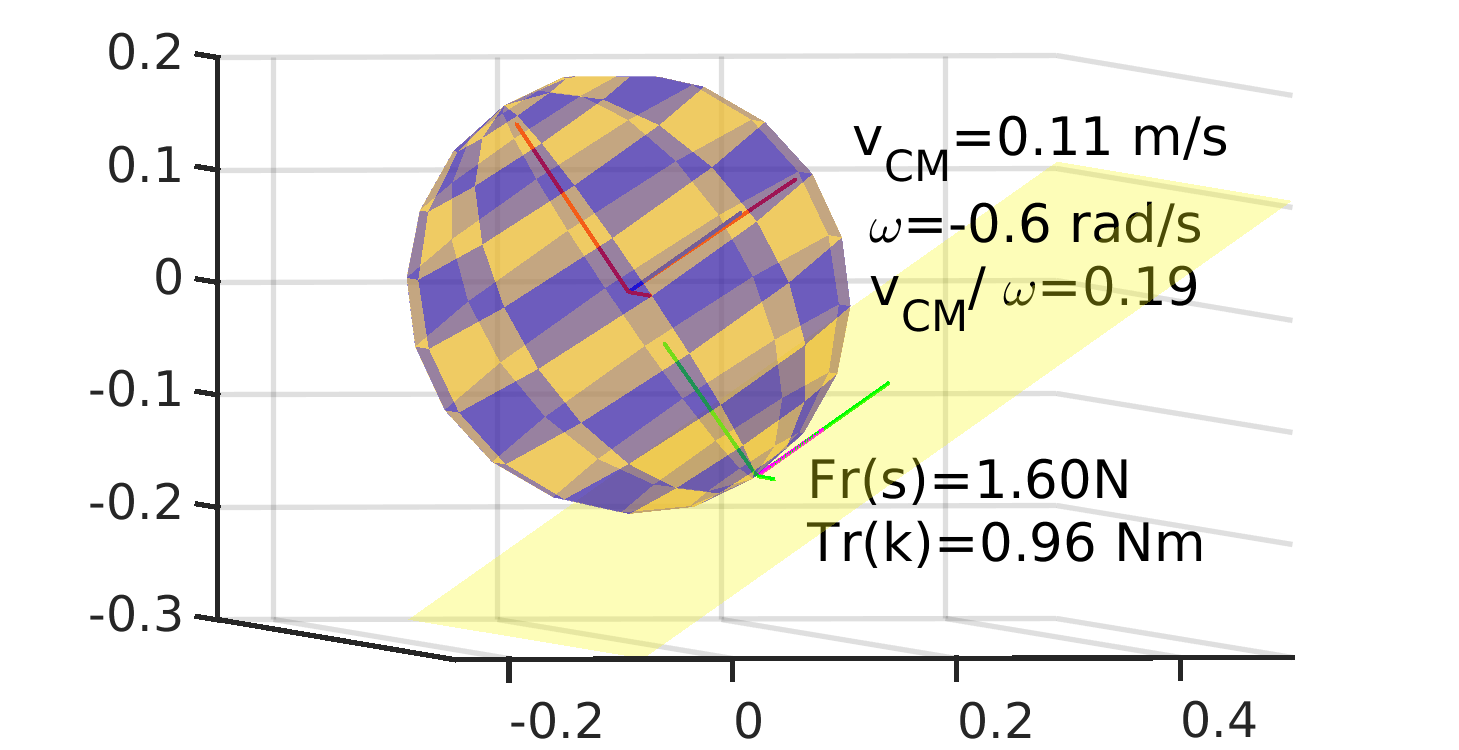} }}%
	\quad
	\subfloat[Time = $0.09  \si{sec}$ (Phase III)]{{\includegraphics[width=0.4\textwidth]{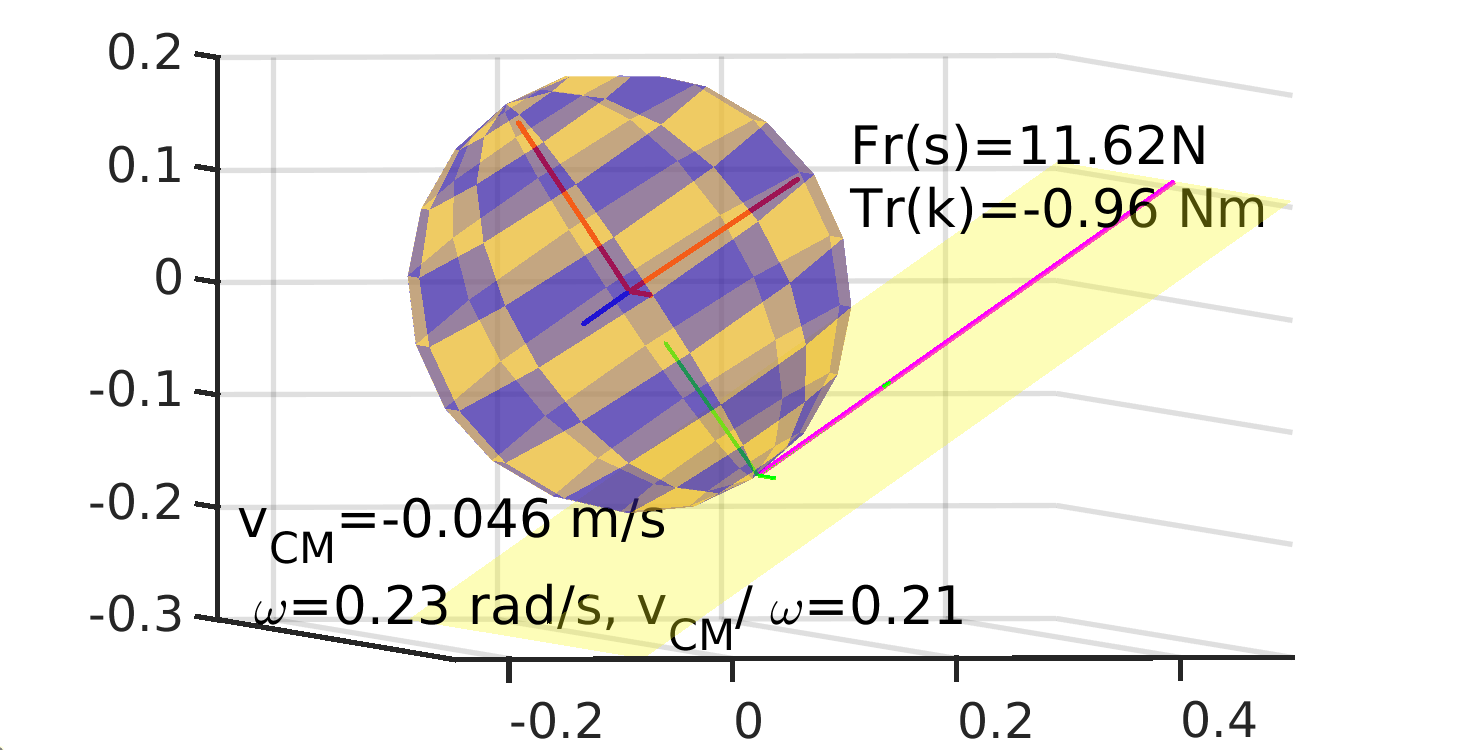} }}%
	\subfloat[Time = $0.2   \si{sec}$ (Phase IV)]{{\includegraphics[width=0.4\textwidth]{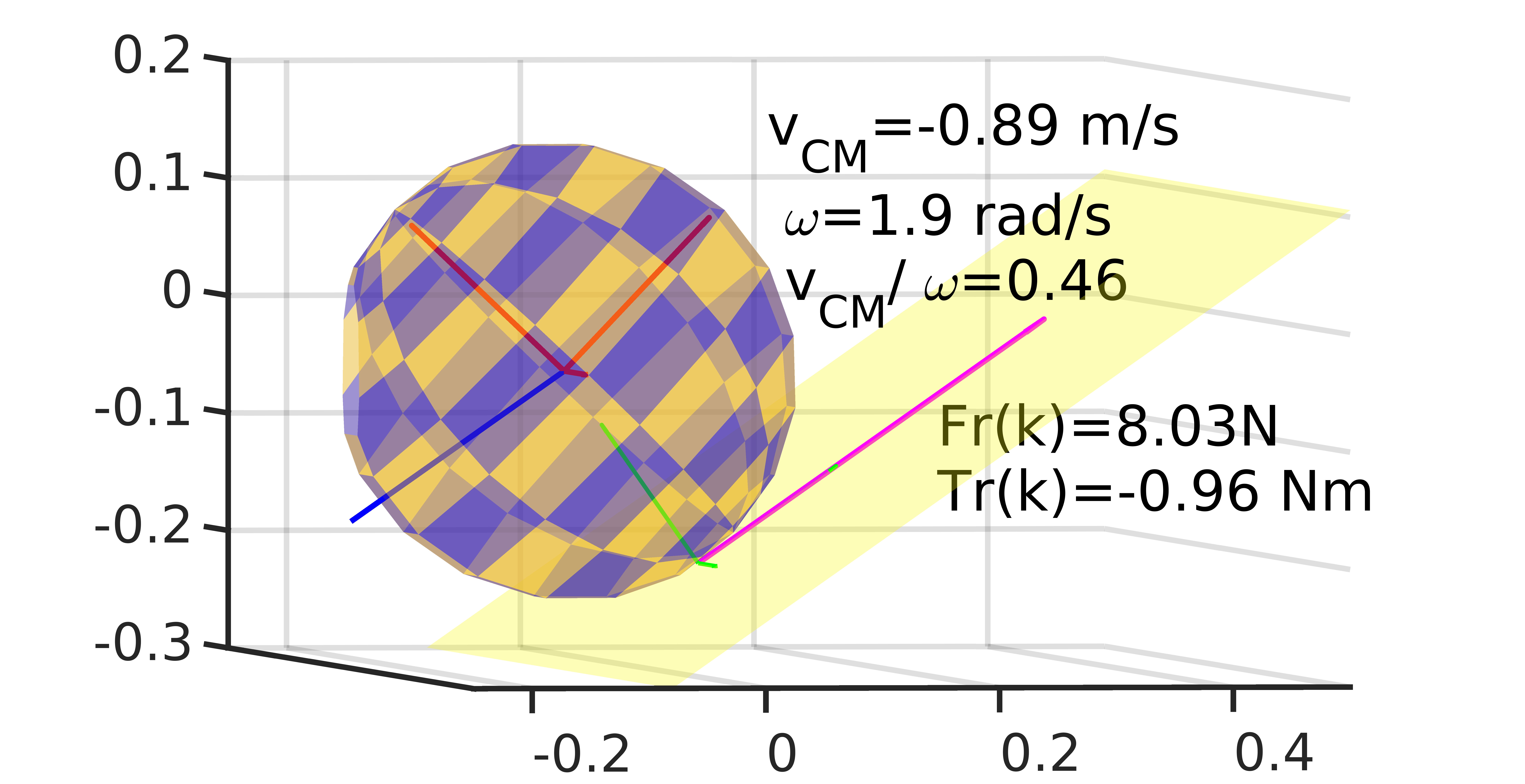} }}%
	\caption{Sphere going up an incline of $35^o$ with $v_0 = 0.5 \si{m/s}$.}
	\label{fig:sphere_up_incline_35deg}
\end{figure}

	A 3D sphere of radius $R = 0.2 \si{m}$ and mass $m = 5 \si{kg}$ starts moving up on an incline with an initial velocity of $0.5 \si{m/s}$ and zero angular velocity. The sphere is already on the incline, and the velocity is parallel with the incline and pointing up. The rolling resistance coefficient between the sphere and incline is $\eta_r = 0.3$. The linear velocity at the center of mass $v_{CM}$, angular velocity $\omega$, and the elastic component of the slide and roll friction loads, $F_E$ and $T_E$, are shown in Fig.~\ref{fig:sphere_up_incline_35_velo}. Snapshots of the simulation are provided in Fig.~\ref{fig:sphere_up_incline_35deg},  where the local reference frame, global contact frame, linear velocity and slide friction force are colored in red, green, blue, and magenta, respectively. Initially, given the nonzero initial velocity, the sphere quickly switches to slip mode in kinetic regime (the static regime last only very briefly). The slide friction force opposes the translational motion and creates a torque that increases $\omega$ (Phase I). Once $\omega$ catches up, the sphere switches to pure rolling mode, where the slide friction is static and the ratio between linear and angular velocity $v_{CM}/\omega$ is close to radius $R=0.2 \si{m}$, indicating pure rolling mode for the sphere (Phase II). After both $v_{CM}$ and $\omega$ decrease to zero, the sphere rolls down the incline (Phase III). Although both $v_{CM}$ and $\omega$ change sign, the sphere remains in pure rolling mode with increasing static friction since the slide micro-deformation is increasing. Eventually, the static friction saturates and the sphere rolls down the incline with slip (Phase IV). In both Phase I and IV, the sphere rolls on the incline in a rolling-with-slip fashion, where both slide and roll loads are in kinetic regime: $F_E = \mu_k N = 8.027 \si{N}$, and $T_E = 2 \eta_r R \mu_k N = 0.963 \si{Nm}$. During Phase II and III, the sphere rolls on the incline without slip, when the slide friction force is in static mode, whereas the roll friction torque is in kinetic mode. Therefore, the elastic part of the slide friction can take any value between $\left[-\mu_s N, \mu_sN  \right] = \left[ - 10.03, 10.03\right]$, while the roll friction is saturated at $T_E = 0.963 \si{Nm}$. \\
	\FloatBarrier
	The same simulation was carried out using different combinations of incline angle $\alpha$ and rolling resistance $\eta_r$. The incline angle $\alpha$ was varied from $1^o$ to $30^o$ by a step of $0.1$; the rolling resistance $\eta_r$ was varied from $0.2$ to $0.5$ by a step of $0.01$. This sweep required $9021=291\times31$ simulations; damping was fixed at $K_D=\sqrt{m K_E}$ and $D_r=\sqrt{I K_r}$. Each simulation captured $1 \si{sec}$ of system dynamics, long enough for the sphere to reach steady state. This steady state was categorized into four scenarios listed in Table~\ref{tab:sphere_up_incline_steady_state}.
\begin{table}[!ht]
	\centering
	\caption{Four steady-state configurations for sphere on incline. Acronyms: ``S'' -- Static; ``PR'' -- Pure Rolling; ``PS'' -- Pure Slip; ``RwS'' -- Rolling with Slipping.}
	\label{tab:sphere_up_incline_steady_state}
	\begin{tabular}{c c c c c}
		\hline
		& S  & PR & PS   & RwS \\ \hline
		kinematic constr.      &  $v_{CM} = 0, \omega = 0$ &$v_{CM} = \omega R$ &  $\omega=0, v_{CM} \neq 0$  & $v_{CM} > \omega R$ \\
		slide mode state                    & static                  & static                & kinetic & kinetic\\
		roll mode state       & static & kinetic & static & kinetic \\
		\hline
	\end{tabular}
	%	\end{center}
\end{table}
\begin{figure}[!ht]
	\centering
	\includegraphics[width=4in]{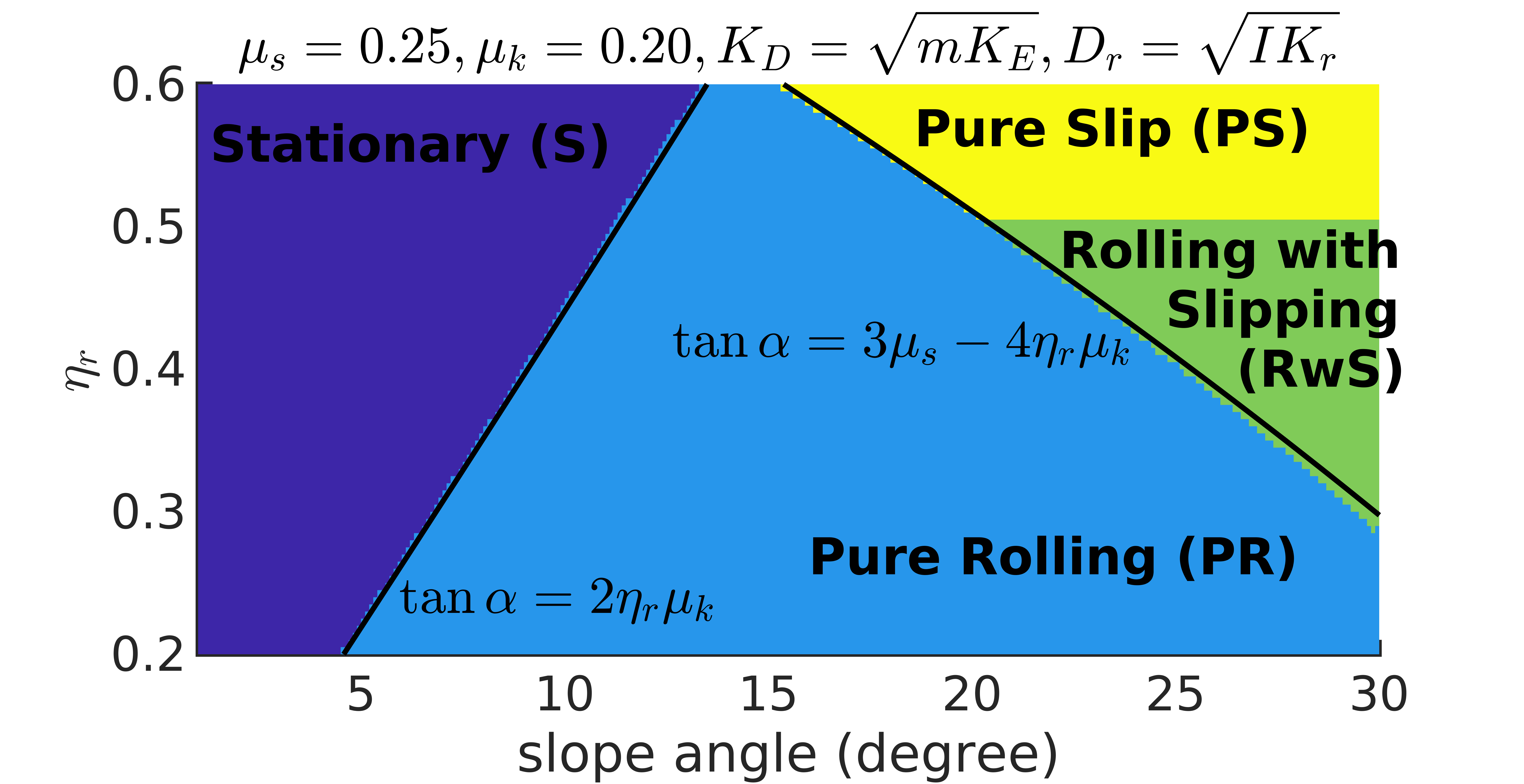}
	\caption{Steady state, captured at end of simulation, for the sphere in section \ref{subsec:sphereIncline}.}	
	\label{fig:sphere_up_incline_various_eta_angle}
\end{figure}
	
	As illustrated in Fig.~\ref{fig:sphere_up_incline_various_eta_angle}, with larger slope angle, the sphere has a tendency to slip at the point of contact. As $\eta_r$ increases, there is more resistance for the sphere to roll, and the sphere is capable of ``freezing'' on the incline for a larger range of angles. When $\eta_r \geq 0.5$, the rolling resistance is so high that eventually the sphere can slide down the steep slope like a brick, with no rolling. The critical state in between different scenarios, shown as the black border, is derived analytically in Appendix \ref{sec:appendix_sphere_incline}. Figure~\ref{fig:sphere_up_incline_various_eta_angle} indicates good agreement between the analytical results and simulation predictions.
	\FloatBarrier
	%%%%%%%
	\subsection{Spinning Sphere}
	\label{subsec:spinningSphere}
\begin{figure}[!ht]
	\centering
	\subfloat[Time = $0 \si{sec}$]{{\includegraphics[width=0.25\textwidth]{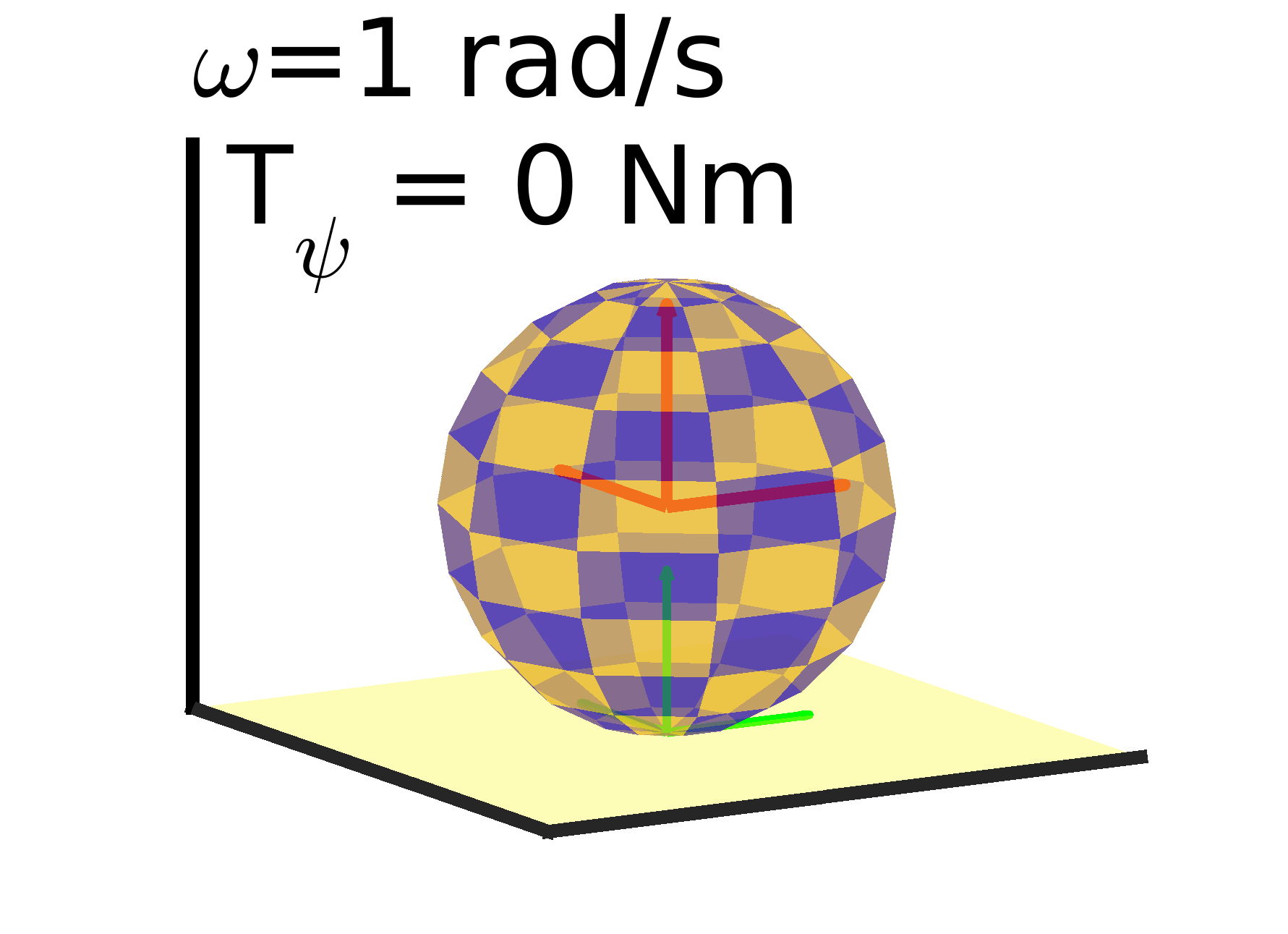} }}%
	\subfloat[Time = $2 \si{sec}$]{{\includegraphics[width=0.25\textwidth]{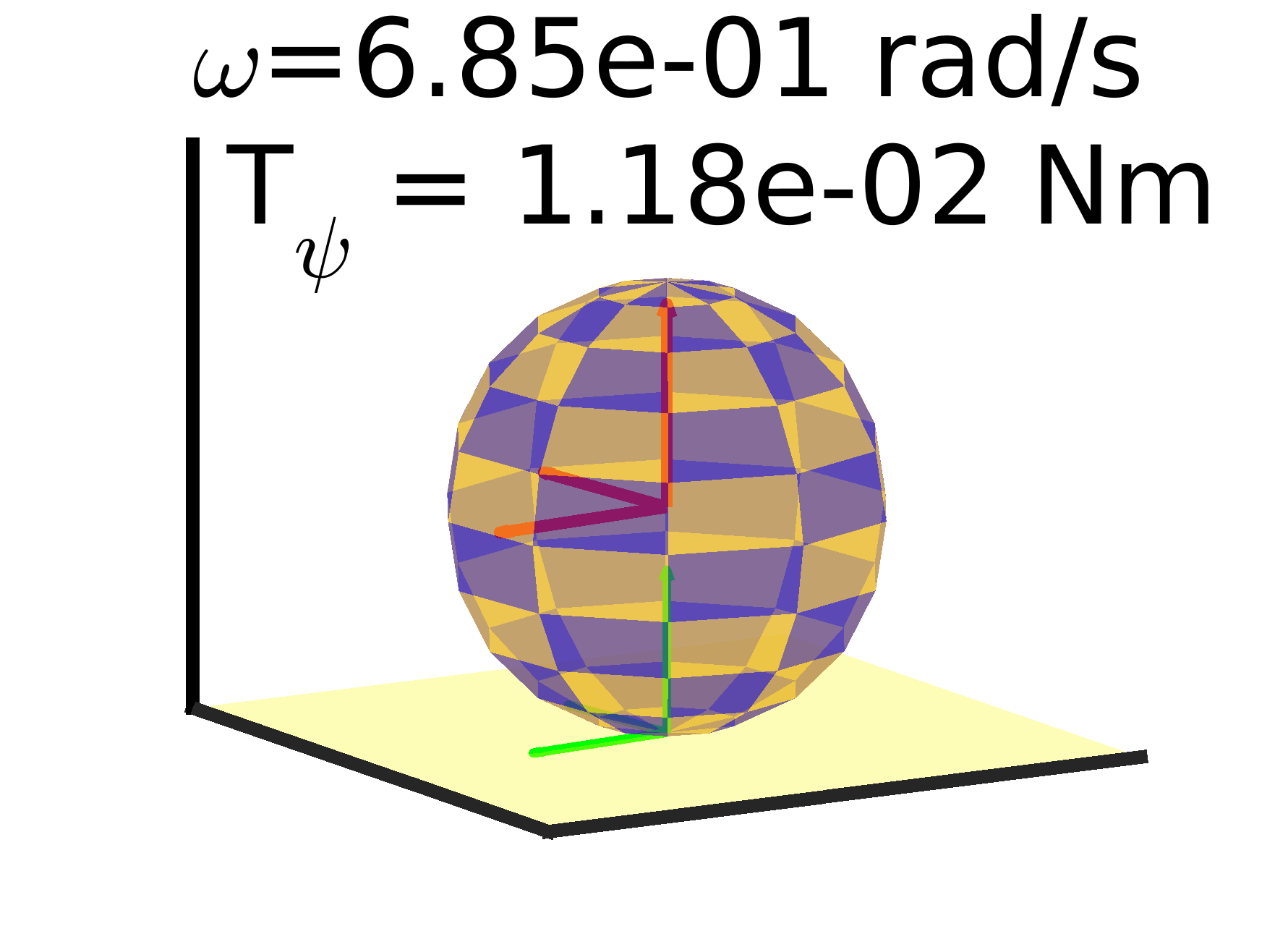} }}%
	\subfloat[Time = $4 \si{sec}$]{{\includegraphics[width=0.25\textwidth]{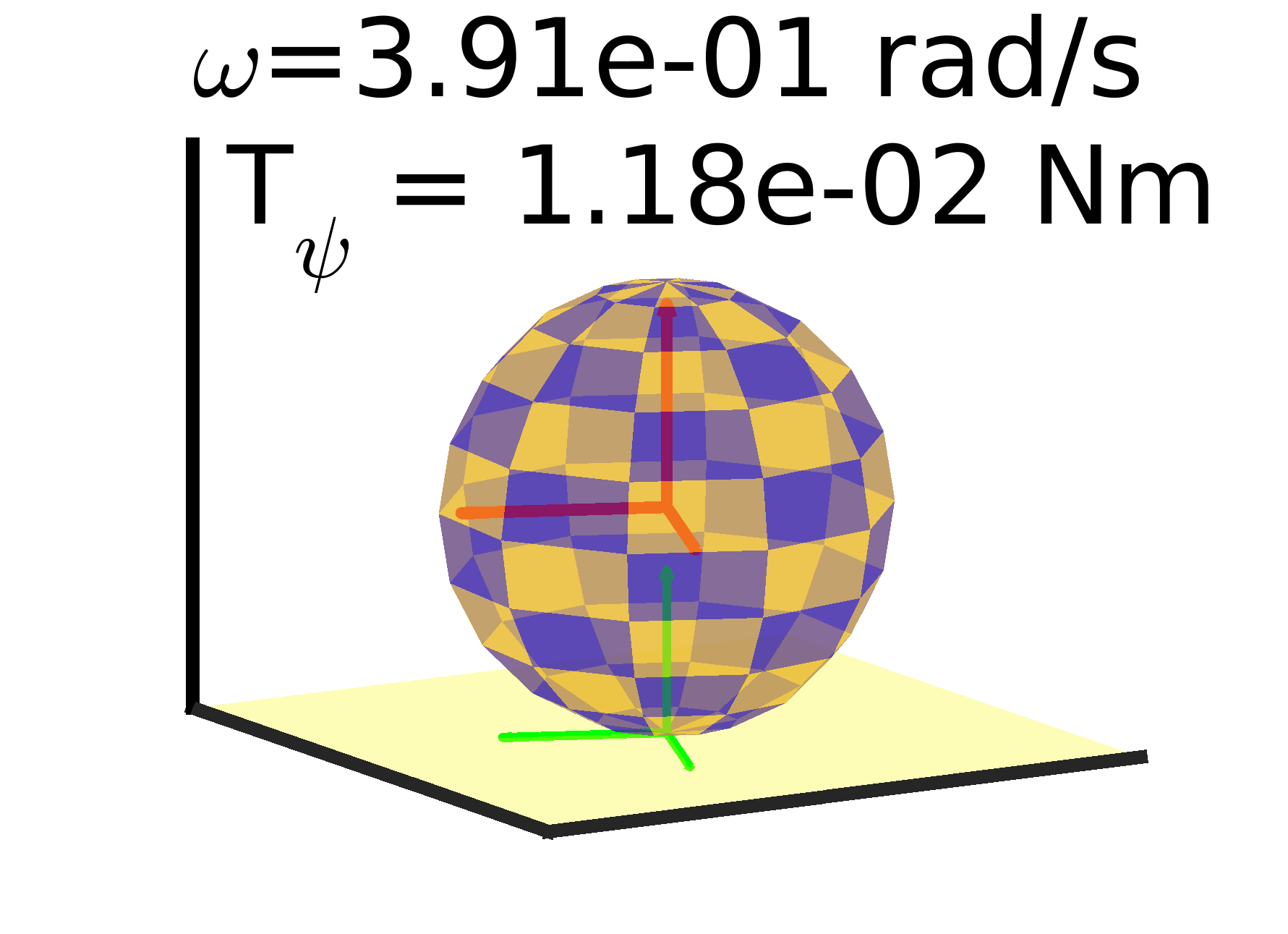} }}%
	\subfloat[Time = $8 \si{sec}$]{{\includegraphics[width=0.25\textwidth]{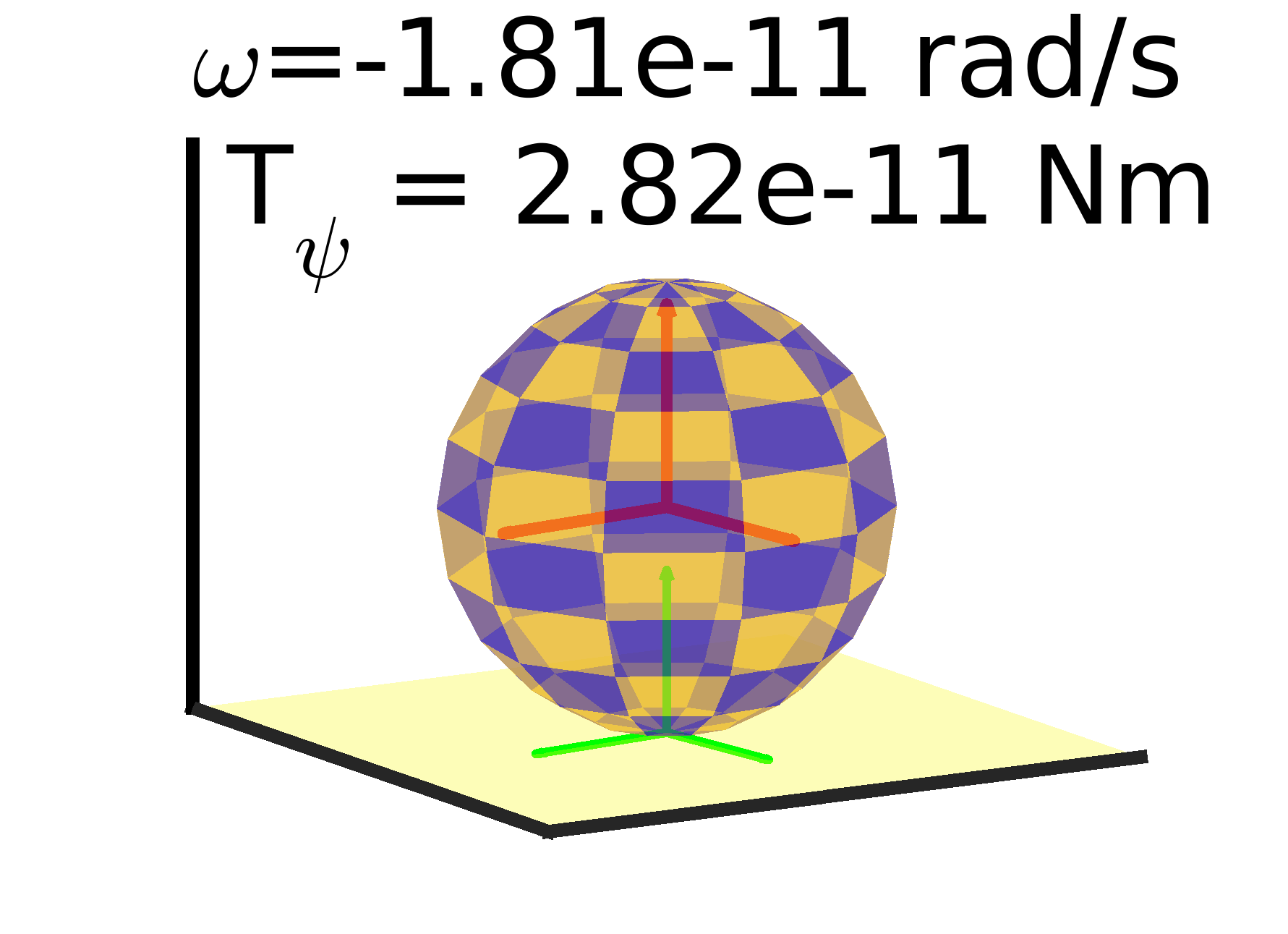} }}%
	\caption{Snapshots, sphere spinning, empirical model; $\eta_{\psi} = 0.006$, $\mathcal{K} = 5$, $\omega_0 = 1 \si{rad/s}$.}	
	\label{fig:sphere_spinning_snapshots}
\end{figure}

	In section~\ref{subsec:spinningFriction}, $\mathcal{K}$ and $\eta_{\psi}$ were specified in two ways: using heuristics, or employing a Hertzian contact theory approach. Therefore, two spin friction tests are performed. For the heuristic approach, the sphere of subsection~\ref{subsec:sphereIncline} is spun on a flat surface with an initial angular velocity $\omega_0 = 1 \si{rad/s}$, spinning resistance coefficient $\eta_{\psi} = 0.006$ and curvature $\mathcal{K} = 5$. The axis of rotation aligns with the gravity. 
	%Assuming that the spin angle $\psi_{ij}$ between two reference frames, $({\nAxis{1}},{\uBarAxis{1}},{\wBarAxis{1}})$ and $({\nAxis{1}},{\uAxis{1}},{\wAxis{1}})$, is minuscule due to the small time step size, the magnitude of $\psi_{ij}$ is approximated as $\psi_{ij} \approx  \sin \psi_{ij} = \| \uBarAxis{1} \times \uAxis{1}  \|$. This avoids numerical calculation of $\cos^{-1}$ or $\sin^{-1}$. The stiffness of spinning friction, $K_{\psi}$, is related with the stiffness of slide friction, $K_E$, through Eqn.~(\ref{eq:gettingKpsi}). 
	The coefficient $D_{\psi}$ is picked for critical damping, see Eqn.~(\ref{eq:criticalDampingSpinCoeff}). The simulation lasts for $8 \si{sec}$, which is long enough for the body to stop spinning. Figure~\ref{fig:sphere_spinning_snapshots} shows snapshots taken at time $t=0$, two during the spinning process (kinetic mode), and one when the sphere settles (static mode). The body frame and global contact frame are colored in red and green, respectively. In kinetic mode, the spinning friction torque $T_{\psi}$ stays constant. The kinetic spin friction torque can be evaluated as $T_{\psi} = \mu_k N \eta_{\psi} R = 0.0118 \si{N m}$.\\
	
\begin{figure}[!ht]%
		\centering
		\includegraphics[width=0.7\textwidth]{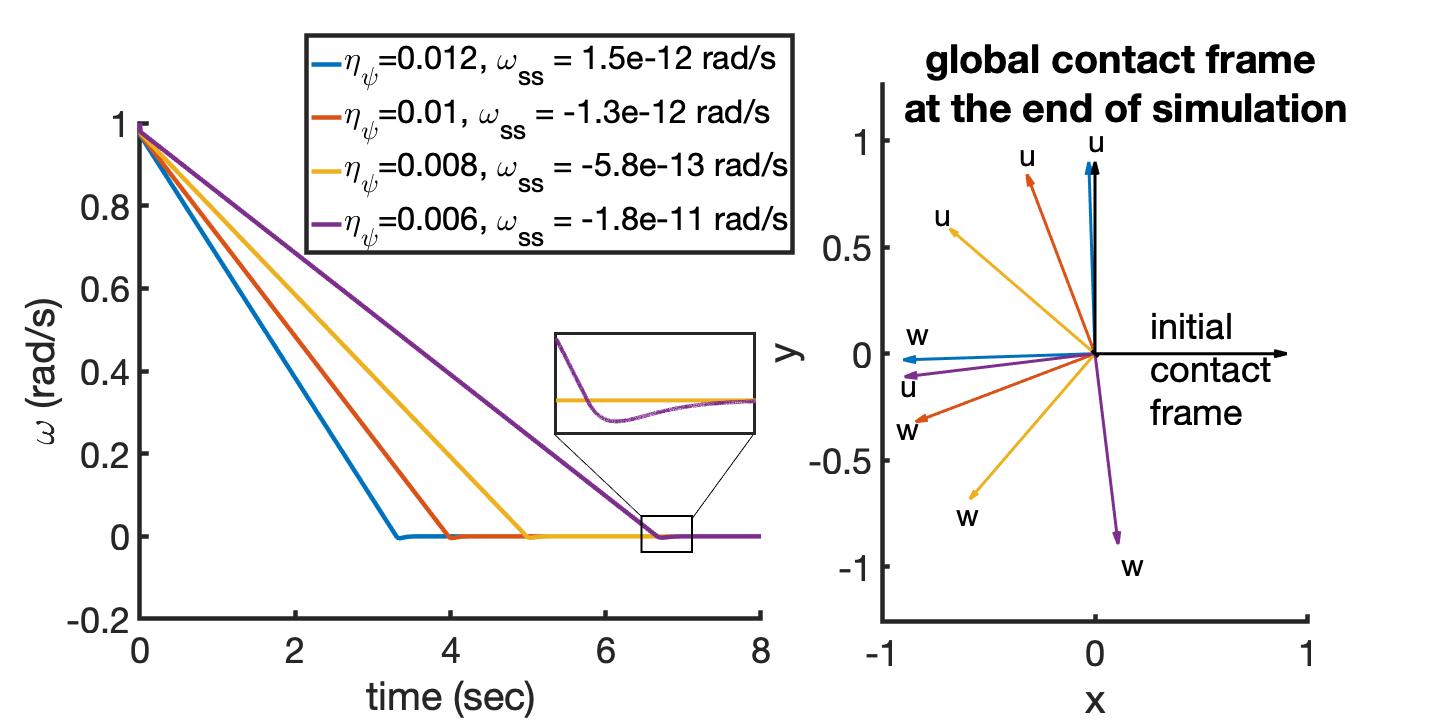}\\
		\caption{Angular velocity $\omega$ over time using various $\eta_{\psi}$(left); Global contact frame when the sphere settles(right).}
		\label{fig:sphere_spinning_various_eta}
	\end{figure}

	The simulation setup described is used in a sensitivity study in which $\eta_{\psi}$ changes from simulation to simulation. Results for four cases; i.e., $\eta_{\psi} = 0.012$, $0.01$, $0.008$ and $0.006$, are reported in Fig.~\ref{fig:sphere_spinning_various_eta}. The evolution of the angular velocity $\omega$ over time and the orientation of the contact frame $(u,w)$ at the end of the simulation ($t = 8 \si{sec}$) are plotted in Fig.~\ref{fig:sphere_spinning_various_eta} using matching colors. The $(u,w)$ contact frame at the beginning of the simulation is shown in black. Note the overshoot in $\omega$, shown in the inset, which indicates that the sphere changes its spinning direction before coming to a stop.\\
	\FloatBarrier
	The second set of numerical experiments simulated the spinning of spheres made of different types of material with curvature $\mathcal{K}$ derived from the Hertzian contact theory, see  Appendix~\ref{sec:appendix_spinning_kappa}. Radius $R$, density $\rho$, Young's modulus $E$ and Poisson ratio $\nu$ are listed in Table~\ref{table:spinning_sphere_physics_based_parameters}. 
	\begin{table}[!ht]
		\centering
		\caption{Sphere parameters, physics-based model (all SI units).}
		\label{table:spinning_sphere_physics_based_parameters}	
		\begin{tabular}{c c c c  c c c c c} 
			\hline
			Test $\#$ & material & $R $ &  $\rho $  & $E $ & $\nu$  \\
			\hline
			1  &   steel  &  $0.02$ &  $8 \times 10^3$ & $2 \times 10^{11}$ & $0.3$  \\ 
			\hline
			2  &  steel  &  $0.04$ &  $8 \times 10^3$ & $2 \times 10^{11}$ & $0.3$  \\ 
			\hline
			3  &  glass &  $0.02$ &  $2.5 \times 10^3$ & $5 \times 10^{10}$ & $0.2$  \\ 
			\hline
		\end{tabular}
	\end{table}
	
	For each test, four values of sliding stiffness $K_E$ are used, $5\times10^5$, $10^6$, $5\times10^6$, and $10^7 \si{N/m}$. Both the spinning history $\Psi_{ij}$ and angular velocity are reported in Fig.~\ref{fig:sphere_spinning_hertzian_contact}. For the same sphere, a smaller $K_E$ results in a larger slide and spin micro-deflection thresholds, $\SlackSsTh$ and $\SlackPsisTh$. Therefore, it takes longer for the sphere to initially switch from static to kinetic mode. A smaller $K_E$ also results in smaller spinning stiffness $K_E$ and damping coefficient $D_{\psi}$, which leads to a larger overshoot of the angular velocity when the sphere comes to a stop. However, $K_E$ does not affect the kinetic spinning torque, therefore, the acceleration during kinetic mode is the same as illustrated by the same slope in each angular velocity plot.
	\begin{figure}[!ht]
	\centering
	\subfloat[Test $\#$1, $R = 0.02\si{m}$, steel ball.]{{\includegraphics[width=0.6\textwidth]{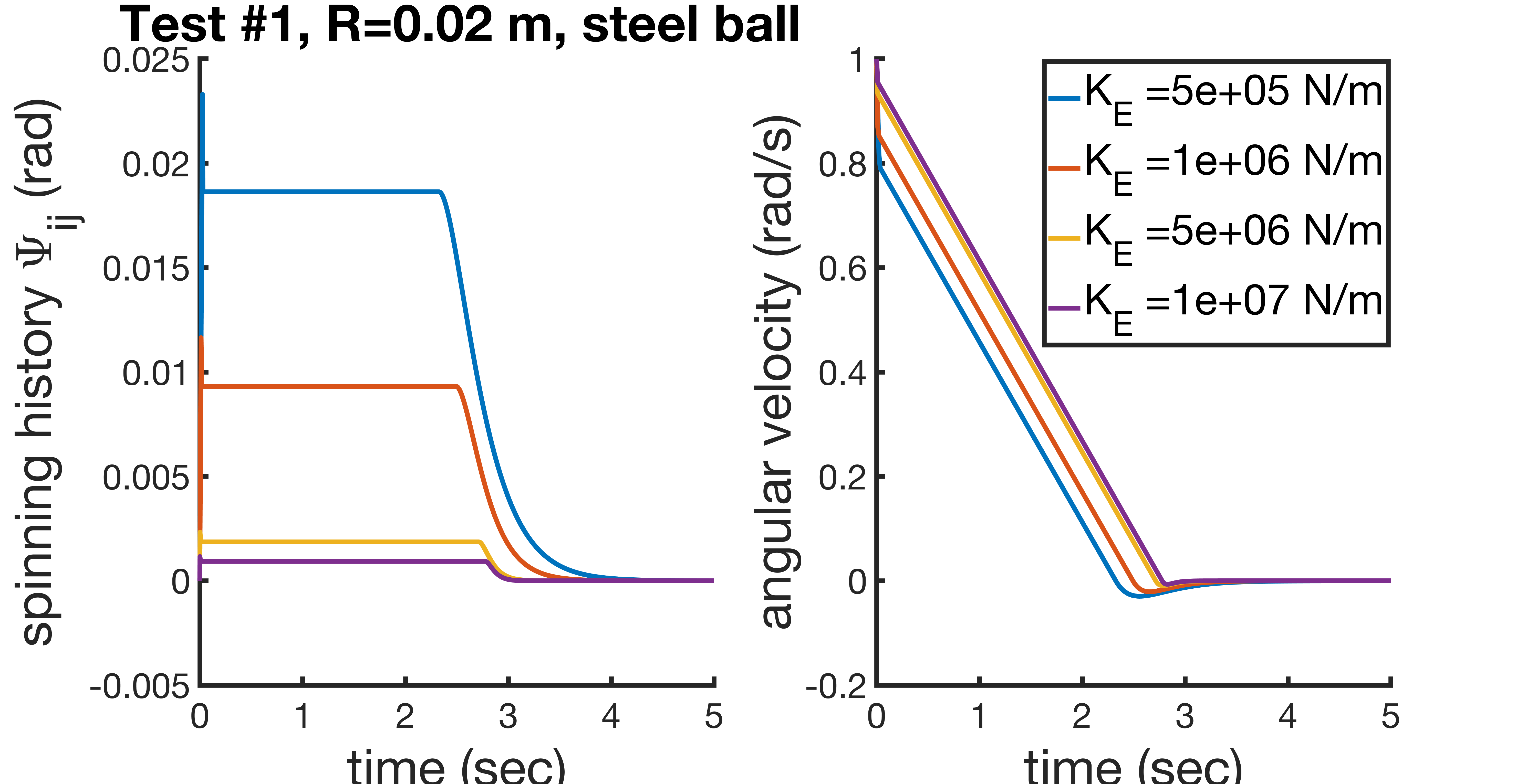} }} 
\end{figure}
\begin{figure}
\ContinuedFloat
\centering	
	\subfloat[Test $\#$2, $R = 0.04\si{m}$, steel ball.]{{\includegraphics[width=0.6\textwidth]{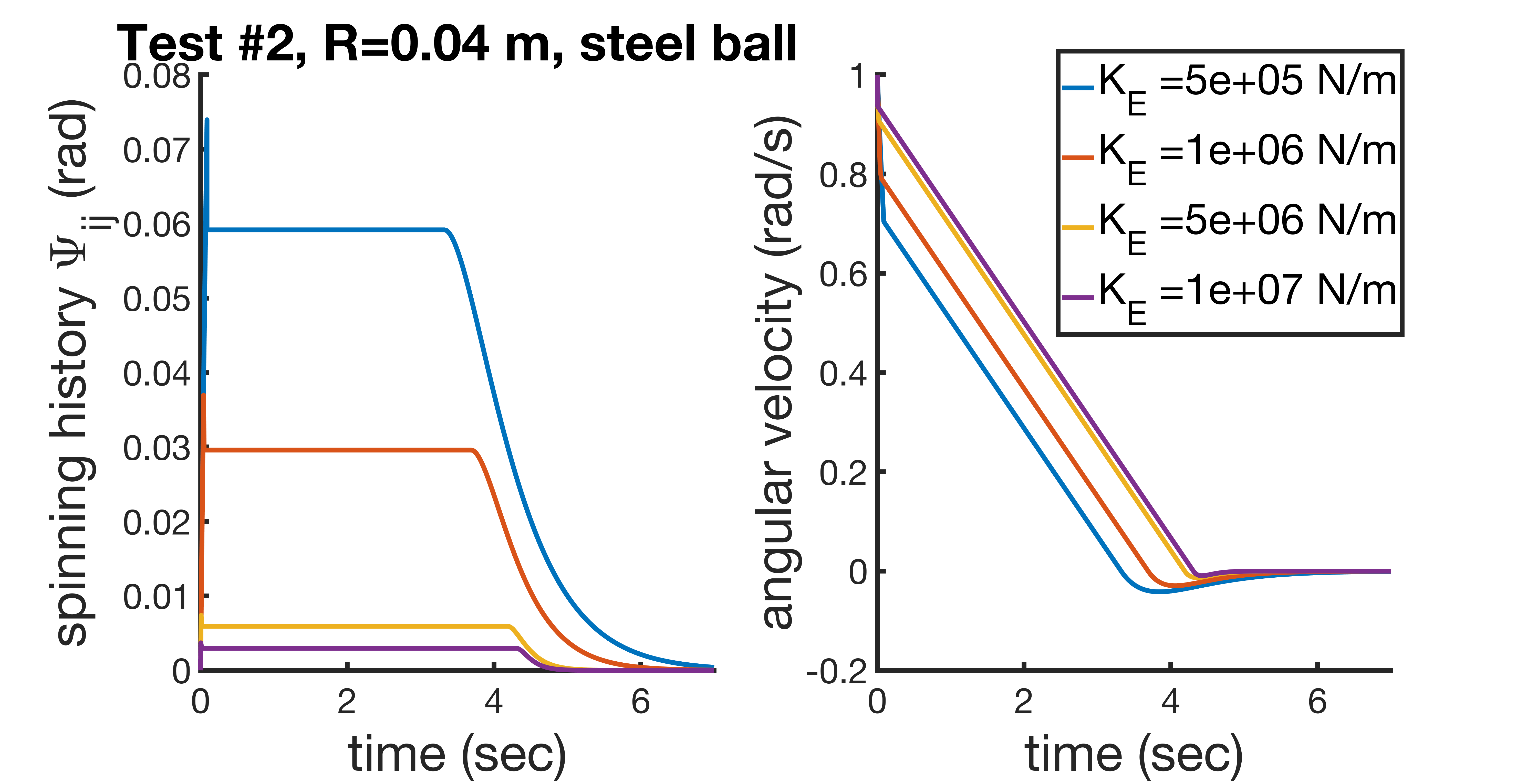} }} \\
	\subfloat[Test $\#$3, $R = 0.02\si{m}$, glass ball.]{{\includegraphics[width=0.6\textwidth]{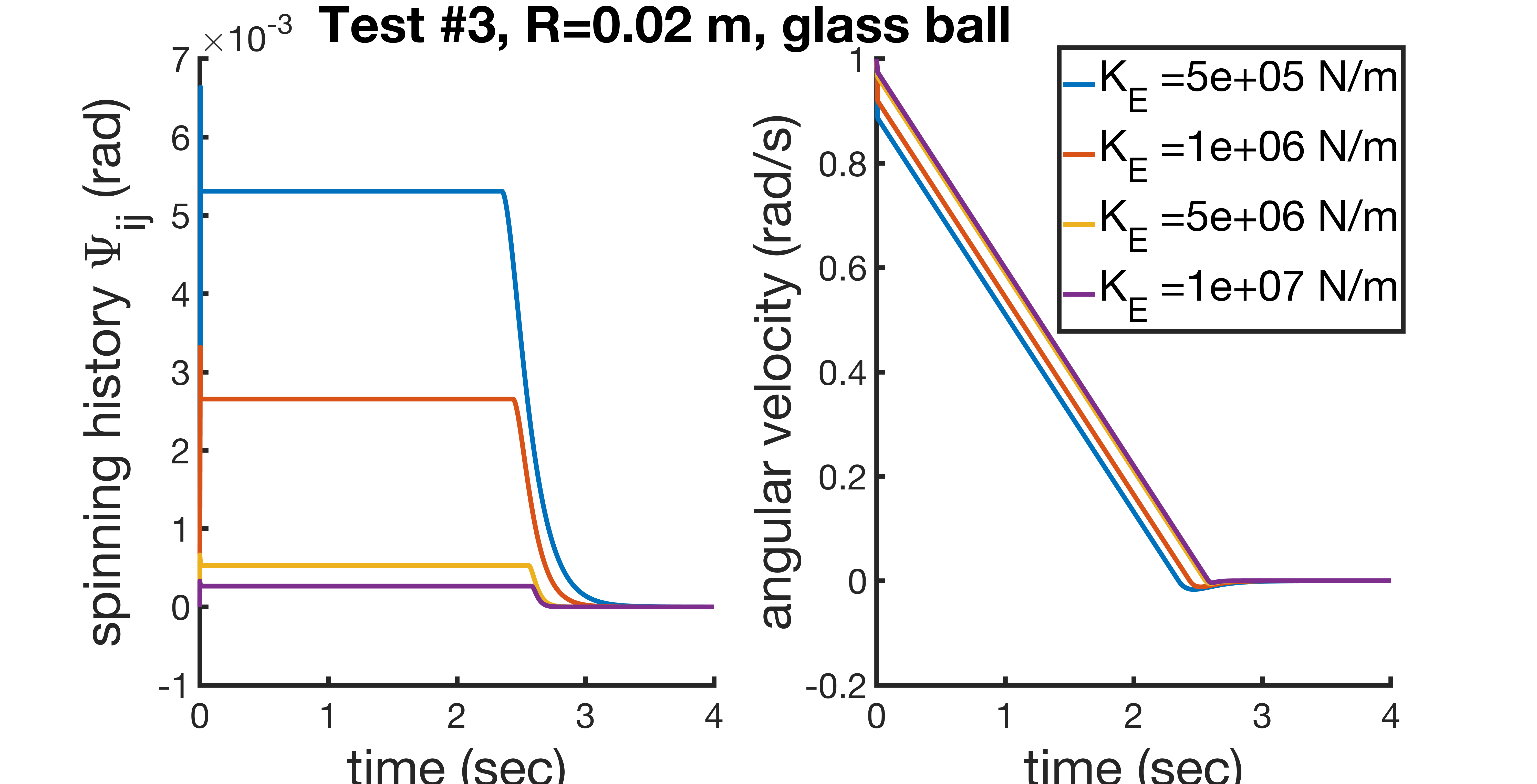} }}%
	\caption{Sphere spinning; physics-based model.}	
	\label{fig:sphere_spinning_hertzian_contact}
\end{figure}
	
	\FloatBarrier
	%%%%%%%
	\subsection{Generalized Motion of an Ellipsoid}
	\label{subsec:generalizedEllipsoid}
	Ellipsoids provide in many cases good approximations for grain shape in granular materials. However, they are not commonly used in DEM most likely owing to nontrivial collision detection and 3D kinematics. Collision detection comes into play when computing normal forces, a task that was bypassed thus far. Indeed, note that in the numerical experiments discussed the {\textit{normal}} contact force, a key ingredient in any friction model, could be calculated analytically. For ellipsoids, an analytical solution is not available for the normal force unless the line of action of gravity passes through the point of contact. Thus, a normal force computational model needs to be brought in. For convenience, a traditional Hookean normal force model \cite{johnson1987contact} is used in section~\ref{subsec:ellipsoidRolling}. Therefore, $N = k_n \delta_n$, where $k_n = 10^7 \si{N/m}$ is the stiffness and $\delta_n$ is the penetration depth, the latter obtained through collision detection. One salient aspect when handling ellipsoids is that due to the change in curvature and normal force, both the stiffness of the roll-/spin-friction models and the threshold micro-deflections associated with slide-/roll-/spin models must be updated at every time step. Additionally, one needs to evaluate $\mathbf{p}_i$ and $\bar{\mathbf{p}}_j$ for computing the relative slide. As illustrated in Fig.~\ref{fig:projProcess}, the previous contact point, $C_i^0$, is projected onto the tangent contact plane $\tau$ to produce $\bar{C}_i^0$ such that vector $\bar{C}_i^0 C_i^1$ has the same length of the geodesic trajectory from $C_i^0$ to the current contact point $C_i^1$.  This quantity is trivial to evaluate when the body has the same curvature on any point of its surface, e.g., a sphere or a plane. However, for an ellipsoid, the length of vector $\bar{C}_i^0 C_i^1$ is approximated as the Euclidean distance between the previous and current contact point, $\|{{\cPoint{i}{0}}{\cPoint{i}{1}}}\|$. In other words, the geodesic is approximated with a straight line that connects the two contact points on the surface of the body of interest. In \cite{TR-2020-03}, this approximation led to no significant changes in the results.
	
	\subsubsection{Ellipsoid with Roll and Slide Friction}
	\label{subsec:ellipsoidRolling}
	An ellipsoid of semi-axes, $a$, $b$ and $c$, and mass $m$ is placed on a flat surface in an upright position, see Fig.~\ref{fig:ellipsoid_rolling_initial}. The initial translational velocity of the center of mass is $v_y = 0.3 \si{m/s}$; there is zero initial angular velocity. The ellipsoid properties and the friction model parameters are given in Table~\ref{table:ellipsoid_parameters}. 
	\begin{table}[!ht]
		\centering
		\caption{Ellipsoid parameters (all SI units), used in subsections \ref{subsec:ellipsoidRolling}.}
		\label{table:ellipsoid_parameters}	
		\begin{tabular}{c c c c  c c  c c  c c  c c} 
			\hline
			$m$ & $a$ & $b$ & $c$ & $\mu_s$ &  $\mu_k$ &$K_E$   & $K_D$   & $\eta_r$ \\ 
			\hline
			5 &  0.2&  0.2& 0.5 & 0.25    & 0.2      &  $10^5$& 1414.21 & 0.2  \\ 
			\hline
		\end{tabular}
	\end{table}
	\FloatBarrier
	Poses in the motion of the ellipsoid are shown in Fig.~\ref{fig:ellipsoid_rolling}. Between time $t=0.9 \si{sec}$ to $t=1.8 \si{sec}$, the rolling changes its direction. Eventually, the ellipsoid settles with a flat pose, see Fig.~\ref{fig:ellipsoid_rolling_end}.
	\begin{figure}[!ht]%
		\centering
		\subfloat[Time = $0 \si{sec}$ \label{fig:ellipsoid_rolling_initial}]{{\includegraphics[width=0.25\textwidth]{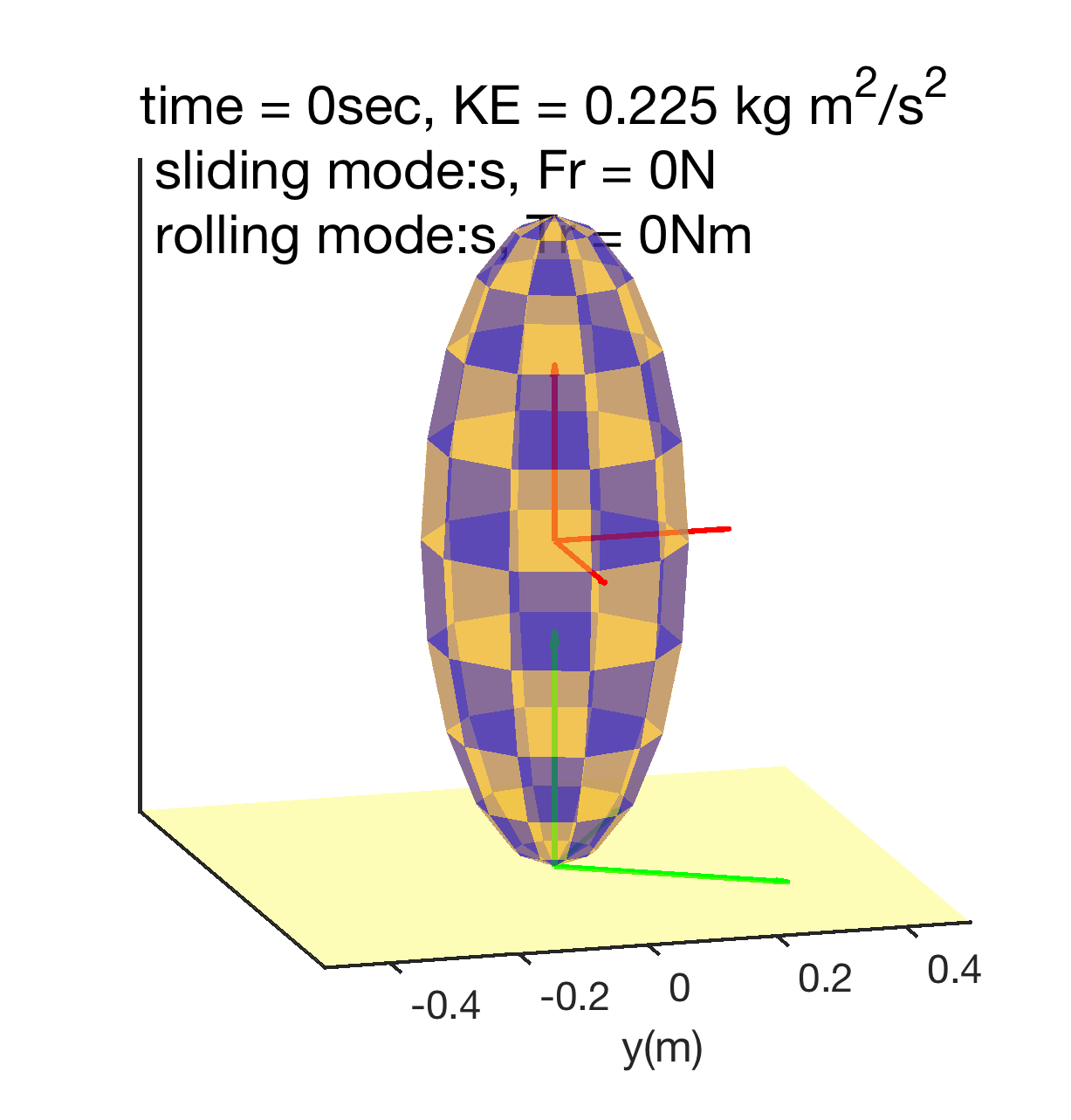} }}%
		\subfloat[Time = $0.75 \si{sec}$]{{\includegraphics[width=0.25\textwidth]{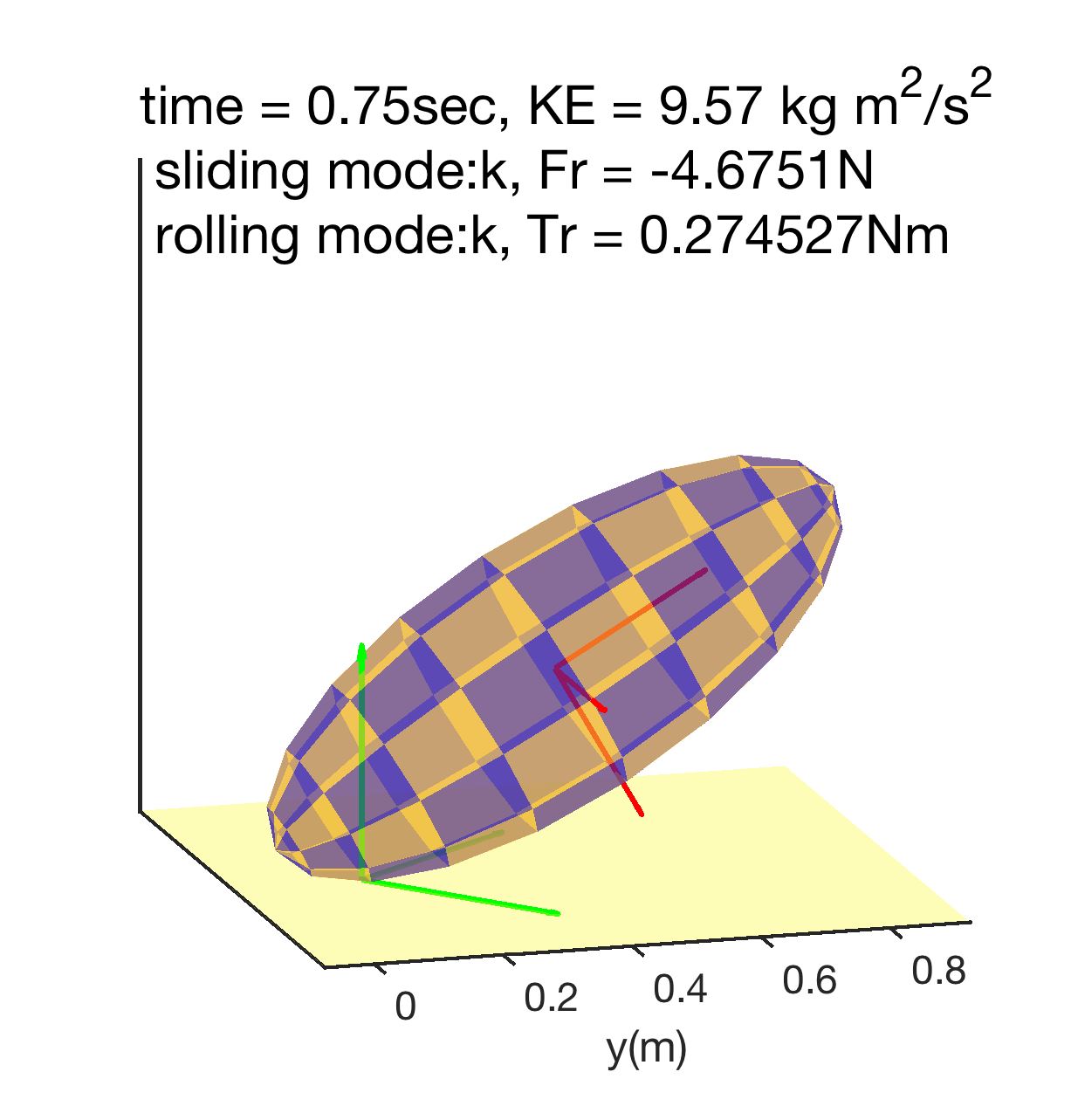} }}%
		\subfloat[Time = $0.9 \si{sec}$]{{\includegraphics[width=0.25\textwidth]{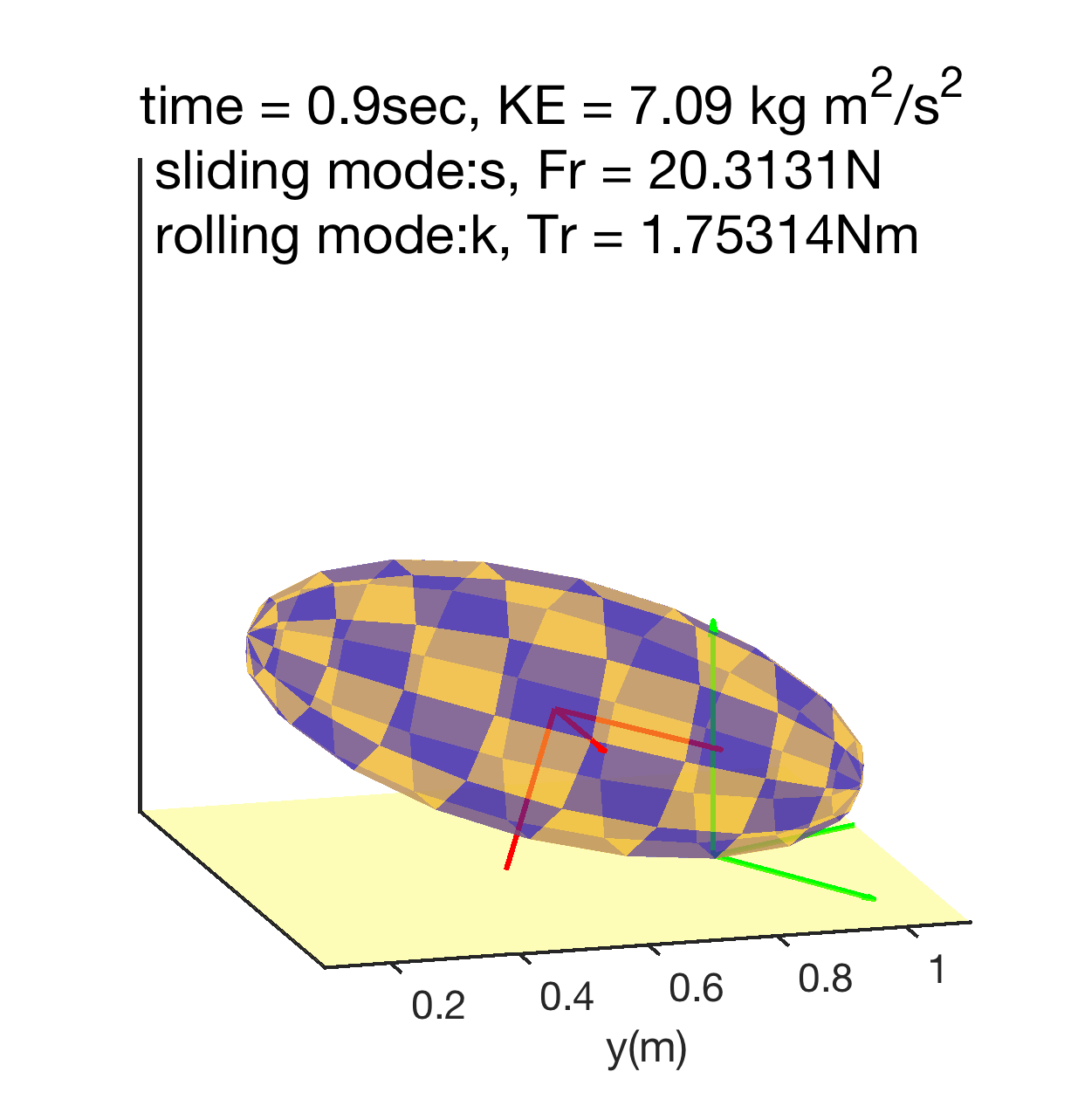} }}
\end{figure}
\begin{figure}
\ContinuedFloat
\centering			
		\subfloat[Time = $1.8 \si{sec}$]{{\includegraphics[width=0.25\textwidth]{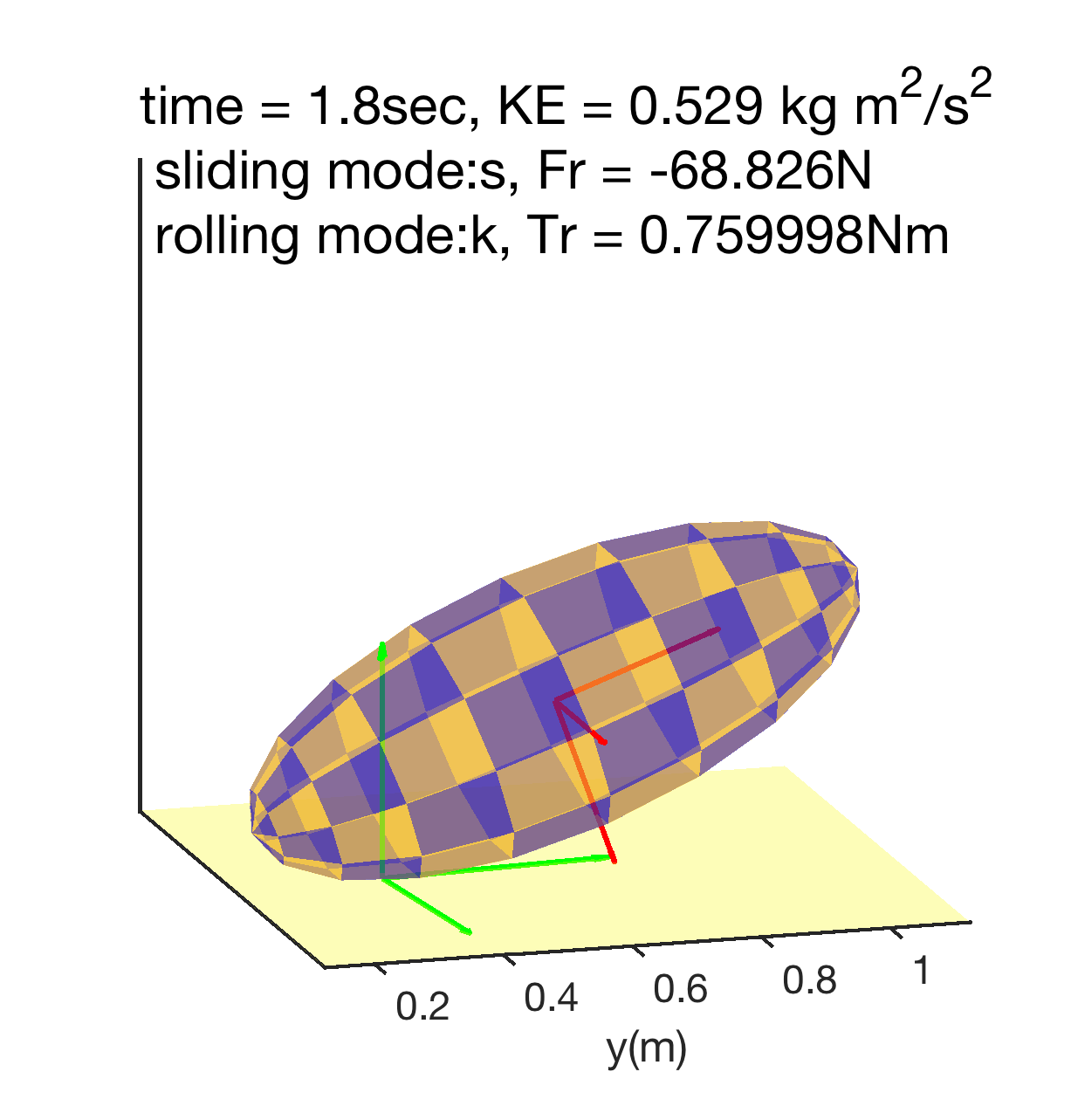} }}%
		\subfloat[Time = $2.1 \si{sec}$]{{\includegraphics[width=0.25\textwidth]{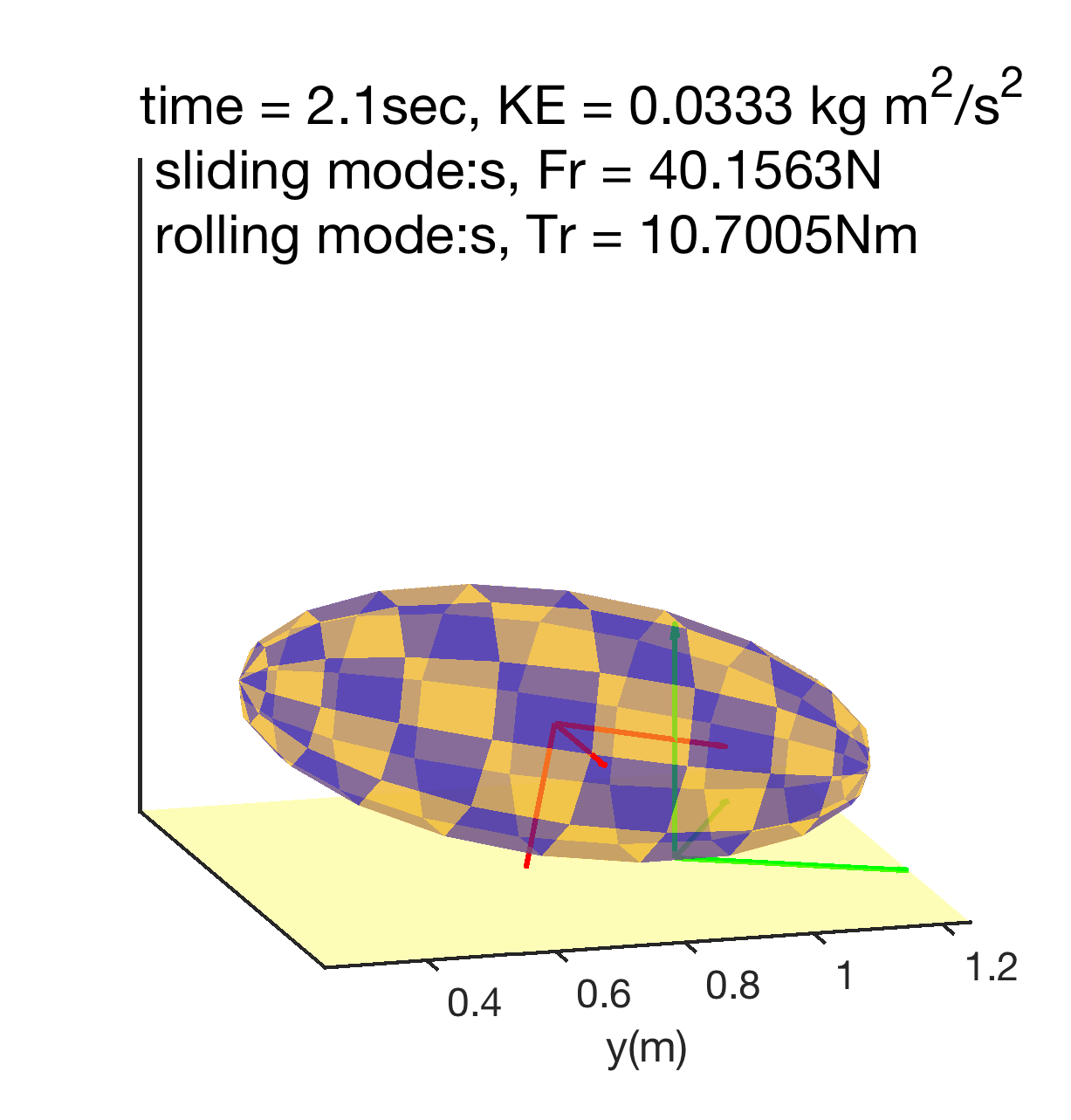} }}%
		\subfloat[Time = $3 \si{sec}$ \label{fig:ellipsoid_rolling_end}]{{\includegraphics[width=0.25\textwidth]{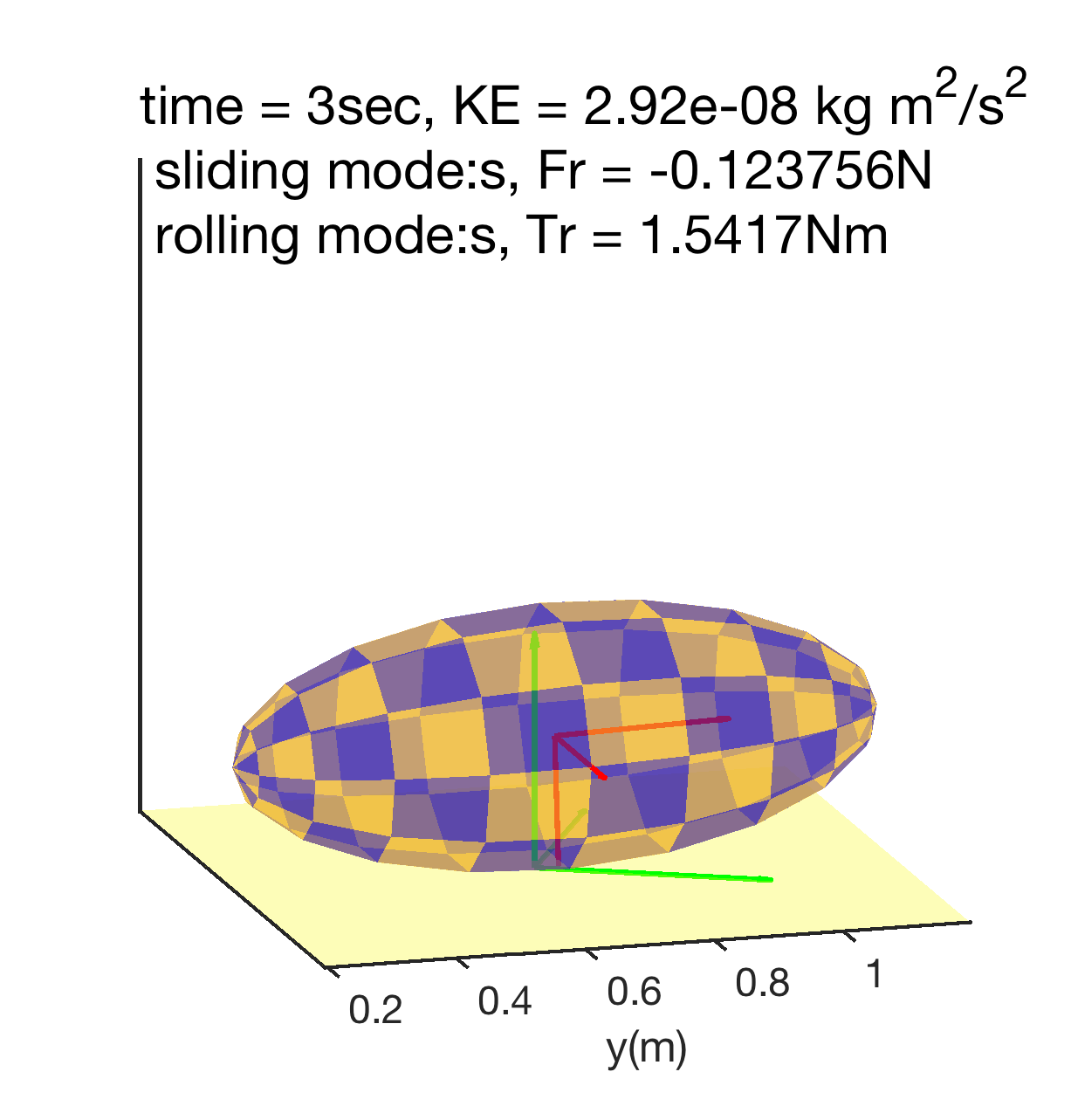} }}%
		\caption{Snapshots of ellipsoid ``falling over''. For closure, a Hookean normal force model was used; the penetration was computed at each time step via collision detection.}
		\label{fig:ellipsoid_rolling}
	\end{figure}
	\FloatBarrier
	\subsubsection{Ellipsoid Spinning}
	\label{subsec:ellipsoidSpinning}
	A steel ellipsoid of semi-axes $a = b = 0.02\si{m}$ and $c = 0.05\si{m}$ is spun with an initial angular velocity $\omega_0 = 1 \si{rad/s}$ about a principal axis; all other angular/translational velocities of the center of mass are zero. Two initial orientations are considered -- upright and flat. In the upright case (Fig.~\ref{fig:ellipsoid_spinning_various_eta_stand}), due to the large curvature at the contact point, the spin stiffness is small, which translates into longer time to rest. When the ellipsoid lays horizontally on the ground, there is more spinning friction due to a smaller curvature, which translates into shorter time to rest (Fig.~\ref{fig:ellipsoid_spinning_various_eta_lay}). In line with expectations, the simulation results suggest that a smaller contact area (higher curvature) leads to less spinning resistance. Table~\ref{tab:ellipsoid_spinning_comparison} outlines the effect of curvature at contact point on spinning-related parameters, such as stiffness $K_{\psi}$, threshold micro-spinning deflection $\mathcal{S}_s^{\psi}$ and kinetic spinning friction $\max (\mathcal{T}_E^{\psi})$, given that the normal force is the same. For comparison, parameters of a sphere with the same volume as the ellipsoid, \textit{i.e.}, $R = \sqrt[3]{abc}$ are listed as well.
	\begin{figure}%[!ht]%
		\centering
		\subfloat[Upright.\label{fig:ellipsoid_spinning_various_eta_stand}]{\includegraphics[width=4.34in]{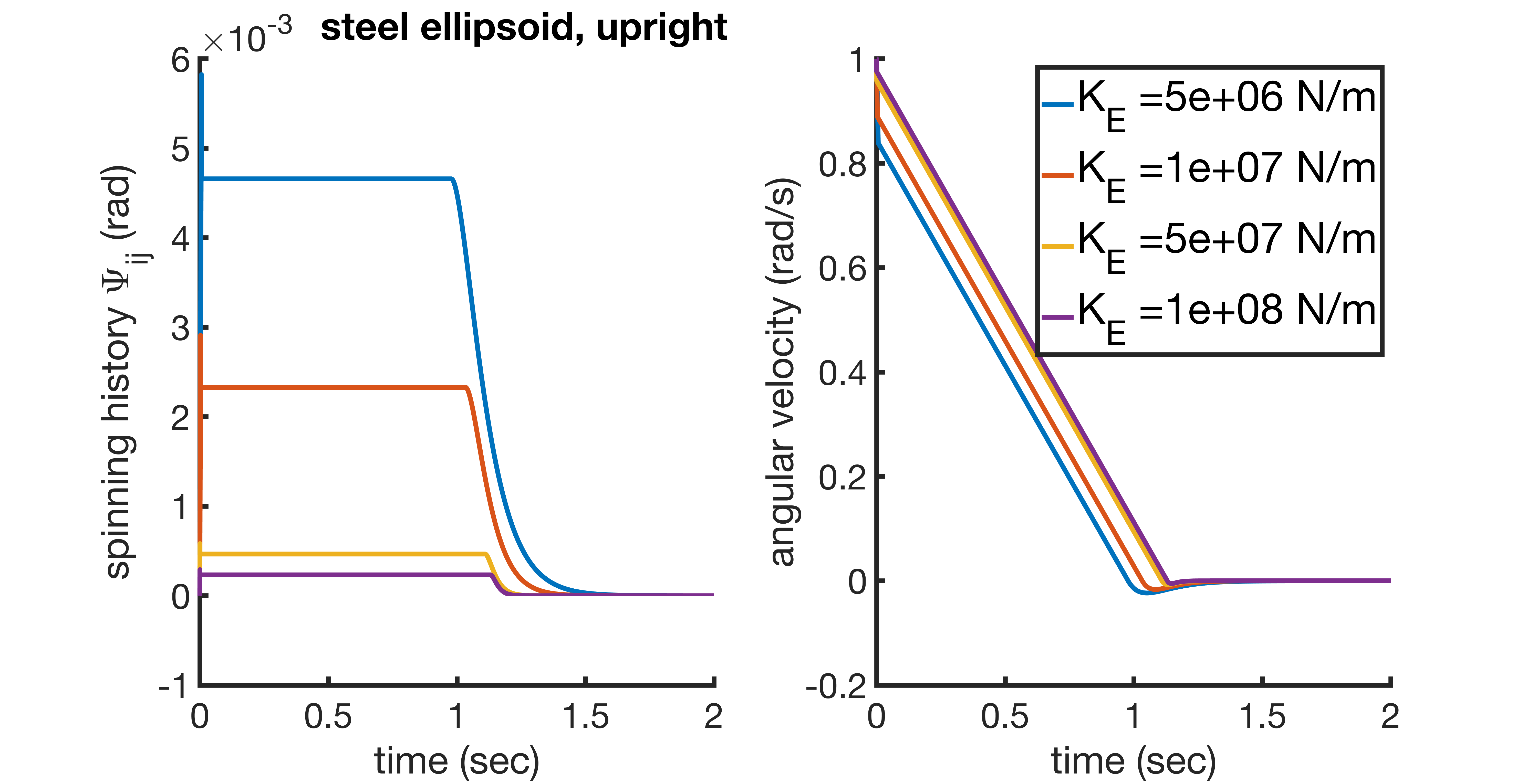}}
\end{figure}
\begin{figure}
\ContinuedFloat
\centering	
		
		\subfloat[Flat.\label{fig:ellipsoid_spinning_various_eta_lay}]{\includegraphics[width=4.34in]{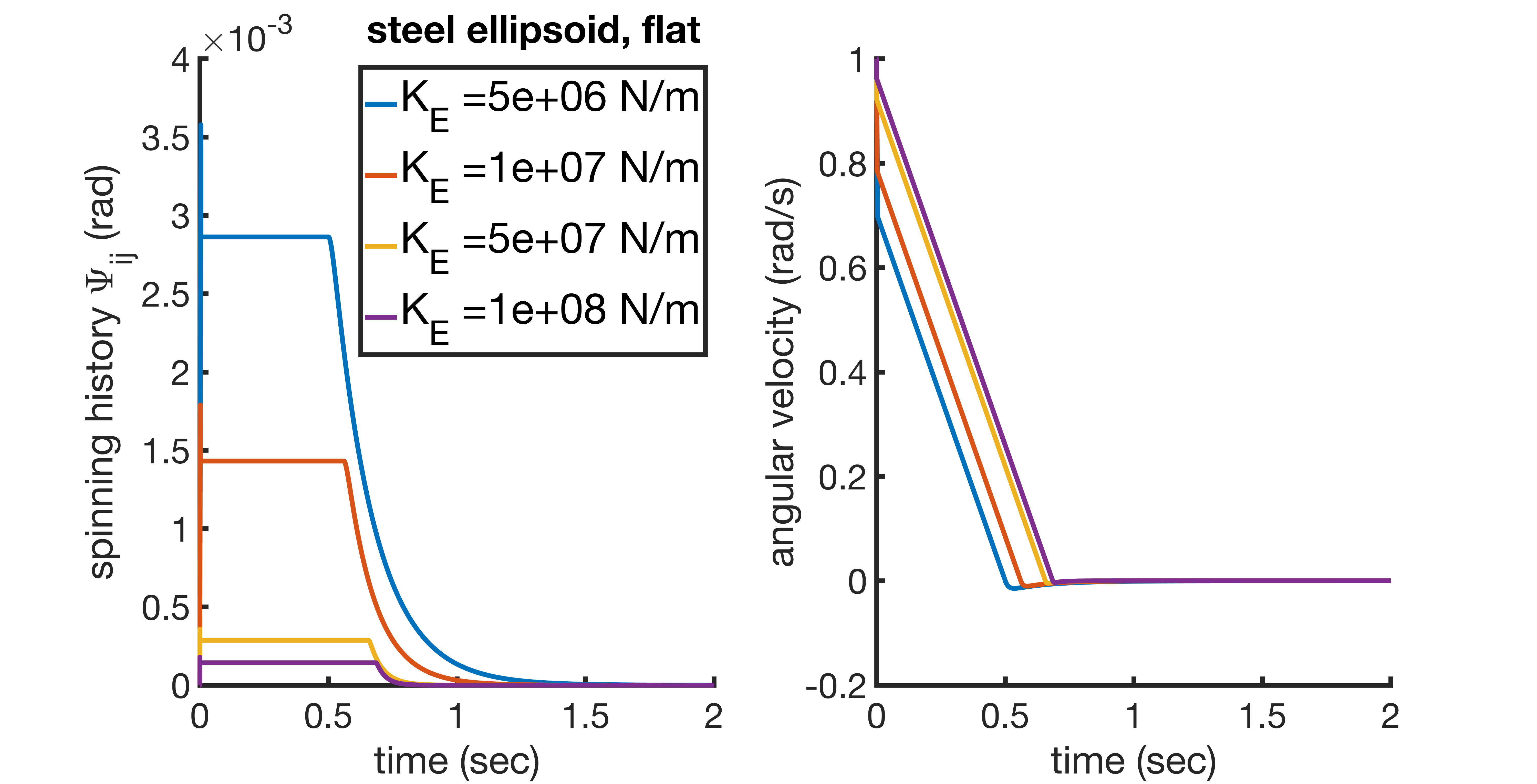}}
		\caption{Ellipsoid spinning.}
	\end{figure}
	\begin{table}[h]
		\caption{Effect of curvature on spinning using $K_E = 5\times10^6 \si{N/m}$.}
		\begin{center}
			\label{tab:ellipsoid_spinning_comparison}
			\begin{tabular}{c l l l}
				\hline
				& sphere & ellipsoid (flat)  & ellipsoid (upright)  \\ \hline
				$\mathcal{K}$                         &36.84   &29                  & 125 \\
				$K_{\psi}$                               &$1.80\times 10^{-2}$&$2.11\times 10^{-2}$                & $7.95 \times 10^{-3}$\\
				$\mathcal{S}_s^{\psi}$            &$3.88\times 10^{-3}$&$3.58\times 10^{-3}$ & $5.82\times 10^{-3}$ \\
				$\max (\mathcal{T}_E^{\psi})$ &$6.96\times 10^{-5}$&$7.53 \times 10^{-5}$              & $4.63\times 10^{-5}$ \\
				\hline
			\end{tabular}
		\end{center}
	\end{table}

	\FloatBarrier
	
	%%%%%%%
	\subsection{Spheres Stacking}
\begin{figure}[!ht]%
		\centering
		\subfloat[Stable.  \label{fig:stack_low_eta_stable}]{{\includegraphics[width=0.45\textwidth]{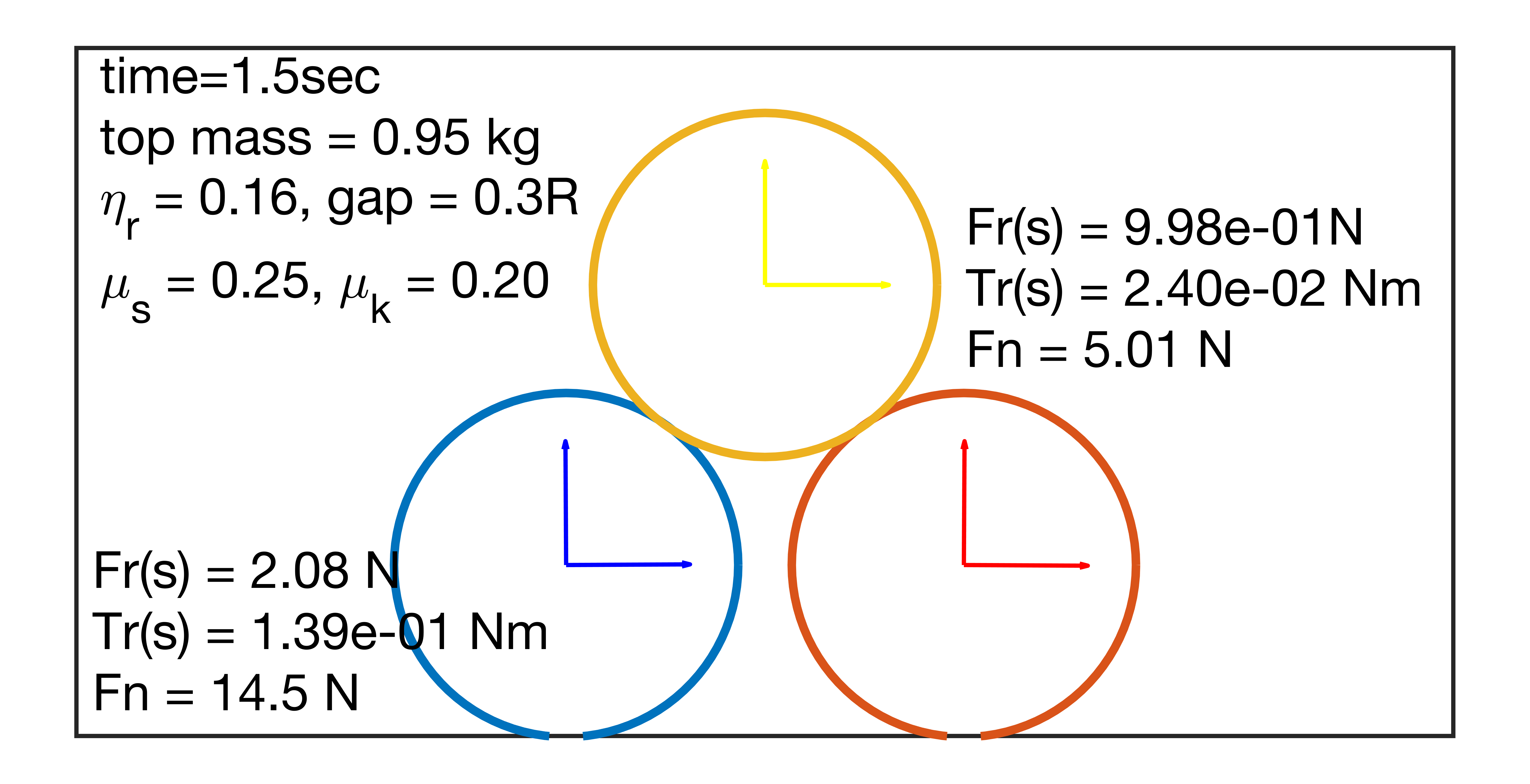} }}
		\subfloat[Roll friction saturates. \label{fig:stack_low_eta_unstable}]{{\includegraphics[width=0.45\textwidth]{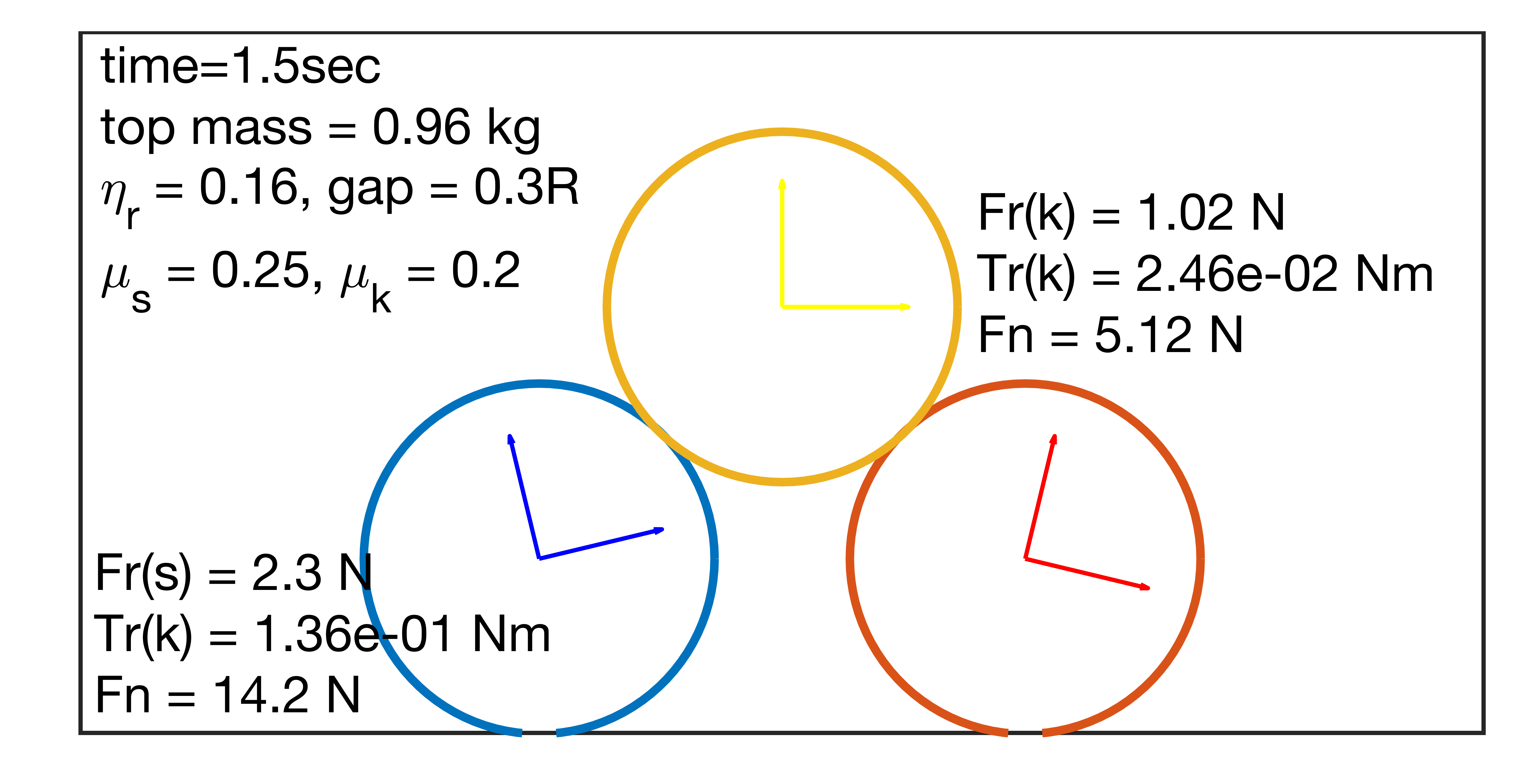} }} \\
		\subfloat[Stable. \label{fig:stack_med_eta_stable}]{{\includegraphics[width=0.45\textwidth]{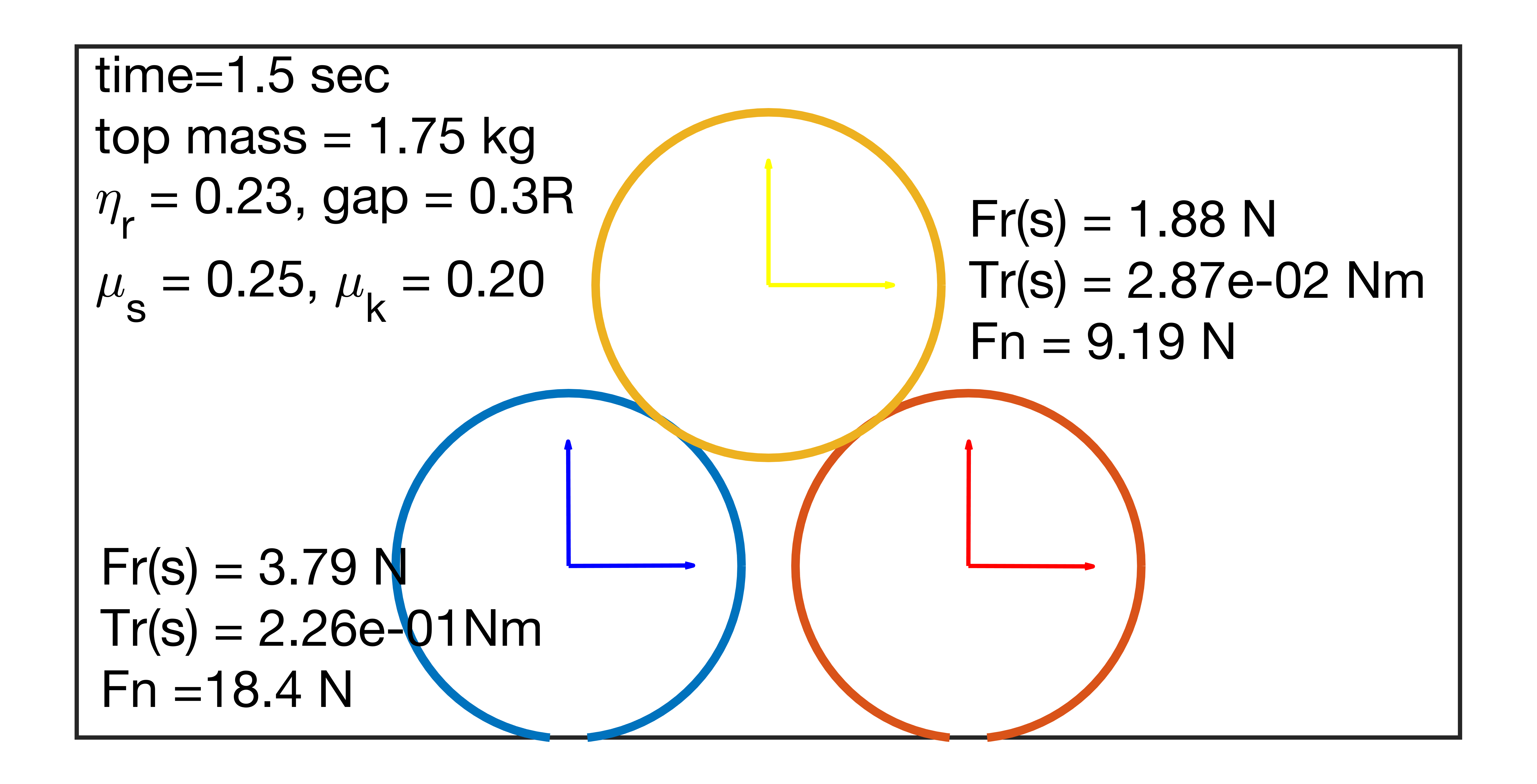} }}
		\subfloat[Slide and roll friction saturate. \label{fig:stack_med_eta_unstable}]{{\includegraphics[width=0.45\textwidth]{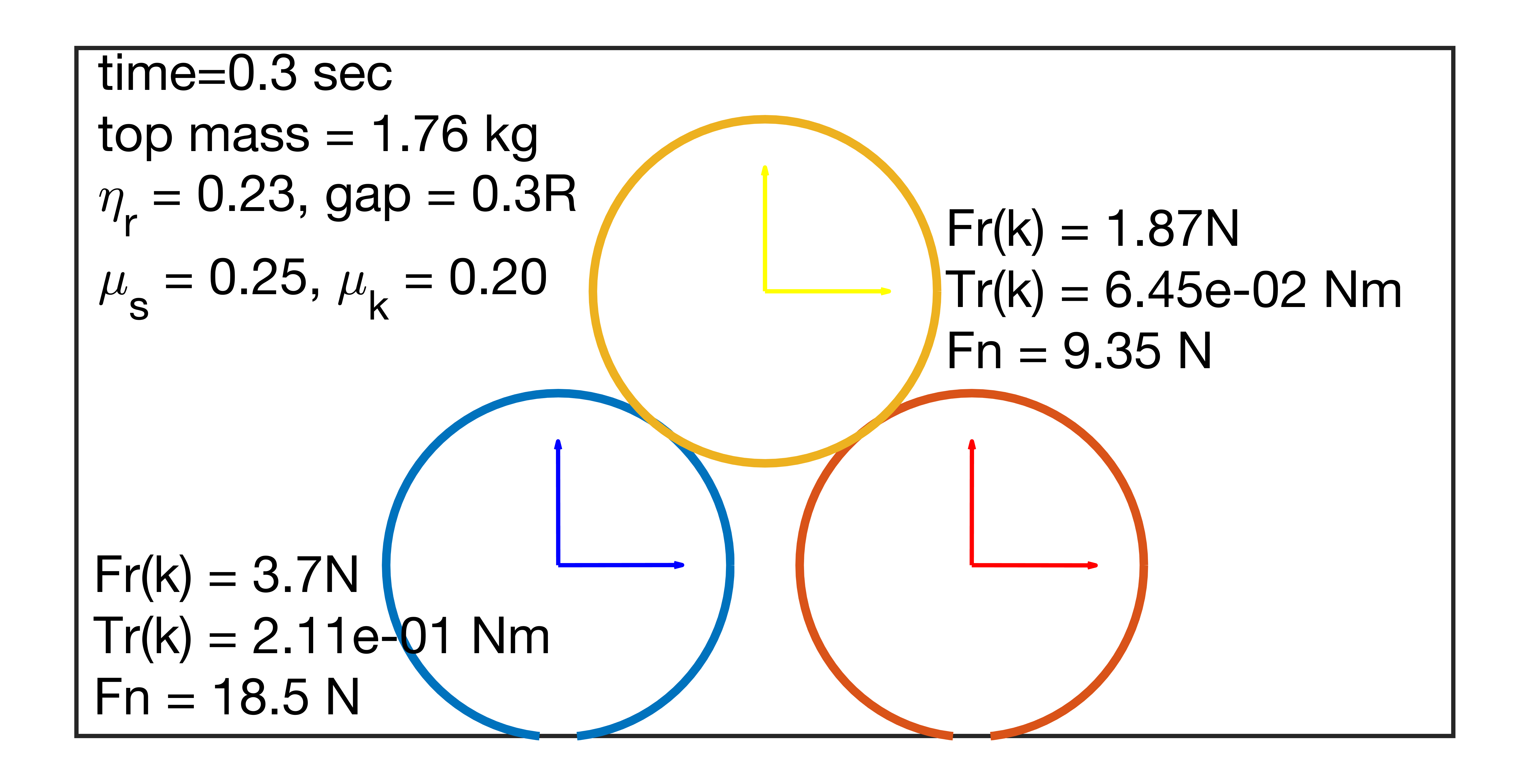} }} \\
		\subfloat[Stable. \label{fig:stack_high_eta_stable}]{{\includegraphics[width=0.45\textwidth]{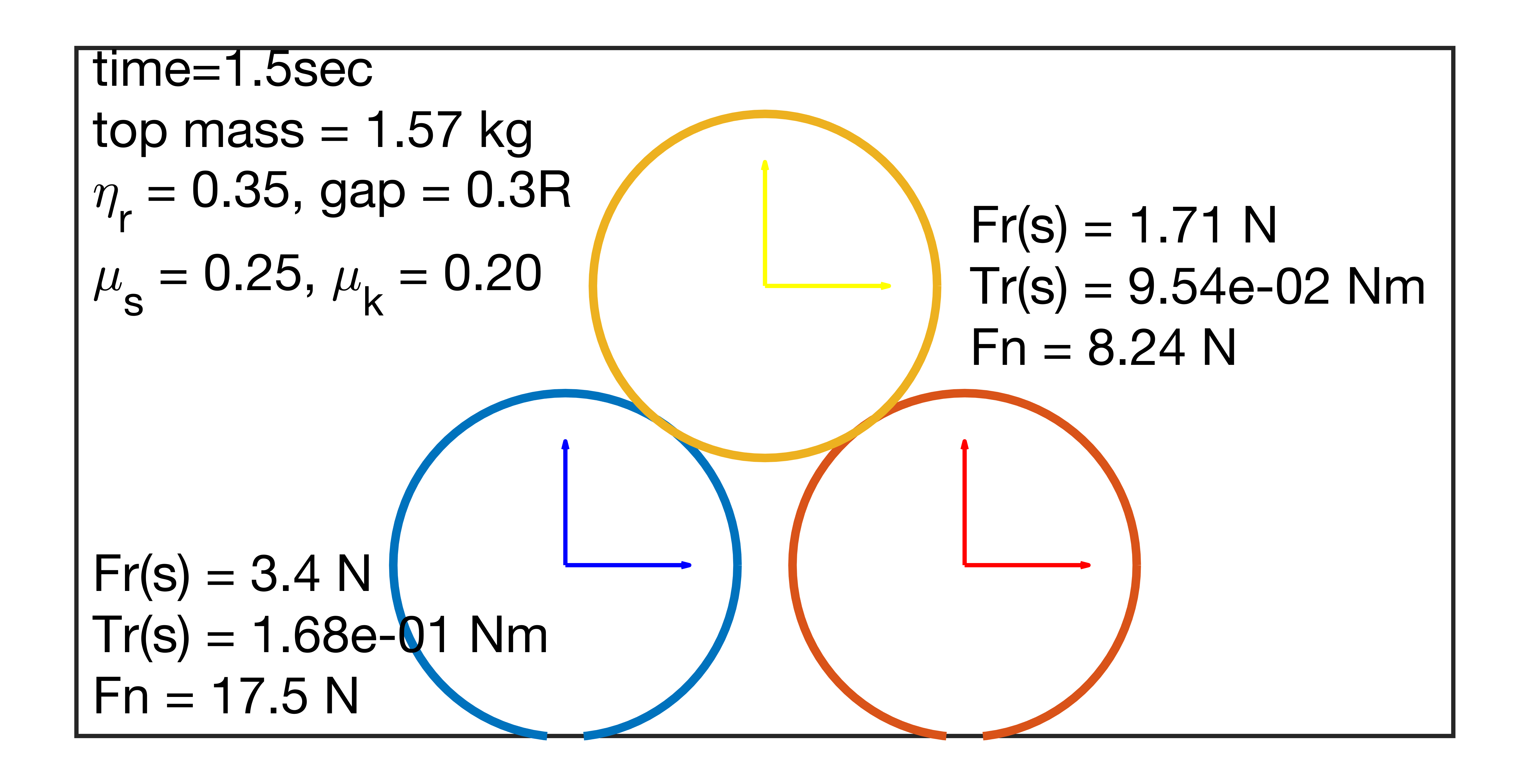} }}
		\subfloat[Slide friction saturates. \label{fig:stack_high_eta_unstable}]{{\includegraphics[width=0.45\textwidth]{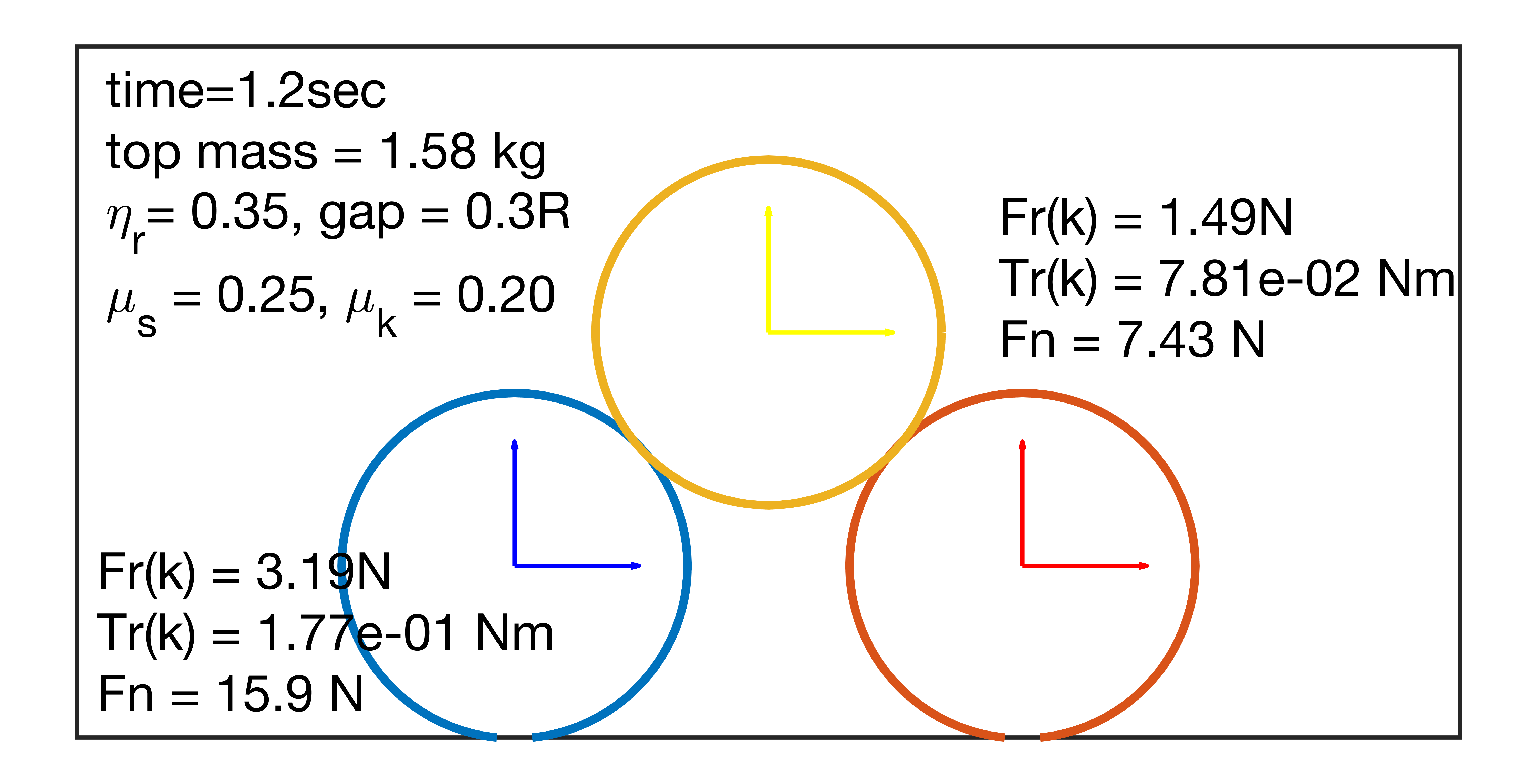} }} \\
		\caption{Snapshots of three-sphere stacking problem.}
		\label{fig:sphere_stacking}
	\end{figure}

	In Fig.~\ref{fig:sphere_stacking}, two identical spheres (blue and red) of mass $m=1 \si{kg}$ and radius $R = 0.15 \si{m}$ are placed on a horizontal flat surface; the distance between their centers is $2.3 R$ ($0.3R$ gap). A third sphere (yellow) of the same radius $R$ but a different mass $m_{top}$ is placed right in-between and above the red and blue spheres. Two scenarios are possible, outcomes of the interplay between the roll and slide frictions: the yellow sphere drops down; or, after it moves a bit, the three-sphere structure stabilizes. To minimize the influence from impact, the top sphere is barely touching the bottom ones when it is released at the beginning of the simulation. The sphere-sphere and sphere-ground contact are described with the same friction parameters, $\mu_s$, $\mu_k$, $K_E$ and $\eta_r$. Since the computation of the normal contact force is nontrivial, e.g., when the bottom spheres roll and/or slide, an open-source physics-based simulation platform, {\CHRONO} \cite{chronoOverview2016,projectChronoWebSite}, is used to perform the test. The normal contact force is computed using a Hertzian contact model \cite{jonJCND2015} available in {\CHRONO}, with Young's modulus $E = 2 \si{MPa}$, Poisson ratio $\nu = 0.3$, and coefficient of restitution $0.4$. The friction loads are computed according to the model discussed herein.
		\begin{figure}[!ht]
		\centering
		\includegraphics[width=4.in]{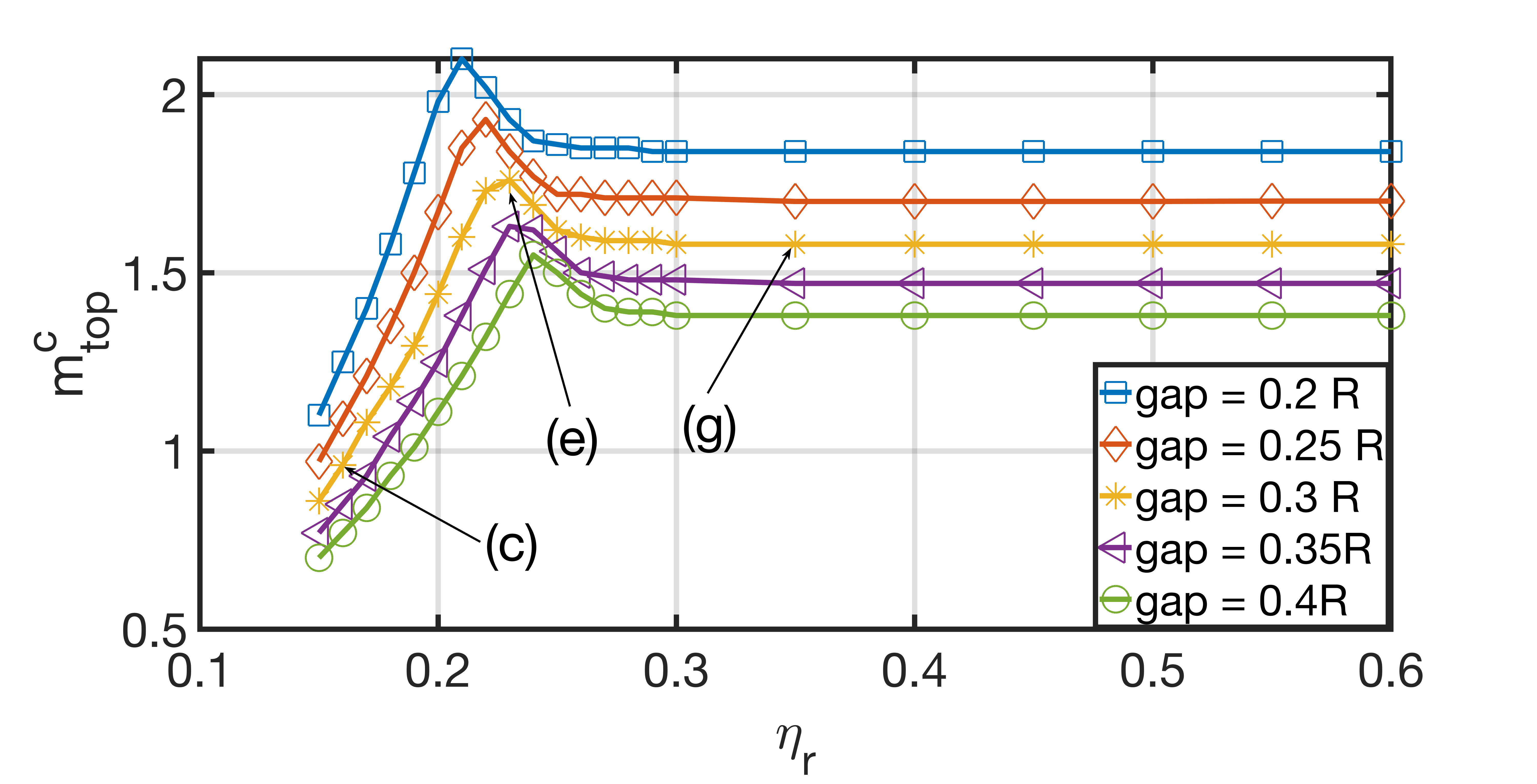}	
		\caption{Plot of $m_{top}^c$; i.e., minimum mass of top sphere for the pile to collapse.}
		\label{fig:stack_top_mass_vs_eta}
	\end{figure}
	
	Figure~\ref{fig:stack_top_mass_vs_eta} reports as a function of $\eta_r$ the smallest mass $m_{top}$ for which the top sphere collapses to the ground; this critical value is called $m_{top}^c$. Several curves are shown, each associated with an initial gap between the red and blue spheres. For all initial gaps, $m_{top}^c$ increases with $\eta_r$ up to a certain point, $\eta_r^c$. For $\eta_r>\eta_r^c$ the critical mass decreases until it reaches a plateau -- increasing $\eta_r$ no longer influences the stability of  the stack. Figures~\ref{fig:stack_low_eta_stable}, ~\ref{fig:stack_med_eta_stable} and~\ref{fig:stack_high_eta_stable} are snapshots of a stable stack when $\eta_r = 0.16$, $0.23$ and $0.35$, respectively, where both sliding and rolling mode of all contacts are static. In comparison, Fig.~\ref{fig:stack_low_eta_unstable},\ref{fig:stack_med_eta_unstable}, and~\ref{fig:stack_high_eta_unstable} capture how the stack starts to collapse with a slightly heavier load (increased by $0.01 \si{kg}$), as the top sphere drops and pushes the bottom ones to the side. When $\eta_r = 0.16$, the rolling mode of the bottom spheres with the ground is kinetic, while the sliding mode remains static. The bottom spheres roll outward, illustrated by the body frame attached, see Fig.~\ref{fig:stack_low_eta_unstable}. When $\eta_r$ increases to $0.23$, more rolling friction restricts the bottom spheres from rolling away, therefore more weight can be supported, until slide friction mode also becomes kinetic. In Fig.~\ref{fig:stack_high_eta_unstable}, $\eta_r = 0.35$, the slide friction force saturates while the rolling mode is still static, implying that when the pile collapses, the bottom spheres slide outward instead of rolling; i.e., there is no angular velocity, only translational velocity. Since slide friction saturates before the roll friction, the amount of weight the stack can support depends on $\mu_s$, rather than $\eta_r$, which explains the flat part of the curve in Fig.~\ref{fig:stack_top_mass_vs_eta}.
\FloatBarrier	
	%=================================================
	\section{Conclusions and Future Work}
	\label{sec:conclusions}
	A phenomenological model is proposed to produce slide, roll and spin friction loads at the interface between two 3D bodies in mutual contact. At a minimum, the model calls for five basic parameters: $\mu_s$, $\mu_k$, $K_E$, $\eta_r$, and $\eta_{\psi}$ -- the last two necessary only if producing roll and spin friction loads is a matter of interest. The model includes more parameters yet heuristics have been discussed in relation to how to choose these second group of parameters from the five basic ones. The simultaneous computation of the three friction loads is anchored by a methodology that produces relevant kinematic information by tracking the motion of the contact point on the surface on the two bodies in mutual contact; in fact, this is the salient contribution of the effort. The force-displacement relations for slide force, roll torque and spin torque are classical -- a spring-dashpot model is used with elongation (micro-deformation) capped and damping acting during stick mode only. In producing the slide/roll/spin loads, the model allows for two distinct friction modes, static and kinetic, through tracking the mode at the previous time step and the micro-deflection dictated by how the contact point moves on the two surfaces. Several numerical experiments were carried out to gauge the predictive attributes of the proposed model. The numerical results have been validated against analytical solutions for several trivial cases that included the dynamics of a sphere, disk, and ellipsoid. A stacking problem was considered to highlight, just as in the case of the ellipsoid, the interplay between the slide and roll friction. The stacking problem required the implementation of the friction model in an open source dynamics engine that provided the normal loads entering the friction computation.
	
	The next step is to implement this model in a simulation engine for granular dynamics ({\CHRONO}::Granular) in order to study problems with millions of degrees of freedom. This is an ongoing project. 
	
	%=================================================
	\section*{Acknowledgments}
	\noindent This work was supported by US Army Research Office grant W911NF1910431.
	%=================================================
	
%\bibliographystyle{unsrt}
%\bibliography{../../SBEL-LaTeX/BibResources/refsMBS,../../SBEL-LaTeX/BibResources/refsDEM,../../SBEL-LaTeX/BibResources/refsCompSci,../../SBEL-LaTeX/BibResources/refsTerramech,../../SBEL-LaTeX/BibResources/refsFSI,../../SBEL-LaTeX/BibResources/refsChronoSpecific,../../SBEL-LaTeX/BibResources/refsSBELspecific}
\def\cprime{$'$}

	\begin{appendices}
		%=================================================
		\section{Discussion of other friction models}
		\label{sec:discussion}
		This section provides a brief overview of literature; and, it allows for a discussion of how the proposed approach to friction load computation compares with other approaches used in the field.
		
		\subsection{Slide-Friction Models}
		\label{sec:appendix_slide_friction_review}
		In \cite{mindlin53}, the authors proposed for the elastic friction force between two identical spheres a tangential force-displacement relation that depends on the loading history and satisfies the saturation condition $F_t \leq \mu F_n$. The tangential force $F_t$ is updated based on the change in the tangential displacement $\delta_s$,
		\begin{equation}
		\label{eq:Mindlin_Ft_increment}
		F_t \coloneqq F_t + K_t \delta_s \; .
		\end{equation}
		The tangential stiffness $K_t$ depends on material properties, radius of contact patch, friction coefficient $\mu$, normal force $F_n$, current tangential force, and the loading-unloading-reloading history. The authors summarized eleven different scenarios for updating $K_t$. This scheme is simplified in \cite{walton1986viscosity,vu1999accurate} by casting the computation of $K_t$ into two and four scenarios, respectively. The friction force saturation is enforced by comparing $\Delta F_t$ and $\mu \Delta F_n$. If $\Delta F_t < \mu \Delta F_n$, the tangential force is incremented by Eqn.~(\ref{eq:Mindlin_Ft_increment}); otherwise, both $K_t$ and $\delta_s$ are adjusted. Their models have been used to predict wave propagation of sphere particles \cite{sadd1993contact}, chute flow of soy beans \cite{zhang2000simulation}, and particle suspension flow in viscous liquids \cite{apostolou2008discrete} in DEM.\\
		Instead of updating the tangential force incrementally, in \cite{cundall79} the tangential displacement is updated through the tangential relative velocity $\bm{v}_t$ from the time when contact initiates, $t_0$, to the current time $t$,
		\begin{equation*}
		\bm{\xi} = \int_{t_0}^{t} \bm{v}_t (t') dt'\;.
		\end{equation*}
		The updated displacement $\bm{\xi}$ enters a linear spring-dashpot model with elastic component constrained by the capping condition,
		\begin{equation*}
		F_t = -\min (\mu |F_n|, K_t |\bm{\xi}| ) \bm{t} - D_t \bm{v}_t.
		\end{equation*}
		To dissipate the energy, $D_t$ is used to mimic critical damping, $D_t = \sqrt{m K_t}$, which is switched off when full sliding occurs. The choice of $K_t$ and $D_t$ are empirical. This approach does not differentiate between static and kinetic friction coefficients; and, it misses the kinematic information required to compute in a 3D setup the roll friction load.\\
		In \cite{tsuji1992lagrangian}, $K_t$ is derived based on Herzian contact theory
		\begin{equation}
		\label{eq:defK_t}
		K_t = 8 G_{eff} \sqrt{R_{eff} \delta_n}\;,
		\end{equation}
		where the effective radius $R_{eff} \equiv R_i R_j/(R_i + R_j)$, $G_{eff}$ is the effective shear modulus and $\delta_n$ is the normal penetration. The damping coefficient $D_t$ is set to be the same as the one in the normal direction $D_n$,
		\begin{equation}
		D_t = D_n = \alpha ({mK_n})^{1/2}\delta_n^{1/4}\;,
		\end{equation}
		where $\alpha$ is an empirical parameter dependent on the coefficient of restitution. Due to its simple implementation and the choice of physics-based parameters, the model in \cite{tsuji1992lagrangian} has been widely used, e.g. in applications such as spout fluidized bed \cite{zhong2006simulation}, sand pile formation \cite{liu2011identification}, 3D printing \cite{parteli2016}, four-bladed mixer \cite{remy2009discrete}, etc. \\
		Another approach to implicitly satisfy the saturation condition is to limit the elongation of the spring $\bm{\xi}$ to $\mu |F_n|/K_t$ for slide-mode friction. In \cite{luding2008cohesive}, a test force is computed first, $F_t^0 = - K_t \bm{\xi} - D_t \bm{v}_t$, and compared to the maximum static friction, $\mu_s |F_n|$. If not saturated, the test force is kept as the friction force, and the spring elongation is updated as $\bm{\xi}' = \bm{\xi} + \bm{v}_t \Delta t$. Otherwise, the friction force is in a kinematic regime, in which $F_t = \mu_k |F_n| F_t^0/|F_t^0|$, and the spring length is adjusted to $\bm{\xi}' = -\frac{1}{K_t} (F_t + D_t \bm{v}_t)$.\\
		The slide-mode friction solution advanced herein uses a linear spring-dashpot model, like in \cite{luding2008cohesive}. However, the spring elongation is updated from the relative motion of the contact point on the two surfaces in mutual contact without resorting to using the tangential velocity $\bm{v}_t$. This opens the door to seamlessly factoring in both the static and dynamic friction coefficients. Moreover, we can access a fine level of contact kinematic detail unavailable to the traditional models in which the gathering of kinematic information is based on the relative velocity of the two bodies at the contact point. Essentially, the approach used herein extracts the required kinematic information using collision detection data instead of explicitly relying on an integration in time of the relative velocity between bodies. As further discussed in the next subsection, this time integration is problematic given the SO(3) Lie group structure of the 3D rotations \cite{MarRat94}.

		\subsection{Rolling Friction Models}
		\label{sec:appendix_rolling_friction_review}
		
		When a body rolls over ground or another body, the normal stress profile in the contact patch has an asymmetric distribution with the front half of the contact patch developing higher stress than the back half. This load offset results in a rolling resistance torque. Additional rolling resistance can arise from plastic deformation in and around the contact patch \cite{johnson1987contact}, viscous hysteresis \cite{brilliantov1996model}, and asperities/surface roughness/adhesion at the interface \cite{rollFrictionAsperities1976}. To model rolling friction numerically, a linear spring with a cutoff is introduced in \cite{tordesillas2002incorporating} for the micropolar model of granular media. In \cite{AleMihaiFriction2013}, a complementarity-based rolling friction model is proposed for in the context of nonsmooth dynamics. Some group accounted the effect of rolling by adjusting the tangential displacement of the slide friction model based on the change of contact area due to rolling \cite{vu1999accurate}. Beer and Johnson \cite{beer1976mechanics} first derived theoretically a constant torque model where the rolling friction $M_r$ is proportional to the normal contact force $F_n$. In \cite{zhou1999rolling}, the relation is applied to simulate sand pile formulation, 
		\begin{align}
		\bm{\omega}_{rel}&= \bm{\omega}_i - \bm{\omega}_j \;, \\
		M_r &= -\frac{\bm{\omega}_{rel}}{\|\bm{\omega}_{rel}\|} \mu_r R_{eff} F_n \; ,  \label{eq:rolling_fric_constant_torque_Mr}
		\end{align}
		where $\mu_r$ is the rolling friction coefficient. Due to its simplicity, the constant torque model has been widely used in practice, e.g., in particle flows \cite{chu2008numerical,mccarthy2010quantitative}, blast furnace \cite{zhou2005discrete}, cohesive powder \cite{thakur2016scaling}, and hopper discharge \cite{schwartzdem2012} etc.\\
		In light of the spring-dashpot model for slide friction, a torsional spring-dashpot model is proposed \cite{iwashita1999mechanics} to model the rolling friction in 2D, which has been applied to investigate the flow of rice\cite{markauskas2011investigation}, and strain localization behavior of soil \cite{mohamed2010comprehensive}. The spring rotation is integrated from relative angular velocity. However, this can not be easily extended to 3D owing to the non-trivial connection between the angular velocity of a body and the SO(3) Lie group structure of the 3D rotations. Therefore, the relative angular velocity can not be directly integrated in time \cite{wang2015rolling} and one needs to invoke other approaches, e.g., Euler parameter representations of the 3D rotation \cite{Haug89} or Lie integrators \cite{hairer2006geometric}.\\
		In \cite{wang2009new}, it was shown that the relative rotation between two coordinate systems can be decomposed into two physically and sequentially independent rotations: one pure axial rotation around the line between the centers of two bodies, and the other rotation with an axis perpendicular to a specified plane. Using this result, in \cite{jiang2015novel,holmes2016bending}, the relative angular velocity was split into spinning and rolling components, $\bm{\omega}_{sp}$ and $\bm{\omega}_{r}$, where the spinning component is along the contact normal $\bm{n}$
		\begin{align}
		\bm{\omega}_i - \bm{\omega}_j &= \bm{\omega}_r + \bm{\omega}_{sp}, \\
		\bm{\omega}_{sp} &= \bm{n} \cdot \left[(\bm{\omega}_i - \bm{\omega}_j)  \bm{n}\right] \; . \label{eq:3D_kinematics_spin}
		\end{align}
		Once decomposed, $\bm{\omega}_r$ enters a rotational capped-spring-dashpot model,
		\begin{align*}
		M_r &= M_r^{k} + M_r^d \\
		M_{r, t + \Delta t} ^k & = M_{r, t}^k - K_r \bm{\omega}_r \Delta t, \quad \|M_{r, t + \Delta t} ^k \|  \leq \mu_r R_{eff} F_n, \\
		M_{r, t + \Delta t}^d &= -C_r \bm{\omega_r}.
		\end{align*}
		The rolling stiffness is related to the tangential stiffness through $K_r = K_t R_{eff}^2$. The damping component $M_r^d$ is switched off when $M_r^k$ saturates, and $C_r = \eta_r C_r^{crit} = \eta_r 2 \sqrt{I_r K_r}$. This 3D model is used to examine the quasi-static behavior of sands through tri-axial and plane-strain compression tests \cite{jiang2015novel}. The results reported therein match the peak strength of the typical sands. \\
		An objective relative rolling velocity is defined in \cite{luding2008cohesive} as
		\begin{equation*}
		\bm{v}_r = -R_{eff} (\bm{n} \times \bm{\omega}_i - \bm{n} \times \bm{\omega}_j) \;,
		\end{equation*}
		which is integrated in time to yield a pseudo rolling resistance force $F_r$ through a translational spring-dashpot model. By calculating the moment of the pseudo-force $F_r$, one can evaluate the rolling torque $M_r$ as follows, 
		\begin{equation*}
		M_r = R_{eff} \mathbf{n} \times F_r,
		\end{equation*}
		which conserves angular momentum of the pair. The pseudo-force itself does not act on the center of the mass. The spring elongation of the quasi rolling force is restrained in the same fashion as its slide friction force counterpart. In \cite{luding2008cohesive}, this approach is used for a tensile test simulation of cohesive powders. The results show that the effect of rolling and spinning is very weak for tensile strength. A similar strategy is used in \cite{holmes2016bending} for a rotating drum test to study the discharge of raw material into blast furnaces.\\
		In our model, the rolling friction applied to body $i$ can be different from the one applied to body $j$ when the rolling mode is static, while for other models, the rolling friction applied to body $i$ is the same as its reaction torque applied to body $j$.

		\subsection{Spinning Friction Models}
		The literature on spin-friction torque models is relatively limited as spin friction is rarely considered in simulations. In \cite{lubkin1951torsion}, the theory of elastic contact between two spherical bodies was used as a basis to derive the relation between spinning torque and spinning angle for spherical contact, which led to an expression for the so-called critical spinning torque necessary to initiate free spinning \begin{equation*}
		M_{sp}^{crit} = \frac{3\pi}{16} \mu F_n a \; ,
		\end{equation*}
		where $a$ is the radius of contact patch. The relation was extended to viscoelastic material in \cite{dintwa2005torsion,van2006discrete}. In \cite{marshall2009discrete,yang2013mechanistic}, the approach employed a capped torsional spring-dashpot mechanism for spinning friction. The relative spinning rate was defined as a scalar value
		\begin{equation*}
		\omega_{sp} = \bm{n} \cdot (\bm{\omega}_i - \bm{\omega}_j) \;,
		\end{equation*}
		and the spinning torque evaluated as
		\begin{equation}
		\label{eq:spinFrictionTorque}
		M_{sp} = -K_{sp} \int_{t_0}^{t} \omega_{sp}(t') dt' - D_{sp} \omega_{sp} \; .
		\end{equation}
		The spinning stiffness $K_{sp}$ and damping coefficient $D_{sp}$ are related to the corresponding slide-friction quantities $K_t$ and $D_t$ through the contact radius $a$:
		\begin{equation*}
		K_{sp} = K_t a^2/2, \quad \quad D_{sp} = D_t a^2/2 \;.
		\end{equation*} 
		Thus, the spin friction torque is given by Eqn.~(\ref{eq:spinFrictionTorque}) but capped by a critical torque value  $M_{sp}^{crit} = 2/3 \mu F_n a$; crossing this value marks the transition from stick to slip mode for spin friction. \\
		The same model with almost identical parameters was used in \cite{jiang2015novel}, with one caveat: the angular displacement of the torsional spring was incremented as a vector as in Eqn.~(\ref{eq:3D_kinematics_spin}). \\
		In \cite{luding2006contact,luding2008cohesive}, the relative spinning velocity was defined  as
		\begin{equation*}
		\bm{v}_o = R_{eff} \left[ \bm{n} \cdot (\bm{\omega}_i - \bm{\omega}_j) \right] \bm{n} \;, 
		\end{equation*}
		and was subsequently integrated to yield a pseudo spinning resistance force $F_{sp}$ through a translational spring-dashpot model. The spinning torque, $M_{sp}$, is then calculated using the pseudo force  
		\begin{equation*}
		M_{sp} = R_{eff} F_{sp}. 
		\end{equation*}
		%%%%%%%
		\section{Finding the Contact Frame}
		\label{sec:appendix_global_frame}
		To simplify the calculation of the global contact reference frame $({\nAxis{1}},{\uAxis{1}},{\wAxis{1}})$, an intermediate configuration $(\hat{\mathbf{n}}_1, \hat{\mathbf{u}}_1, \hat{\mathbf{w}}_1)$ is created such that 
		\begin{equation*}
		\hat{\mathbf{n}}_1 = \mathbf{R}\nAxis{1} = [0,0,1]^T, \quad \hat{\mathbf{u}}_1 = \mathbf{R}\uAxis{1}, \quad \hat{\mathbf{w}}_1 = \mathbf{R}\wAxis{1},
		\end{equation*}
		where $\mathbf{R}$ is the transformation matrix. Since $\hat{\mathbf{u}}_1$ and $\hat{\mathbf{w}}_1$ are orthogonal and perpendicular to $\hat{\mathbf{n}}_1$, one can assume that $\hat{\mathbf{u}}_1 = [\sin \theta, \cos \theta, 0]^T$, and $\hat{\mathbf{w}}_1 = [-\cos \theta, \sin \theta, 0]^T$. Let $\uAxTwo{0}{i,1} = [a_x, a_y, a_z]$ and $\wAxTwo{0}{i,1} = [b_x, b_y, b_z]$, then the cost function can be formulated as, 
		\begin{equation*}
		f(\theta) = \uAxTwo{0}{i,1} \cdot \hat{\mathbf{u}}_1 + \wAxTwo{0}{i,1}\cdot \hat{\mathbf{w}}_1 = (a_x + b_y) \sin \theta + (a_y - b_x) \cos \theta,
		\end{equation*}
		which yields a unique global minima/maxima in the range of $[0, 2\pi]$. The global maxima is reached when the gradient of the cost function is zero, and the solution is evaluated as follows,
		\begin{equation*}
		\theta^{\star} = \begin{cases}
		\pi/2 \quad \mbox{or} \quad 3\pi/2, & \mbox{if }  a_y - b_x = 0, \\
		\tan^{-1} \frac{a_x + b_y}{a_y - b_x} \quad \mbox{or} \quad \tan^{-1} \frac{a_x + b_y}{a_y - b_x} + \pi & \mbox{if } a_y - b_x \neq 0,
		\end{cases}
		\end{equation*}
		whichever gives a larger $f(\theta^{\star})$. Now the global tangential contact frame at current time step can be written as $\mathbf{u}_1 = \mathbf{R}^{T} \hat{\mathbf{u}}_1$ and $\mathbf{w}_1 = \mathbf{R}^{T} \hat{\mathbf{w}}_1$, with $\hat{\mathbf{u}}_1 = [\sin \theta^{\star}, \cos \theta^{\star}, 0]^T$, and $\hat{\mathbf{w}}_1 = [-\cos \theta^{\star}, \sin \theta^{\star}, 0]^T$.
		\begin{figure}[!ht]
			\centering
			\includegraphics[width=2.5in]{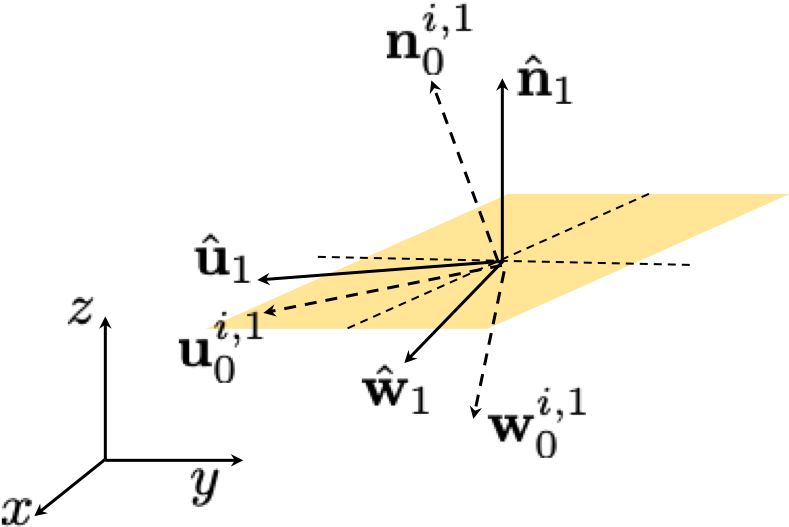}
			\caption{Evaluate the global contact frame at current time step.}
			\label{fig:globa_frame_opt}
			\vspace{-15pt}
		\end{figure}

		\FloatBarrier
		
		%%%%%%%
		
		\section{Derivation of Spinning Stiffness $K_{\psi}$ Based on Hertzian Contact Theory }
		\label{sec:appendix_spinning_kappa}
\begin{figure}[!ht]
			\centering
			\includegraphics[width=1.8in]{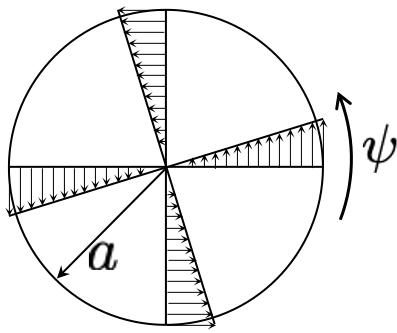}
			\caption{Spinning rotation of the contact area and the distribution of micro-sliding.}
			\label{fig:spinning_surface}
			\vspace{-15pt}
		\end{figure}

		In Hertzian contact theory, when two spheres touch, the contact area is a 2D circular shape with a radius of $a = \sqrt{R_{eff}\delta_n}$, where $R_{eff}$ is the effective radius and $\delta_n$ is the normal penetration, see Eqn.~(\ref{eq:defK_t}). When micro-deformation in slide-mode occurs, assuming it takes on the same value $x$ at any point of the contact area, the potential energy per unit area is $\frac{1}{2} K_E x^2$. When micro-deformation of angle $\psi$ occurs in spin-mode between two surfaces, the micro-deformation is distributed linearly along the radius $a$, with zero at the center of the contact patch, and $a \psi$ at the outer edge of the contact area, see Fig.~\ref{fig:spinning_surface}.
		
		For a surface element $r d\theta dr$ located at a distance of $r$ from the contact patch center, the energy due to a micro-deformation of $r \psi$ is $\frac{1}{2} K_E \left(r \psi \right)^2$. Therefore, the total energy over the contact patch due to spinning can be evaluated as
		\begin{align}
		\int_{0}^{2 \pi} \frac{1}{\pi a^2}  \int_{0}^{a} \frac{1}{2} K_E \left(r \psi \right)^2  r dr d\theta = \frac{a^2}{4} K_E \psi ^2.
		\end{align} 
		The stiffness $K_{\psi}$ is then defined such that 
		\begin{equation*}
		\frac{1}{2} K_{\psi}\psi^2 = \frac{1}{4} a^2 K_E \psi ^2 \;.
		\end{equation*}
		Therefore, the spin stiffness is tied to the slide stiffness as
		\begin{equation}
		\label{eq:K_psi_K_E_relation}
		K_{\psi} = \frac{1}{2} a^2 K_E.
		\end{equation}
		When the micro-sliding reaches its threshold $\SlackSsTh$, at the outer edge of the contact area, the spin angle should also saturate, transitioning from static to kinetic spin mode. Therefore,
		\begin{equation}
		a \SlackPsisTh = \SlackSsTh, \quad a \SlackPsikTh = \SlackSkTh.
		\end{equation}
		The spinning torque threshold can be derived as
		\begin{equation*}
		M_{\psi}^{max} = K_{\psi} \SlackSsTh = \frac{a^2}{2}  K_E  \frac{\SlackSsTh} {a} = 0.5 a K_E \SlackSsTh = 0.5 a \mu_s F_n.
		\end{equation*}
		Compared with Eqn.(\ref{eq:gettingKpsi}), this mechanics-based model is identical to the empirical approach when the curvature $\mathcal{K} = 1/a$ and $\eta_{\psi}=0.5$.  In Hertzian elastic contact theory, the normal contact force $F_n$ is related to penetration through a physics-based stiffness $k_{Hz}$,
		\begin{equation}
		\label{eq:hertz_contact_fn_k}
		F_n = k_{Hz} \delta_n ^ {\frac{3}{2}}, \quad 	k_{Hz} = \frac{4}{3} E_{eff} \sqrt{R_{eff}},
		\end{equation}
		where $E_{eff} = E_i^{\star} E_j^{\star}/(E_i^{\star} + E_j^{\star})$, $E_{i,j}^{\star} = E_{i,j} / (1 - \nu_{i,j}^2)$, $E_{i,j}$ and $\nu_{i,j}$ being Young's modulus and Poisson ratio of sphere $i$ and $j$, respectively. Substituting Eqn.(\ref{eq:hertz_contact_fn_k}) into Eqn.(\ref{eq:K_psi_K_E_relation}), one can derived spinning stiffness as
		\begin{equation*}
		K_{\psi} = 0.5 R_{eff} \delta_n K_E = 0.5 R_{eff} (\frac{F_n}{k_{Hz}})^{\frac{2}{3}} K_E = 0.5  \left(  \frac{3 F_n R_{eff}}{4 E_{eff}} \right) ^{\frac{2}{3}} K_E.
		\end{equation*}
		%For a sphere spinning on the ground scenario, it comes to a stop sooner when it is larger, heavier, less stiff or has more friction among the contact surface. For example, a steel ball of $R = 0.02 m$, $\rho = 8 \times 10^3 kg/m^3$, $E = 200 \si{GPa}$, and $\nu = 0.3$ will have the following properties,  (\SBELfeedbackONE{Luning: change this into a table once simulations are done, need it here for double-checking the math, or move this to the numerical result part.})
		%\begin{align}
		%k_{Hz} &= \frac{4}{3} \frac{E}{1- \nu^2} \sqrt{R} = 4.14 \times 10^{10} \si{N/m}, \\
		%F_n &= m g = \rho V g = \rho \frac{4}{3} \pi R^3 g =  2.627 \si{N}, \\
		%\delta &= \left(\frac{F_n}{k_{Hz}}\right) ^{\frac{2}{3}} = 1.59 \times 10^{-7} \si{m}, \\
		%a &= \sqrt{Rd} = 5.64 \times 10^{-5} \si{m} \\
		%K_{\psi} &= 1.59 \times 10^{-9} K_E \\
		%M_{\psi}^{max} &= 0.5 a \mu_s F_n = 1.852 \times 10^{-5} \si{Nm} 
		%\end{align}
		\begin{figure}[!ht]
			\centering
			\includegraphics[width=2.5in]{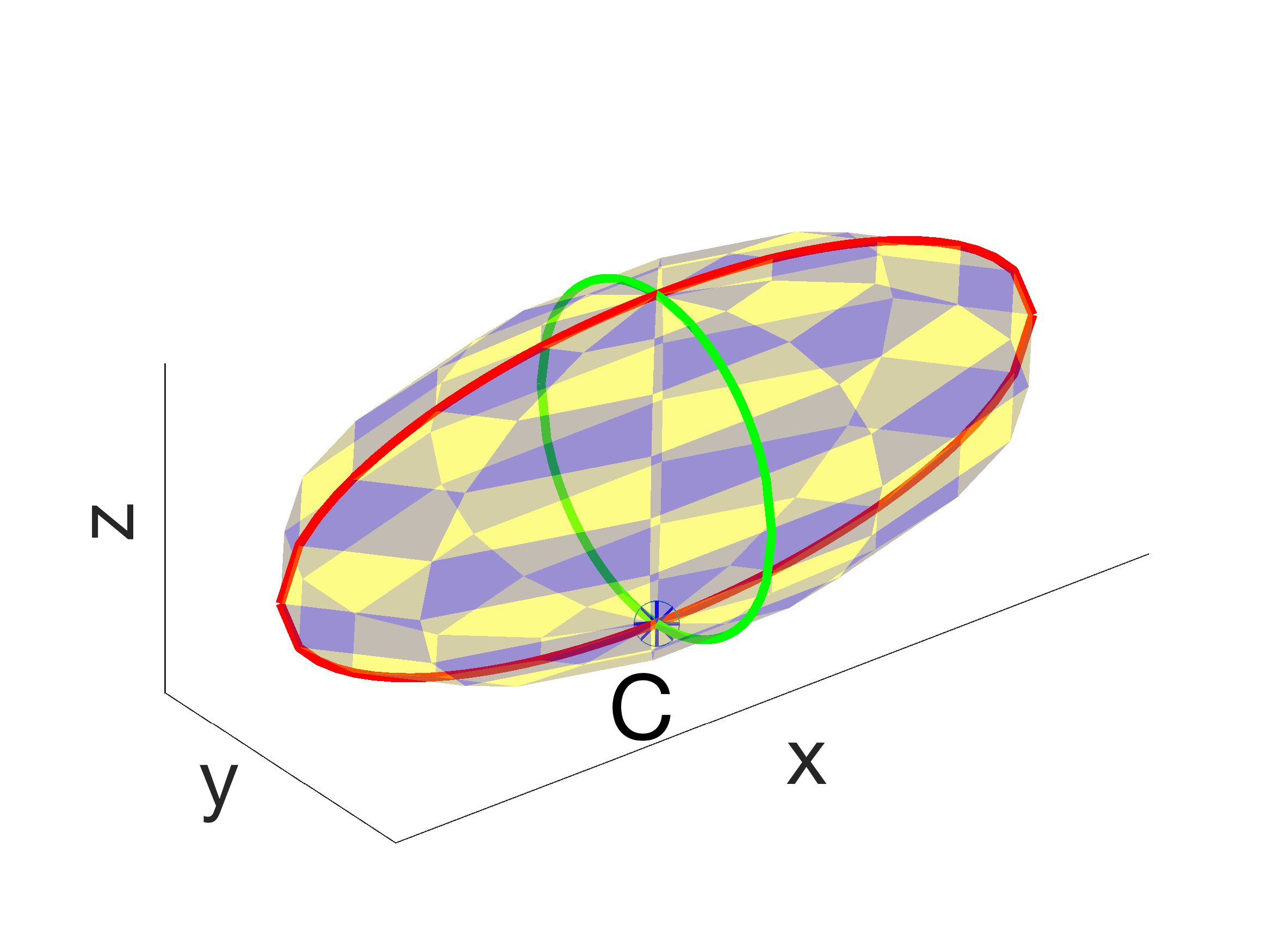}
			\caption{Spinning curvature $\mathcal{K}$ of an ellipsoid.}
			\label{fig:ellipsoid_spinning_curvature}
			\vspace{-15pt}
		\end{figure}
		
		For an ellipsoid, $\mathcal{K}$ is evaluated as the mean curvature at contact point $C$ of two contours in princial direction, \textit{i.e.}, $\mathcal{K} = (\mathcal{K}_C^{xz} + \mathcal{K}_C^{yz})/2$, see an example in Fig.~\ref{fig:ellipsoid_spinning_curvature}.
		\section{Analytical Solution of Sphere Rolling on an Incline}
		\label{sec:appendix_sphere_incline}
		For a sphere of radius $R$ and mass $m$ rolling on an incline of angle $\alpha$, the maximum static slide friction force and rolling frictin torque are evaluated as follows, 
		\begin{align*}
		F_r^{max} &= \mu_s N = \mu_s m g cos \alpha, \\
		T_r^{max} &= K_R \Theta_i = 2 \eta_r R \mu_s N = 2 \eta_r R \mu_s m g \cos \alpha.
		\end{align*}
		When the final steady-state switches from stationary to pure rolling, the slide friction remains static, balancing out the gravity down the slope, while the rolling friction torque reaches its maximum,
		\begin{equation*}
		F_r = m g \sin \alpha \leq F_r^{max}, \quad F_r R = m g \sin \alpha R > T_r^{max}.
		\end{equation*}
		This indicates that for the sphere to be stationary on the incline, $\alpha \leq \tan^{-1}(2 \eta_r \mu_s)$. \\
		As the final steady-state transitions to rolling with slipping mode, the slide friction force saturates, $F_r = \mu_s m g \cos \alpha$, while the rolling friction mode remains kinetic, $T_r = 2 \eta_r R \mu_k m g \cos \alpha$. The acceleration at the center of mass, $\ddot{x}$, and the angular acceleration, $\ddot{\theta}$, can be derived as
		\begin{equation*}
		\ddot{x} = (m g \sin \alpha - F_r^{max}) /m, \quad \ddot{\theta} = (R F_r - T_r) / (0.4 m R^2).
		\end{equation*}
		Substituting the accelerations into the kinematic constraint during pure rolling $\ddot{x} = R \ddot{\theta}$, one can derive the slope angle for sphere to roll down the incline without slipping as $\alpha \leq \tan^{-1} (3.5 \mu_s - 5 \eta_r \mu_k)$.\\
		When the steady-state of the sphere switches from rolling with slipping to sliding without rolling, the slide friction mode remains kinetic. Since no rolling occurs, $\ddot{\theta} = 0$, the rolling friction balances out with the moment generated from slide friction force, therefore, one can derive
		\begin{equation*}
		F_r = \mu_k m g \cos \alpha, \quad T_r =R F_r = R \mu_k m g \cos \alpha.
		\end{equation*}
		The rolling friction mode transitions from kinetic to static, indicating $T_r$ smaller than the kinetic rolling friction torque, $T_r < 2 \eta_r R \mu_k m g \cos \alpha$, which yields the condition for pure sliding $\eta_r \geq 0.5$.
	\end{appendices}

\end{document}